\documentclass[aps,floatfix,pra,10pt,twocolumn,nolongbibliography,showkeys]{revtex4-2}
\usepackage[utf8]{inputenc}
\usepackage[pdftex]{graphicx}
\usepackage{amsfonts}
\usepackage{amssymb}
\usepackage{amsmath}
\usepackage{mathtools}  
\usepackage{dsfont}
\usepackage{array}
\usepackage{physics}
\usepackage{bm} 
\usepackage{mathrsfs} 
\usepackage{pifont} 
\usepackage{multirow} 
\usepackage[dvipsnames]{xcolor}
\usepackage[pdftex,
            pdftitle={Gaussian boson sampling for binary optimization},
            pdfauthor={Jean Cazalis, Tirth Shah, Yahui Chai, Karl Jansen, Stefan Kuehn},
            bookmarks,
            colorlinks,
            linkcolor=myblue,
            citecolor=mymagenta,
            menucolor=black,
            urlcolor=myblue,
            plainpages=false,
            pdfpagelabels,
            hypertexnames=false]{hyperref}
\usepackage{verbatim} 
\usepackage{graphicx}
\usepackage[export]{adjustbox}
\usepackage{cleveref} 

\definecolor{mymagenta}{RGB}{200, 0, 100}
\definecolor{myblue}{RGB}{45, 48, 146}

\DeclareMathOperator{\Tor}{Tor}

\DeclareMathOperator*{\argmin}{arg\,min}

\newcommand{\enstq}[2]{\left\{#1\mathrel{} : \mathrel{}#2\right\}}

\newtheorem{example}{Example}

\begin{document}

\title{Gaussian boson sampling for binary optimization}

\author{Jean Cazalis}
\thanks{These authors contributed equally to this work.}
\affiliation{Q.ANT GmbH, Handwerkstraße 29, 70565 Stuttgart, Germany}
\affiliation{Deutsches Elektronen-Synchrotron DESY, Platanenallee 6, 15738 Zeuthen, Germany}

\author{Tirth Shah}
\thanks{These authors contributed equally to this work.}
\affiliation{Q.ANT GmbH, Handwerkstraße 29, 70565 Stuttgart, Germany}

\author{Yahui Chai}
\affiliation{Deutsches Elektronen-Synchrotron DESY, Platanenallee 6, 15738 Zeuthen, Germany}

\author{Karl Jansen}
\affiliation{Deutsches Elektronen-Synchrotron DESY, Platanenallee 6, 15738 Zeuthen, Germany}
\affiliation{Computation-Based Science and Technology Research Center, The Cyprus Institute, 20 Kavafi Street,
2121 Nicosia, Cyprus}

\author{Stefan Kühn}
\affiliation{Deutsches Elektronen-Synchrotron DESY, Platanenallee 6, 15738 Zeuthen, Germany}

\date{\today}

\begin{abstract}
    Binary optimization is a fundamental area in computational science, with wide-ranging applications from logistics to cryptography, where the tasks are often formulated as Quadratic or Polynomial Unconstrained Binary Optimization problems (QUBO/PUBO). In this work, we propose to use a parametrized Gaussian Boson Sampler (GBS) with threshold detectors to address such problems. We map general PUBO instance onto a quantum Hamiltonian and optimize the Conditional Value-at-Risk of its energy with respect to the GBS ansatz. In particular, we observe that, when the algorithm reduces to standard Variational Quantum Eigensolver, the cost function is analytical. Therefore, it can be computed efficiently, along with its gradient, for low-degree polynomials using only classical computing resources. Numerical experiments on 3-SAT and Graph Partitioning problems show significant performance gains over random guessing, providing a first proof of concept for our proposed approach.
\end{abstract}

\keywords{Gaussian Boson Sampling, Variational Quantum Eigensolver, Combinatorial Optimization}

\maketitle

\tableofcontents

\section{Introduction}

Quantum computing promises to offer computational capabilities superior to those of conventional computers. To unlock this potential, reach quantum utility~\cite{kim_evidence_2023} and ultimately achieve quantum advantage, it is not only required to make progress in the development of quantum hardware, but also in the creation of more efficient quantum algorithms~\cite{dalzell_quantum_2023}.

Combinatorial optimization is a key application area~\cite{abbas_quantum_2023}, alongside quantum simulation~\cite{bauer_quantum_2020}, quantum machine learning~\cite{schuld_machine_2021, gujju_quantum_2024} and cryptography~\cite{pirandola_advances_2020}. Many combinatorial optimization problems can be reformulated as polynomial unconstrained binary optimization (PUBO) problems, or quadratic unconstrained binary optimization (QUBO) problems, which is a special class of PUBO problems. The state-of-the-art quantum algorithms able to run on near-term quantum hardware and designed to address them fall within the scope of the Quantum Approximate Optimization Algorithm and the Variational Quantum Eigensolver (VQE) families~\cite{abbas_quantum_2023}. They both belong to the broader category of Variational Quantum Algorithms (VQA), which usually follow a common framework consisting of three building blocks~\cite{cerezo_variational_2021, bharti_noisy_2022}: a cost function to minimize, a parametrized quantum circuit (often referred to as the \emph{ansatz}) and a parameter optimization scheme. VQAs are generally implemented using hybrid algorithms that exploit the strengths of both quantum and classical computing. A quantum computer estimates the cost function or its gradient, after which classical optimization methods, such as gradient descent, are employed to iteratively update the circuit parameters to reduce the value of the cost function.

The efficiency of VQAs is closely linked to the synergy between its core components—the cost function, the ansatz, and the optimization scheme—and how these elements interact with the problem being addressed~\cite{cerezo_variational_2021}. The main challenges of the VQA approach concern efficiency, which is the ability to estimate the cost function (and gradients when necessary) with reasonably low resources, and trainability, which can be considerably hindered by barren plateaus~\cite{mcclean_barren_2018,arrasmith_equivalence_2022, Larocca2024}. These plateaus often result from noise, a lack of problem-specific tailoring, or the nature of the cost function itself~\cite{Cerezo2021,Wang2021,Ortiz2021,cerezo_variational_2021}.

Despite their potential, VQAs are mainly regarded as heuristic methods. This perception stems from the current inability to theoretically guarantee their success, which therefore requires empirical experiments to validate their effectiveness and identify the conditions under which they shine. It underlines the importance of developing new VQA approaches and in particular new types of ansatz, broadening the scope of problems that can be effectively addressed.

This work introduces a novel approach by employing, instead of a digital quantum computer, a parametrized Gaussian Boson Sampler (GBS) with threshold detectors as ansatz. We combine it with the Conditional Value-at-Risk Variational Quantum Eigensolver (CVaR-VQE)~\cite{barkoutsos_improving_2020} cost function. Then, we focus on applying this framework to general PUBO and QUBO problems, although it could be easily extended to address other classes of binary optimization problems.

A GBS is a continuous variable and non-universal quantum protocol, seen as a strong candidate for achieving quantum advantage in a near future~\cite{hamilton_gaussian_2017, kruse_detailed_2019, hangleiter_computational_2022}. It operates by preparing squeezed states, entangling them within a linear interferometer and measuring them with quantum photon detectors. The motivation for developing VQA approaches for GBS is twofold: it can process information much faster than other types of quantum computers, and it is also relatively easier to build and scale up with current technology. Experiments have shown that it is possible to make these devices on a large scale~\cite{zhong_experimental_2019,zhong_quantum_2020,zhong_phase-programmable_2021,madsen_quantum_2022,deng_gaussian_2023}, but there are still challenges to be overcome before they can surpass conventional computers~\cite{oh_classical_2023}.

The use of CVaR to solve binary problems in the context of VQA has been first proposed and tested in~\cite{barkoutsos_improving_2020}. Notably, the use of the CVaR-VQE cost function was then implemented to address the Flight-Gate Assignment problem in~\cite{chai_towards_2023} and subsequently in the conference paper~\cite{cazalis_gaussian_2023}, with the present paper presenting an extended version of the latter.

\begin{figure}[ht]
    \centering
    \includegraphics[width=0.8\columnwidth]{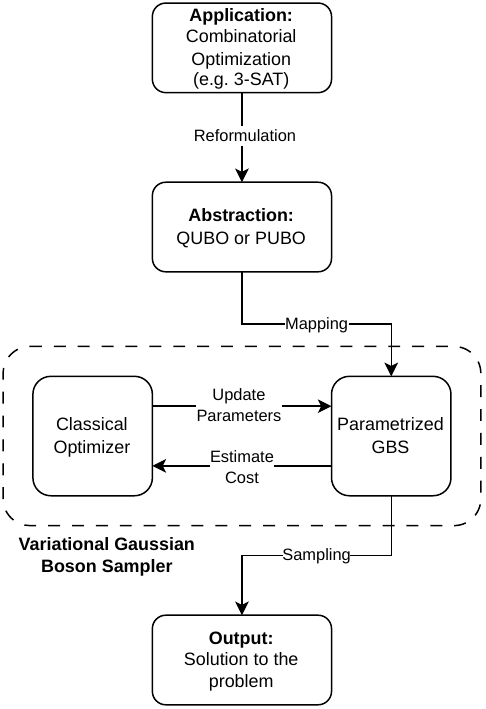}
    \caption{Diagram of the algorithm framework. A combinatorial optimization problem is reformulated as a QUBO/PUBO problem. It is then mapped to a variational GBS, comprising a parametrized GBS and a classical optimizer. Finally, several samples are collected before outputting the solution.}
    \label{fig:vgbs}
\end{figure}

Our goal is to create a flexible approach, improving on the ideas first proposed in~\cite{banchi_training_2020}. In particular, we employ a more general GBS ansatz and cost function, the latter being computed analytically in the case of VQE. We also provide two distinct parametrizations, making it easier to choose the parameters to be optimized according to the problem at hand. The diagram in Fig.~\ref{fig:vgbs} gives an overview of our approach and Fig.~\ref{fig:goldotter} provides more details on the algorithm itself.

The paper is structured as follows: Section~\ref{sec:gbs} and Section~\ref{sec:unconstrained_binary_optimization} give brief introductions to GBS and unconstrained binary optimization respectively. In Section~\ref{sec:vqe_cvar}, we describe the CVaR-VQE cost function and explain how to estimate it. Section~\ref{sec:mapping} presents the general framework of our approach, including the mapping of QUBO/PUBO problems to GBS. We also detail how to compute the cost function analytically in the VQE case. In Section~\ref{sec:simulations}, we share the results of our simulations, and we conclude in Section~\ref{sec:conclusions}. Additionally, we provide three appendix sections that give extra details on how to train our system more effectively, discuss the technical aspect of limited squeezing, and present supplementary numerical experiments.

\section{Gaussian boson sampling with threshold detectors}
\label{sec:gbs}

In this section, we describe the Gaussian Boson Sampling (GBS) protocol, which was introduced primarily to demonstrate quantum advantage~\cite{hamilton_gaussian_2017, kruse_detailed_2019, hangleiter_computational_2022}.

\begin{figure}[ht]
    \centering
    \includegraphics[width=\columnwidth]{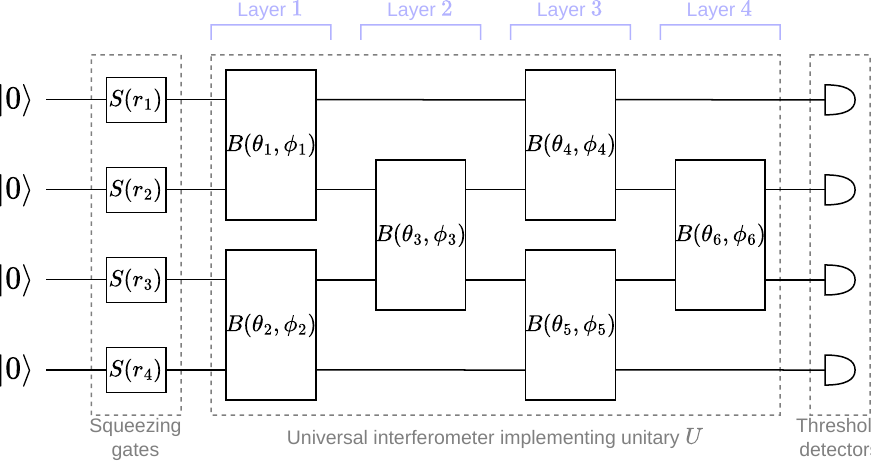}
    \caption{A schematic of representation of the GBS protocol considered in this paper. A universal interferometer, implemented with mesh of Mach-Zehnder interferometers using the Clements' layout~\cite{clements_optimal_2016} and implementing the unitary transformation $U$, is fed with squeezed vacuum states of light. Each interferometer is made of two balanced beamsplitters and two to four (depending on the implementation) phases shifters controlled by two phase shifts, $\theta$ and $\phi$. Finally, the state is measured using threshold detectors. Notice that for the interferometer to effectively implement any unitary $U$, the circuit must include local rotation gates prior to measurement. However, since the output distribution of the measurements is independent of these gates, we have opted to exclude them from the schematic.}
    \label{fig:gbs}
\end{figure}

The original scheme consists of preparing a pure $\ell$-modes Gaussian state without displacement then sample all its modes in the Fock basis, which can be achieved on a photonic platform by using Photon Number-Resolving (PNR) detectors~\cite{hamilton_gaussian_2017}. Another setup uses Threshold Detectors (TDs), which record a \emph{click} when one or more photons are detected and \emph{no click} otherwise~\cite{quesada_gaussian_2018, bulmer_threshold_2022}. A schematic of this protocol is presented on Fig.~\ref{fig:gbs}. These detectors are typically cheaper and more robust to noise than PNR detectors, which make them experimentally easier to handle. Furthermore, since they output \emph{click patterns} as binary strings, there exists a direct correspondence with binary variable assignments. This makes them convenient for addressing binary optimization problems such as QUBO or PUBO. For these reasons, we will only focus on the GBS with threshold detection in the following, even though our algorithm could be easily extended to other similar setups.

Despite its non-universality, the use of a GBS with TDs on photonic platforms has two advantages over other quantum hardware for quantum information processing. First, GBS with photonic platforms feature high processing speed, reaching several MHz of sampling rate~\cite{thekkadath_experimental_2022}. These regimes seem inaccessible for other quantum computing platforms such as those based on trapped ions, superconducting qubits or cold atoms~\footnote{Google Sycamore's repetition rate can reach a maximum of 15 kHz, which provides an order of magnitude for the upper limit of sampling rates for non-photonic quantum computers, since superconducting qubits have the shortest gate time in this category.}. Additionally, due to the relative ease of its experimental implementation, it is more likely to be available at a sufficient scale for practical use sooner than other platforms. Medium to large-scale experiments on optical tables have already been carried out~\cite{zhong_experimental_2019,zhong_quantum_2020,zhong_phase-programmable_2021,madsen_quantum_2022,deng_gaussian_2023}, but those are still too lossy to provide a substantial quantum advantage~\cite{oh_classical_2023}. Besides, scalable implementations on chips have already been demonstrated~\cite{bell_testing_2019, paesani_generation_2019, arrazola_quantum_2021, thekkadath_experimental_2022}.

In the following parts of this section, we describe the physical and computational properties of GBS in more detail.

\subsection{State preparation}
\label{sec:state_preparation}

The first step is to prepare a pure $\ell$-mode Gaussian state with no displacement. The class of Gaussian states can be efficiently represented by their first two statistical moments~\cite{weedbrook_gaussian_2012,serafini_quantum_2023}. Physically, the state is prepared from the vacuum by applying squeezing gates with parameters $\bm{r} = (r_1,\dots,r_\ell) \in \mathbb{R}_+^\ell$ followed by a universal passive interferometer, whose action on creation/annihilation operators $\hat{\bm{\zeta}} = (\hat{a}_1,\dots,\hat{a}_\ell,\hat{a}^{\dagger}_1,\dots,\hat{a}^{\dagger}_\ell)$ reads
\begin{align*}
    \hat{\bm{\zeta}} \longmapsto \begin{pmatrix}
                                     \bm{U} & 0 \\ 0 & \bm{U}^{*}
                                 \end{pmatrix} \hat{\bm{\zeta}} \, .
\end{align*}
Here $\bm{U}$ can be any $\ell\times \ell$ unitary and $\bm{U}^{*}$ denotes its element-wise complex conjugate. We gather all these circuit parameters in a vector
\begin{align*}
    \bm{\theta} = (\bm{r}, \bm{U}) \in \mathbb{R}^{\ell}_+ \times U_\ell(\mathbb{C}) \eqqcolon \mathcal{P}_C \, ,
\end{align*}
and denote by $\hat{\rho}_{\bm{\theta}}=\dyad*{\psi_{\bm{\theta}}}{\psi_{\bm{\theta}}}$ the output Gaussian state before measurement. In experiments, the universal multiport interferometer implementing $\bm{U}$ is decomposed into a mesh of beamsplitters and phase shifters~\cite{clements_optimal_2016}, as in Fig.~\ref{fig:gbs}. This parametrization will be referred later as the\emph{ Wigner parametrization}. We refer to~\cite{yao_design_2023} where a general framework for optimizing Gaussian states in $\mathcal{P}_C$ is proposed.

The squeezing parameters control the mean number of photons injected into the circuit
\begin{align}
    \label{eq:mean_photon}
    \overline{n} \coloneqq \sum_{j=1}^{\ell} \expval{\hat{N}_j}_{\bm{\theta}} = \sum_{j=1}^\ell \frac{\tanh(r_j)^{2}}{1 - \tanh(r_j)^{2}} \, ,
\end{align}
where $\hat{N}_j$ is the photon number operator in the mode $j$ and $\expval*{\hat{A}}_{\bm{\theta}}$ denotes the expectation value of an observable $\hat{A}$ in the quantum state $\hat{\rho}_{\bm{\theta}}$. Since there is no displacement, the output state is solely characterized by its complex Wigner covariance matrix $\bm{\sigma}_{\bm{\theta}}\in \mathbb{C}^{2\ell\times2\ell}$, defined by (hereafter, we set $\hbar=2$)
\begin{align*}
    (\bm{\sigma}_{\bm{\theta}})_{ij} = \frac{1}{2}\expval{\anticommutator{\Delta\hat{\zeta}_i}{\Delta\hat{\zeta}_j^\dagger}}_{\bm{\theta}} \, , \quad \forall i,j=1,\dots,2\ell\, ,
\end{align*}
where $\{\cdot,\cdot\}$ denotes the anticommutator, $\Delta \hat{\xi} \coloneqq \hat{\xi} - \expval*{\hat{\xi}}_{\bm{\theta}}$ for any observable $\hat{\xi}$, and $\hat{\zeta}_i$ runs through the ladder operators collected in $\hat{\bm{\zeta}}$. The Wigner covariance matrix is related to the Husimi covariance matrix $\bm{\Sigma}_{\bm{\theta}}\in \mathbb{C}^{2\ell\times2\ell}$ through
\begin{align*}
    \bm{\Sigma}_{\bm{\theta}} = \tfrac{1}{2}\bm{\sigma}_{\bm{\theta}} + \tfrac{1}{2} \mathbb{I}_{2\ell}\, ,
\end{align*}
where $\mathbb{I}_{2\ell}$ denotes the identity matrix of size $2\ell$. The covariance matrix $\bm{\Sigma}_{\bm{\theta}}$ characterizes entirely the $Q$ function of the system
\begin{multline*}
    Q_{\bm{\theta}}(\bm{\alpha}) \coloneqq \frac{1}{\pi^\ell} \qty|\innerproduct{\psi_{\bm{\theta}}}{\bm{\alpha}}|^2 \\
    = \frac{1}{\pi^\ell\sqrt{\det(\bm{\Sigma}_{\bm{\theta}})}} \exp \qty(-\tfrac{1}{2}\bm{\beta}\bm{\Sigma}_{\bm{\theta}}^{-1}\bm{\beta}^\dagger)\, ,
\end{multline*}
where $\ket{\bm{\alpha}}$ is the coherent state with complex displacement amplitudes $\bm{\alpha} \in \mathbb{C}^{\ell}$, and where $\bm{\beta} = (\bm{\alpha}, \bm{\alpha}^*)$.

\begin{figure}[ht]
    \centering
    \includegraphics[width=0.65\columnwidth]{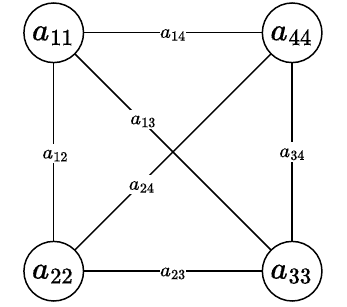}
    \caption{\textbf{Bargmann parametrization.} A pure Gaussian state without displacement can be represented by an undirected and complex-weighted graph that parametrizes its $Q$ function.}
    \label{fig:bargmann}
\end{figure}

The Husimi covariance matrix is related to the parameters $\bm{\theta}=(\bm{r, U}) \in \mathcal{P}_C$ as follows~\cite{hamilton_gaussian_2017, kruse_detailed_2019}:
\begin{multline}
    \label{eq:wigner_to_bargmann}
    \bm{\Sigma}_{\bm{\theta}} = \begin{pmatrix} \mathbb{I}_\ell & \bm{A}_{\bm{\theta}}^{*} \\ \bm{A}_{\bm{\theta}} & \mathbb{I}_\ell \end{pmatrix}^{-1} \\ 
    \qq*{with} \bm{A}_{\bm{\theta}} = \bm{U}
    \begin{pmatrix}
        \tanh(r_1) &        & 0      \\
            & \ddots &        \\
        0   &        & \tanh(r_\ell)
    \end{pmatrix}  \bm{U}^{T} \, .
\end{multline}
The symmetric matrix $\bm{A}_{\bm{\theta}}$ parametrizes the stellar function (or Bargmann function) of the quantum state~\cite{gagatsos_efficient_2019}. Since $\bm{A}_{\bm{\theta}}$ can be viewed as the adjacency matrix of complex-weighted graph (see Fig.~\ref{fig:bargmann}), it enables the embedding of several graph problems into a GBS~\cite{bromley_applications_2020}. We refer to \emph{Bargmann parametrization} when it is the coefficients of $\bm{A}_{\bm{\theta}}$ that are optimized.

The Wigner and the Bargmann parametrizations are equivalent in the sense there is a one-to-one correspondence between them, provided by the right side of Eq.~\eqref{eq:wigner_to_bargmann}:
\begin{align*}
    \mathcal{P}_C \simeq \mathcal{P}_B \coloneqq \qty{\bm{A} = \bm{A}^{T} \in \mathcal{M}_\ell(\mathbb{C}) ~|~\norm*{\bm{A}} < 1 } \, ,
\end{align*}
where $\norm*{\bm{A}}$ denotes the maximum singular value of the matrix $\bm{A}$. Given $\bm{A}$, one can retrieve $\bm{\theta}=(\bm{r, U}) \in \mathcal{P}_C$ using the Takagi-Autonne decomposition~\cite{houde_matrix_2024}.

However, depending on the problem to address, it may be more advantageous to employ either one or the other parametrization. For example, in Wigner parametrization, the number of layers (see Fig.~\ref{fig:gbs}) making up the interferometer can be directly truncated, thus reducing loss effects. The Bargmann parametrization, on the other hand, is better suited to representing the structure of graph problems in certain cases (see numerical results for $\alpha=1$ in Section~\ref{sec:results_comments}).

In the following, we will use the subscript $\bm{\theta}$ to denote the ansatz parameters, regardless of the underlying representation of the object in question, which should be evident from the context.

\subsection{Sampling with threshold detectors}
\label{sec:gbs_sampling}

After being prepared, the photons are sent to the detectors. The outcome measurements are described by the measurement operators
\begin{align}
    \label{eq:measurement_operators}
    \hat{\Pi}^{(0)}_j = \ket{0_j}\bra{0_j} \quad \text{and} \quad \hat{\Pi}^{(1)}_j = \hat{\mathbb{I}} - \hat{\Pi}^{(0)}_j\, ,
\end{align}
where $\hat{\mathbb{I}}$ and $\ket{0_j}$ are respectively the identity operator and the vacuum on the $j^{\mathrm{th}}$ optical mode~\cite{bulmer_threshold_2022}.

In the sequel, it will be useful to introduce the following notation: for a matrix
\begin{align*}
    \bm{A} =
    \begin{pmatrix}
        \bm{A}_1 & \bm{A}_2 \\ \bm{A}_3 & \bm{A}_4
    \end{pmatrix}\in \mathbb{C}^{2n\times 2n}
\end{align*}
with $n\times n$ blocks $\{\bm{A}_1,\bm{A}_2,\bm{A}_3,\bm{A}_4\}$, and for $J\in \mathcal{J}_n$ where $\mathcal{J}_n$ is the power set of $\{1,\dots,n\}$, we define the matrix $\bm{A}_{(J)}$ from $\bm{A}$ by keeping for each block $\bm{A}_j$ the rows and columns whose indices are in $J$.

The probability to observe the click pattern $\mathbf{x} = (x_1,\dots,x_\ell)\in \{0,1\}^\ell$ stems from the inclusion-exclusion principle~\cite{bulmer_threshold_2022} and reads
\begin{align}
    \label{eq:proba_td}
    \mathbb{P}(\mathbf{x}; \hat{\rho}_{\bm{\theta}}) = \tr\qty[\qty(\bigotimes_{j\in J_{\mathbf{x}}} \hat{\Pi}_j^{(1)} \bigotimes_{j\notin J_{\mathbf{x}}} \hat{\Pi}_j^{(0)}) \hat{\rho}_{\bm{\theta}}]= \frac{\Tor(\bm{O}_{(J_{\mathbf{x}})})}{\sqrt{\det(\bm{\Sigma}_{\bm{\theta}})}}  \, ,
\end{align}
where
\begin{align}
    \label{eq:proba_td_notations}
    J_{\mathbf{x}} \coloneqq \{j\in [\ell] ~|~x_j=1\} \, , \quad
    \bm{O} \coloneqq \mathbb{I}_{2\ell} - \bm{\Sigma}_{\bm{\theta}}^{-1}\in \mathbb{C}^{2\ell\times2\ell} \, ,
\end{align}
and where $\Tor(\bm{A})$ denotes the \emph{Torontonian} of a matrix $\bm{A}\in \mathbb{C}^{2\ell \times 2\ell}$. It is defined as~\cite[Eq. (12)]{quesada_gaussian_2018}
\begin{align*}
    \Tor(\bm{A}) \coloneqq \sum_{J\in \mathcal{J}_\ell} (-1)^{\abs{J}} \frac{1}{\sqrt{\det \qty(\mathbb{I}_{2\abs{J}} - \bm{A}_{(J)})}} \, .
\end{align*}
Computing the Torontonian of a $k$-clicks event exactly involves the computation of $2^k$ determinants. The state-of-the-art algorithms for this task scale exponentially with $k$ in the worst-case~\cite{bulmer_boundary_2022,kaposi_polynomial_2022}.

Sampling patterns according to the distribution probability defined by Eq.~\eqref{eq:proba_td} is closely related to the GBS original scheme with PNR detectors in the no-collision regime~\cite{quesada_gaussian_2018}, which is thought to be classically intractable in the worst-case~\cite{hamilton_gaussian_2017,hangleiter_computational_2022}. Indeed, the existence of a polytime classical algorithm would imply the collapse of the polynomial hierarchy to its third level, which would have daunting consequences in complexity theory. Under additional conjectures, the same conclusion would hold for approximate sampling, up to additive error~\cite{hangleiter_computational_2022}. For GBS under a general local noise model, it has been shown that worst-case sampling should also be classically difficult~\cite{deshpande_quantum_2022}.

\section{Unconstrained binary optimization}
\label{sec:unconstrained_binary_optimization}

Combinatorial optimization (CO) involves optimizing a function in a discrete space~\cite{papadimitriou_combinatorial_1982, nemhauser_integer_1988, korte_combinatorial_2018}. This field has become highly critical in various industrial sectors such as logistics, finance, energy, and cryptography where most problems can be formulated within this framework. In many cases, finding the optimal solutions is known to be NP-hard~\cite{korte_combinatorial_2018}.

Special cases of CO problems are Polynomial Unconstrained Binary Optimization (PUBO) and Quadratic Unconstrained Binary Optimization (QUBO) problems where a multivariate polynomial is minimized over a space of binary variable assignments. These abstractions encompass many relevant CO problems~\cite{kochenberger_unconstrained_2014, glover_quantum_2022}, although a polynomial overhead in the problem size may be paid during reformulation. Due to their importance, QUBO/PUBO problems have been widely studied in the literature and numerous classical metaheuristics to address them have been proposed, see for instance~\cite{kochenberger_unconstrained_2014} and the references therein.

Furthermore, QUBO problems are known to be equivalent to classical Ising models. We refer to~\cite{lucas_ising_2014} where explicit constructions of numerous CO problems as Ising problems are conducted.

In the context of quantum algorithms, PUBO formulations are particularly relevant. Indeed, although any PUBO can be reduced to a QUBO, the reduction procedure requires several ancillary variables~\cite{babbush_resource_2013}. However, in many quantum algorithms proposed in the literature, there is a correspondence between binary variables and quantum resources, such as qubits or qumodes. Given the current limited availability of these resources, it may be crucial to avoid such an overhead in certain situations~\cite{stein_evidence_2023, salehi_unconstrained_2022, tabi_quantum_2020, chai_towards_2023,chai_simulating_2023}.

In more precise terms, a PUBO instance is determined by a set of $\ell$ binary variables and a multivariate polynomial $H$ with degree at most $k\leq \ell$:
\begin{align}
    \label{eq:pubo_hamiltonian}
    H(x_1,\dots,x_\ell) = \sum\limits_{J\in\mathcal{J}_\ell} h_J x_J \, ,
\end{align}
where $\mathcal{J}_\ell$ is the power set of $\{1,\dots,\ell\}$, for all $J\in\mathcal{J}_\ell$ we have defined $x_J = \prod_{j\in J} x_j$ and where the degree $k$ is the maximal order of a single term
\begin{align*}
    k = \max \enstq{\abs{J}}{ x_J \neq 0} \, .
\end{align*}
A QUBO instance is a special case of PUBO where the polynomial is quadratic, that is $k=2$. Solving a PUBO instance involves finding the binary variable assignments which minimize the PUBO energy:
\begin{align}
    \label{eq:pubo_problem}
    \argmin_{(x_1,\dots,x_\ell) \in \{0,1\}^\ell} H(x_1,\dots,x_\ell) \, .
\end{align}
For the sake of brevity, we will also denote $H(\mathbf{x}) = H(x_1,\dots,x_\ell)$ for a click pattern $\mathbf{x} = (x_1,\dots,x_\ell)$.

Two examples of QUBO/PUBO problems are described below. They will be used later as application examples of the algorithm described in this paper.

\begin{example}[Graph Partitioning]
    \label{ex:graph_partitioning}
    A specific instance of the Graph Partitioning problem is, given an undirected graph $G=(V,E)$, to partition its vertices into two subsets $V_{0}$ and $ V_{1}$ of the same size such that the number of edges between these two subsets is minimal, see Fig.~\ref{fig:graph_partitioning}. This problem is known to be NP-hard~\cite{buluc_recent_2016}. Graph Partition has many interesting applications, including clustering and finding cliques in social networks.

    \begin{figure}[ht]
        \centering
        \includegraphics[width=0.8\columnwidth]{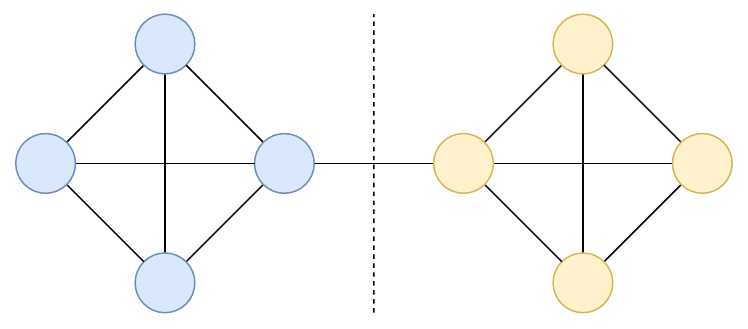}
        \caption{\textbf{Graph Partitioning.} The graph is partitioned into two sub-graphs, represented in this diagram by the sets of blue and yellow vertices respectively, of equal size with the smallest number of edges between the two partitions.}
        \label{fig:graph_partitioning}
    \end{figure}

    Such problem can be recast as a QUBO without any ancillary variable. We denote by $V=(1,\dots,\ell)$ the vertices of $G$ and place on each of them a binary variable $x_{v}\in\{0,1\}$, interpreted as follows:
    \begin{align*}
        x_{v}= j \in \{0,1\} \Longleftrightarrow v\in V_{j} \, .
    \end{align*}
    Then, for $c_1, c_2 >0$, the minimizers of the following QUBO energy
    \begin{gather*}
        H = c_1 H_1 + c_2 H_2 \, , \\
        H_1(x_1,\dots,x_\ell) = \left( \ell/2 - \sum\limits_{v=1}^{\ell} x_{v}\right)^{2}\, ,\\
        H_2(x_1,\dots,x_\ell) = \sum\limits_{(u,v)\in E} (x_{u}+x_{v} - 2x_{u}x_{v}) \, ,
    \end{gather*}
    encode the solutions to the initial Graph Partitioning problem. The term $H_1$ penalizes unbalanced partitions while $H_2$ accounts for the number of edges between the two partitions. For details on selecting the ratio $c_1/c_2$, we refer the reader to~\cite{lucas_ising_2014}.
\end{example}

\begin{example}[3-SAT problem]
    \label{ex:3sat}
    A constrained assignment problem can be represented by a Boolean formula, whose truth interpretation determines whether the problem is feasible or not, i.e. whether there is an assignment that satisfies all the constraints. This problem is known as the Boolean Satisfiability Problem, also called SAT.

    A Boolean formula can always be written in conjunctive normal form~\cite{howson_logic_1997}, i.e.\ as a conjunction of clauses, formed as the disjunction of variables $x$ or their negation ¬$x$. When all clauses have a fixed length $k$, the SAT problem is named $k$-SAT problem, which is known to be NP-hard for $k\geq 3$~\cite{karp_reducibility_2010}.

    We will consider the 3-SAT problem whose inputs are Boolean formulas of the form
    \begin{multline*}
        \phi = \bigwedge_{k=1}^{M} (\mathtt{l}_{k_1}\vee \mathtt{l}_{k_{2}} \vee \mathtt{l}_{k_{3}}) \\
        = (\mathtt{x}_1\vee \mathtt{x}_2 \vee \mathtt{x}_3)\wedge(\mathtt{x}_1\vee\neg \mathtt{x}_4\vee \mathtt{x}_5)\wedge\dots\, ,
    \end{multline*}
    where we have denoted by $M$ the number of clauses, by $\{\mathtt{x}_1,\dots,\mathtt{x}_\ell\}$ the set of variables, and by $\mathtt{l}_k \in \{\mathtt{x}_k,\neg \mathtt{x}_k\}$ the associated literals. Given a binary variable assignment $(x_1,\dots,x_\ell)$, the number of unsatisfied clauses is computed by the following 3-degree polynomial
    \begin{align*}
        H(x_{1},\dots, x_{\ell}) = \sum\limits_{k=1}^{M} (1 - l_{k_{1}})(1 - l_{k_{2}})(1 - l_{k_{3}})\, ,
    \end{align*}
    where
    \begin{align*}
        l_{k_{j}} = \begin{cases}
                        x_{k_{j}}     & \text{if } l_{k_{j}}=x_{k_{j}} \, ,      \\
                        1 - x_{k_{j}} & \text{if } l_{k_{j}}=\neg x_{k_{j}} \, .
                    \end{cases}
    \end{align*}
    Hence, solving the 3-SAT problem with the input formula $\phi$ amounts to solving the PUBO problem for $H$, where the minimum value is 0 if and only if the formula is satisfiable.
    
\end{example}

\section{VQE using the Conditional Value-at-Risk}
\label{sec:vqe_cvar}

The \emph{Variational Quantum Eigensolver} (VQE) is one of the most widespread approaches in the Noisy Intermediate-Scale Quantum era~\cite{cerezo_variational_2021, tilly_variational_2022}. The goal is, given Hamiltonian $\hat{H}$, to approximate its ground state and its ground state energy. To this aim, an ansatz state $\bm{\theta}\in\mathbb{R}^p \mapsto \hat{\rho}_{\bm{\theta}} = \dyad{\psi_{\bm{\theta}}}$ is prepared using a parametrized quantum circuit. The Hamiltonian expectation value in this state
\begin{align}
    \label{eq:vqe_cost_function}
    \mathcal{C}_1(\bm{\theta})=\ev{\hat{H}}{\psi_{\bm{\theta}}}
\end{align}
is then minimized with respect to $\bm{\theta}$ using a classical optimizer. Usually, the quantum device is used to estimate terms in the quantum energy and a classical computer adds them together before updating the circuit parameters. Common strategies rely on gradient-based optimization routines. There, the gradients may be estimated with the quantum computer, using for instance the parameter-shift rule~\cite{mitarai_quantum_2018,schuld_evaluating_2019}. VQE has been primarily utilized for quantum simulation in quantum chemistry or material science~\cite{cao_quantum_2019}. More recently, VQE has been proposed to address CO problems~\cite{nannicini_performance_2019}, although feeble performance was observed.

To improve the efficiency, it was argued in~\cite{barkoutsos_improving_2020} that employing a more general cost function, namely the \emph{Conditional Value-at-Risk} (CVaR), would be beneficial. The CVaR cost function reads
\begin{align}
    \label{eq:cvar_cost_function}
    \mathcal{C}_\alpha(\bm{\theta}) = \mathrm{CVaR}_\alpha(Y(\hat{H},\bm{\theta})), \quad \alpha\in (0,1] \, ,
\end{align}
where $Y(\hat{H},\bm{\theta})$ is the distribution of the observable $\hat{H}$ in the quantum state $\hat{\rho}_{\bm{\theta}}$, and
\begin{align*}
    \mathrm{CVaR}_\alpha(Y) = \mathbb{E}[Y~|~Y \leq F_Y^{-1}(\alpha)]
\end{align*}
denotes the CVaR with $\alpha$-left tail of a random variable $Y$, and $F_Y$ is the cumulative density function of $Y$. Following~\cite{barkoutsos_improving_2020}, we call this approach CVaR-VQE. Notice that when $\alpha=1$, we retrieve the standard VQE cost function Eq.~\eqref{eq:vqe_cost_function}. In the limit $\alpha\to 0$, the CVaR cost function corresponds to the minimum possible outcome of the observable $\hat{H}$ in the quantum state $\hat{\rho}_{\bm{\theta}}$.

In the experiments, the cost function can be estimated by performing $K$ measurements of $\hat{\rho}_{\bm{\theta}}$, computing the corresponding energies $E_1\leq\dots\leq E_K$ and averaging over the $\lceil\alpha K \rceil$ the lowest values
\begin{align*}
    \mathcal{C}_\alpha(\bm{\theta}) \simeq \frac{1}{\lceil\alpha K \rceil}\sum_{k=1}^{\lceil\alpha K \rceil} E_k \, .
\end{align*}
To understand why the use of the CVaR cost function should be advantageous, one can observe that, contrary to quantum physics, Hamiltonians for CO problems are generally diagonal in the computational basis. Consequently, the ground states can be written as single computational basis states, that can be directly sampled by the quantum circuit. Hence, it should be more efficient to train the model on the lowest energy states only, which is what CVaR does. However, in the limit case $\alpha\to 0$, the cost function typically lacks smoothness, making the optimization hard to perform. In this regard, the hyperparameter $\alpha$ smoothens the cost landscape. However, it is important to note that the CVaR cost function does not yield any reward for increasing the fidelity with the ground state beyond $\alpha$. Thus, in practice, one chooses a fixed value of $\alpha$ that  is small but still reasonably large to be able to reveal the solution with a realistic number of measurements~\cite{barkoutsos_improving_2020}.

A systematic numerical investigation on VQE-based approaches for CO problems conducted in~\cite{diez-valle_quantum_2021} supports the utilization of CVaR. Enhancements are also observed in~\cite{chai_towards_2023} where CVaR is used to address randomly generated Flight-Gate Assignment problem instances. See also~\cite{chai_simulating_2023} where the same approach is carried out on a trapped ion quantum computer.

In the following, we consider $\alpha\in (0,1]$, but treat the case $\alpha=1$ (standard VQE) separately.

\section{Mapping QUBO/PUBO problem to GBS and optimization methods}
\label{sec:mapping}

In this section, we present the general framework of our algorithm and describe its main subroutines in more detail.

\begin{figure}[ht]
    \centering
    \includegraphics[width=\columnwidth]{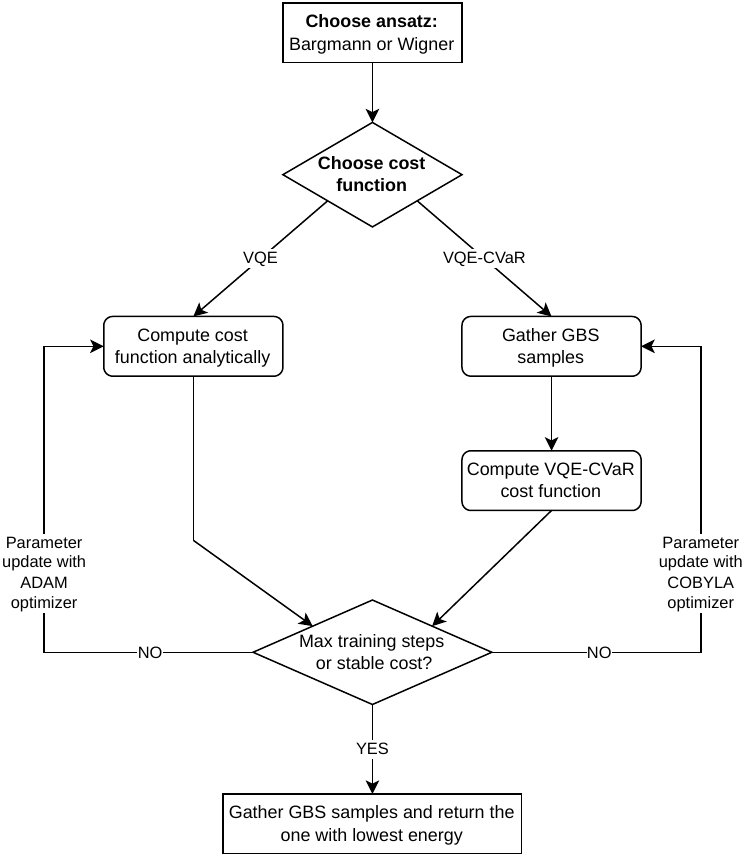}
    \caption{Flowchart of the inner workings of the variational GBS algorithm introduced in this paper. The parameter update and cost function evaluation subroutines depend on the choice of cost function and of the ansatz parametrization. Once the training has been performed, several samples are collected and the algorithm outputs the one with the lowest energy.}
    \label{fig:goldotter}
\end{figure}

Our algorithm consists of the CVaR-VQE approach described in Section~\ref{sec:vqe_cvar} where the ansatz is a GBS, parametrized either in the Wigner parametrization with $\bm{\theta} \in \mathcal{P}_C$ or in the Bargmann parametrization with $\bm{A}_{\bm{\theta}} \in \mathcal{P}_B$ (see Section~\ref{sec:state_preparation}). The algorithm cost function is parametrized by $\alpha\in (0,1]$ which determines its type: conventional VQE for $\alpha=1$ and CVaR-VQE for $\alpha <1$. The diagram in Figure~\ref{fig:vgbs} gives an overview of our algorithm framework and the flowchart in Figure~\ref{fig:goldotter} describes its inner structure.

\subsection{Mapping}

To address the QUBO/PUBO problem in Eq.~\eqref{eq:pubo_problem} with the CVaR-VQE approach described in Sec.~\ref{sec:vqe_cvar}, the original QUBO/PUBO Hamiltonian in Eq.~\eqref{eq:pubo_hamiltonian} must be promoted to a quantum observable $\hat{H}$. We encode $H$ into a diagonal Hamiltonian
\begin{align*}
    \hat{H} = \sum_{\mathbf{x}} H(\mathbf{x}) \hat{P}_{\mathbf{x}} \, ,
\end{align*}
where $\sum_{\mathbf{x}} \hat{P}_{\mathbf{x}} = \hat{\mathbb{I}}$ is such that
\begin{align*}
    \tr(\hat{H} \rho_{\bm{\theta}})= \sum_{\mathbf{x}} H(\mathbf{x}) \tr(\hat{P}_{\mathbf{x}} \rho_{\bm{\theta}})  = \sum_{\mathbf{x}} H(\mathbf{x}) \mathbb{P}(\mathbf{x},\rho_{\bm{\theta}}) \, ,
\end{align*}
where $\mathbb{P}(\mathbf{x},\rho_{\bm{\theta}})$ is given by Eq.~\eqref{eq:proba_td}. Therefore, we require that for any multivariate polynomial $H$ of the form~\eqref{eq:pubo_hamiltonian} the quantum expectation value of $\hat{H}$ in the state $\rho_{\bm{\theta}}$ is equal to the classical expectation of $H$ with respect to output distribution of the GBS. This directly implies that
\begin{align*}
    \hat{P}_{\mathbf{x}} = \bigotimes_{j \in J_{\mathbf{x}}} \hat{\Pi}^{(1)}_j \bigotimes_{j \notin J_{\mathbf{x}}} \hat{\Pi}^{(0)}_j \, ,
\end{align*}
where $\hat{\Pi}^{(1/0)}_j$ are the projection operators defined in Eq.~\eqref{eq:measurement_operators} and the set $J_{\mathbf{x}}$ is defined in Eq.~\eqref{eq:proba_td_notations}. In addition, by taking $H=x_J$, we compute
\begin{multline*}
    \hat{H} = \sum_{\enstq{\mathbf{x}}{x_j = 1, \forall j\in J}} \hat{P}_{\mathbf{x}} \\
    = \bigotimes_{j\in J} \hat{\Pi}^{(1)}_j \bigotimes_{j \notin J}  \qty(\hat{\Pi}^{(1)}_j + \hat{\Pi}^{(0)}_j)
    = \bigotimes_{j\in J} \hat{\Pi}^{(1)}_j \, .
\end{multline*}
From this, we deduce the promotion rule
\begin{align*}
    x_J \longmapsto \hat{x}_J = \bigotimes_{j\in J} \hat{\Pi}^{(1)}_j \, ,
\end{align*}
from which one can straightforwardly promote any $H$.

\subsection{The VQE case}
\label{sec:vqe_case}

In this section, we study more precisely the case $\alpha=1$, corresponding to the standard VQE cost function.

Our key observation is that the cost function $\mathcal{C}_1(\bm{\theta})$ admits an \emph{analytical form} in this setting. We note it was implicitly pointed out in Refs.~\cite{conti_training_2021,conti_variational_2022} for PNR detectors, and that it still holds for TDs. The important consequence is that, in this case, training can be carried out efficiently on a classical machine before multiple samples are generated by the GBS to recover the solution, see Fig.~\ref{fig:goldotter}. Although sampling is still necessary, this procedure requires far fewer quantum resources (quantified by the number of samples) than standard VQE. As a result, it is not clear whether the case $\alpha=0$ could bring any quantum advantage over classical methods.

In what follows, we detail this observation and its consequences. For $J\in \mathcal{J}_\ell$ a subset of binary variables (or a click pattern), we denote by $\hat{\rho}_{\bm{\theta}}^{J}$ the quantum state remaining after tracing out the modes \emph{not} in $J$. Its Husimi covariance matrix is simply given by $\bm{\Sigma}_{\bm{\theta}}^{J} \coloneqq (\bm{\Sigma}_{\bm{\theta}})_{(J)}$, where $\bm{\Sigma}_{\bm{\theta}}$ is the covariance matrix of $\hat{\rho}_{\bm{\theta}}$ and the subscript $(J)$ means we keep the rows and columns corresponding to $J$ (see Section~\ref{sec:gbs_sampling}). Then, we can write
\begin{multline}
    \label{eq:analytical_expression}
    \expval*{\hat{x}_J}_{\bm{\theta}}
    = \mathbb{P}(1,\dots,1; \hat{\rho}_{\bm{\theta}}^{J}) \\
    = \frac{\Tor\qty(\mathbb{I}_{2\abs*{J}} - (\bm{\Sigma}_{\bm{\theta}}^{J})^{-1})}{\sqrt{\det(\bm{\Sigma}^{J}_{\bm{\theta}})}} \\
    = \sum\limits_{J'\subset J} (-1)^{|J'|} \det(\bm{\Sigma}^{J'}_{\bm{\theta}})^{-1/2} \, ,
\end{multline}
where the second equality follows from Eq.~\eqref{eq:proba_td}, and the third is proven in Eq.~\eqref{eq:more_efficient_0} in Appendix~\ref{app:more_efficient}. Notice this expression is indeed analytical in the parameters $\bm{\theta}\in \mathcal{P}_C$ since $\Sigma_{\bm{\theta}}^{J} > \tfrac{1}{2}\mathbb{I}_{2\abs{J}}$ for any $J\in \mathcal{J}_\ell$. In view of Eq.~\eqref{eq:wigner_to_bargmann}, the same holds for the Bargmann parametrization $\bm{A}\in \mathcal{P}_B$.

Recall that the computational complexity of computing Eq.~\eqref{eq:analytical_expression} scales exponentially with the size of $J$ since computing the Torontonian involves the computation of $2^{\abs{J}}$ determinants. Nevertheless, it remains tractable for PUBO problems with slowly increasing degree (for instance for $k=\mathcal{O}(\ln(\ell))$). In particular, the cost for computing $\mathcal{C}_1(\bm{\theta})$ for QUBO problems is in the worst-case quadratic with the number of binary variables.

We deduce that for small values of $k$ the VQE case is perfectly efficient since the expectation values in the cost function can be quickly and accurately computed by classical methods. This observation also enables the use of mature automatic differentiation libraries, such as \emph{TensorFlow}~\cite{tensorflow2015-whitepaper}, to optimize the parameters of the quantum circuit. In Appendix~\ref{app:more_efficient}, we provide an expression which leverages batch processing to compute $\mathcal{C}_1(\bm{\theta})$ even more efficiently.

In the implementations, we make use of Adaptive Moment Estimation (ADAM), an extension of gradient descent optimization methods that combines aspects of both momentum and adaptive learning rate techniques~\cite{kingma_adam_2017}. ADAM adjusts the learning rate for each parameter individually and maintains moving averages of past gradients and squared gradients to adaptively scale the updates.

We notice that other variational boson sampling schemes trained on the standard VQE cost function have already been addressed in the literature.

Br\'{a}dler and Wallner have proposed to use a Fock Boson Sampler as ansatz instead of a GBS~\cite{bradler_certain_2021}. This scheme is similar to GBS, except that the input states are Fock states. As their generation is not deterministic, it seems difficult to scale up such a system. Indeed, the most successful current experiments~\cite{brod_photonic_2019} are one order of magnitude smaller than the largest GBS experiments~\cite{madsen_quantum_2022, deng_gaussian_2023}.

In~\cite{banchi_training_2020}, the authors proposed to use the so-called WAW parametrization to train GBS distributions, the advantage lying in the ability to express the gradient of a class of cost functions as expectation values over the GBS output distribution, enabling the use of efficient stochastic gradient methods for training the distribution. This class includes conventional VQE (but not CVaR-VQE for $\alpha<1$) and QUBO/PUBO problems. This work is the closest to the present paper. However, as mentioned previously, there is no need to statistically estimate the cost function value or its gradient in this case, since they admit analytical forms. Our proposal therefore significantly expands the current state-of-the-art.

\subsection{The CVaR-VQE case}

When $\alpha\in (0,1)$, the objective function cannot be expressed as an easily-computable differentiable functions, making traditional gradient-based methods impractical. Indeed, one would need to compute the entire cost landscape of $H$ and, in the worst case, the probability of all possible patterns, which is computationally out of reach. Standard gradient estimation schemes~\cite{mari_estimating_2021}, such as the parameter-shift rule, would need a substantial number of measurements to accurately estimate expectation values and gradients \cite{mitarai2018,schuld2019evaluating}. They are therefore not applicable either. 

Therefore, following Refs.~\cite{barkoutsos_improving_2020, chai_towards_2023}, we opt to use the Constrained Optimization BY Linear Approximations (COBYLA) method, a derivative-free optimization algorithm designed to solve constrained optimization problems~\cite{powell_direct_1994}.

\section{Simulation results}
\label{sec:simulations}

\begin{figure*}[htbp]
    \centering
    \includegraphics[width=\textwidth]{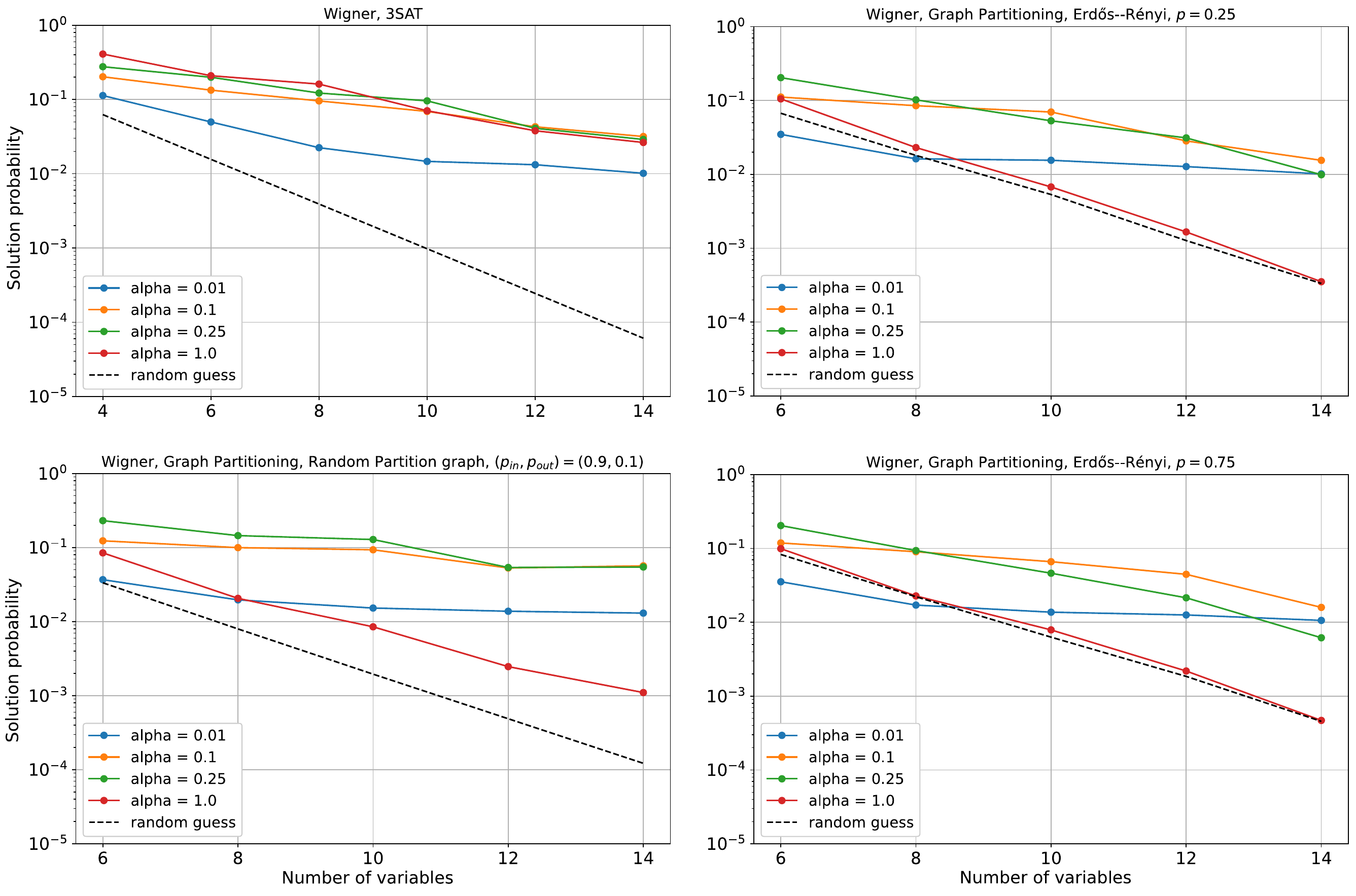}
    \caption{\textbf{Wigner parametrization.} Simulation results for the binary optimization problem solver presented in this article. Two types of optimization problems are considered: 3-SAT and Graph Partitioning, encoded as a PUBO and QUBO problem respectively. The Graph Partitioning instances are generated using three random graph generators: Erd\H{o}s--R\'enyi graphs with connection probabilities $p = 0.25$ or $p = 0.75$, and Random Partition Graphs with intra-cluster and inter-cluster probabilities $p_\mathrm{in} = 0.9$ and $p_\mathrm{out} = 0.1$ respectively. For a comprehensive description of the problem instances, we refer to Sec.~\ref{sec:problems_considered}. For each problem instance, and for each CVaR parameter $\alpha \in \{0.01, 0.1, 0.25, 1\}$, 50 (3-SAT case) or 100 (Graph Partitioning case) random instances are generated. The model is optimized using the Wigner parametrization methods, as detailed in Section~\ref{sec:methods}. The probability of sampling a solution, calculated according to Eq.~\eqref{eq:proba_success}, is averaged over the generated instances and plotted. The dashed line indicates the probability of randomly guessing a solution, averaged over the instances. Notice that some instances have several solutions. We use a logarithmic scale for $y$-axis.    
    }
    \label{fig:result_main_wigner}
\end{figure*}

\begin{figure*}[htbp]
    \centering
    \includegraphics[width=\textwidth]{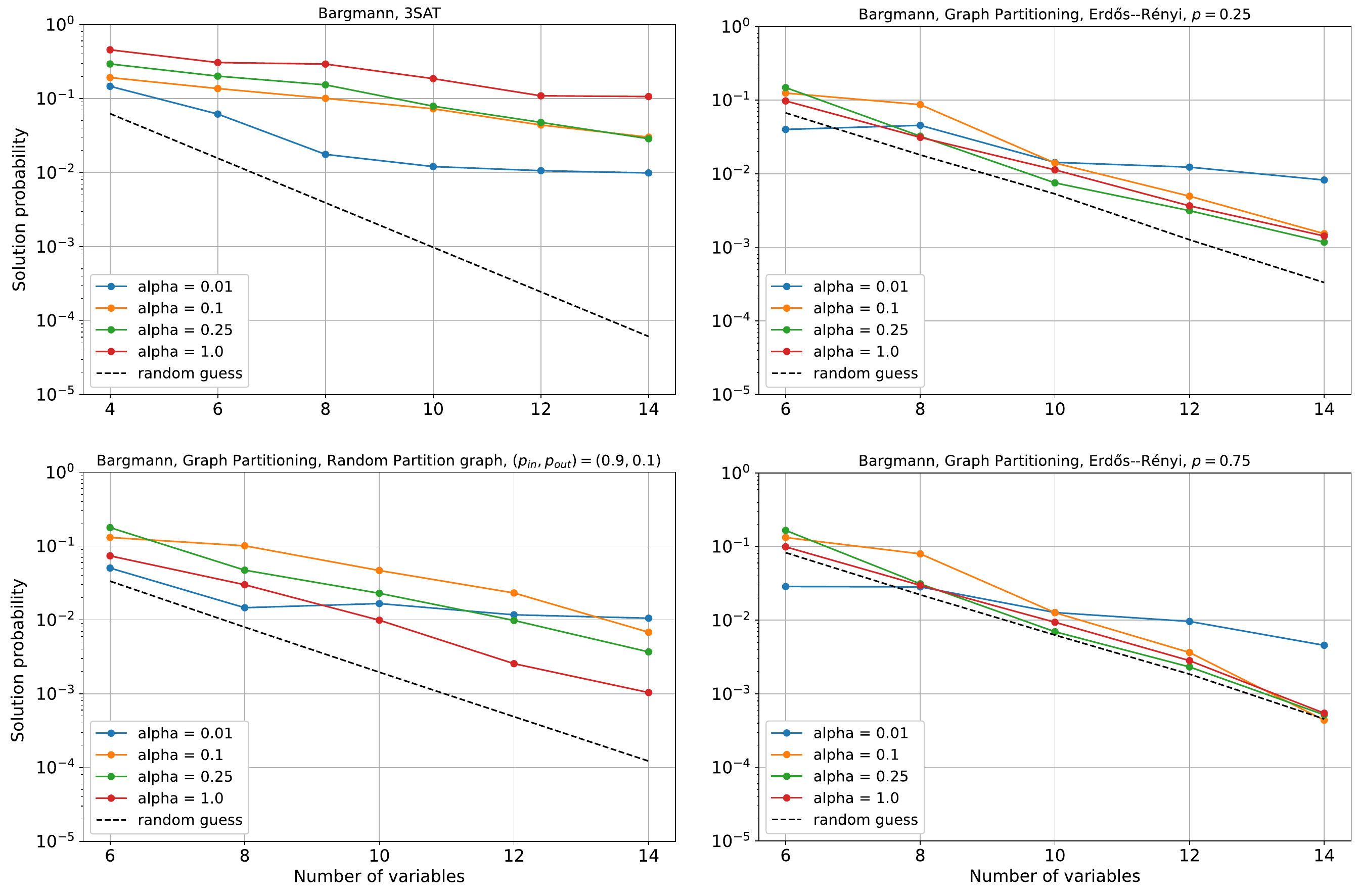}
    \caption{\textbf{Bargmann parametrization.} Simulation results for the binary optimization problem solver presented in this article. Two types of optimization problems are considered: 3-SAT and Graph Partitioning, encoded as a PUBO and QUBO problem respectively. The Graph Partitioning instances are generated using three random graph generators: Erd\H{o}s--R\'enyi graphs with connection probabilities $p = 0.25$ or $p = 0.75$, and Random Partition Graphs with intra-cluster and inter-cluster probabilities $p_\mathrm{in} = 0.9$ and $p_\mathrm{out} = 0.1$ respectively. For a comprehensive description of the problem instances, we refer to Sec.~\ref{sec:problems_considered}. For each problem instance, and for each CVaR parameter $\alpha \in \{0.01, 0.1, 0.25, 1\}$, 50 (3-SAT case) or 100 (Graph Partitioning case) random instances are generated. The model is optimized using the Bargmann parametrization methods, as detailed in Section~\ref{sec:methods}. The probability of sampling a solution, calculated according to Eq.~\eqref{eq:proba_success}, is averaged over the generated instances and plotted. The dashed line indicates the probability of randomly guessing a solution, averaged over the instances. Notice that some instances have several solutions. We use a logarithmic scale for $y$-axis.
    }
    \label{fig:result_main_bargmann}
\end{figure*}

This section is devoted to the simulation results. We will denote by $H$ the QUBO/PUBO polynomial to be minimized and by $\bm{\theta}_*$ the parameters after training the variational GBS.

\subsection{Problems considered}
\label{sec:problems_considered}

We will consider the QUBO and PUBO formulations of the Graph Partitioning problem and the 3-SAT problem respectively. These were introduced in Section~\ref{sec:unconstrained_binary_optimization}. For both problem types, we will consider problems with $\ell$ binary variables for $\ell\in\{6,8,10,12,14\}$.

To guarantee the hardness of the 3-SAT instances, we generate them following the procedure proposed by Selman et al.~\cite{selman_generating_1996}. We consider the random 3-SAT model where $M=4.3\ell$ clauses are produced by randomly selecting three distinct variables according to the uniform distribution on the set of $\ell$ available, and negating each of them with probability $0.5$. Notice this procedure may generate redundant clauses, which are simplified later on. The statistical analysis carried on by Selman et al. suggests that choosing the ratio $M/\ell\simeq 4.3$ produces instances hard to solve. For each value of $\ell$, we generate 50 instances.

A Graph Partitioning instance is solely determined by its input graph. To test our algorithm, we generate Erdős–Rényi graphs $G(\ell,p)$ with $p=0.25$ (non-dense case) and $p=0.75$ (dense case). We also generate Random Partition graphs, and more precisely graphs of two communities with sizes $\ell/2$ where the nodes in the same group are connected with probability $p_{\mathrm{in}}=0.9$ and nodes of different groups are connected with probability $p_{\mathrm{out}}=0.1$. For each graph type and each $\ell$, we generate 100 instances using NetworkX Python library~\cite{Hagberg2008}.

\subsection{Performance metrics}

As argued by Barkoutsos et al., a pertinent benchmark metric for this approach is the probability $p_{\bm{\theta}_*}$ of sampling one minimizer of $H$ after training the ansatz~\cite{barkoutsos_improving_2020}. The motivation is that knowing the complete ground state of $\hat{H}$, as required in quantum simulation problems, is not crucial in the context of quantum approaches for combinatorial optimization. Indeed, we are only interested in sampling a solution to the problem with a sufficiently high probability.

The probability $p_{\bm{\theta}_*}$ can be computed as follows. We denote $\{\mathbf{x}_1,\dots,\mathbf{x}_L\}$ the minimizing clicks patterns of $H$. Then, since the corresponding projections $\{\hat{P}_{\mathbf{x}_{1}},\dots,\hat{P}_{\mathbf{x}_{L}} \}$ are mutually orthogonal, we have
\begin{align}
    \label{eq:proba_success}
    p_{\bm{\theta}_*} = \sum_{k=1}^{L} \mathbb{P}(\mathbf{x}_k; \hat{\rho}_{\bm{\theta}_*})\, ,
\end{align}
where each term can be computed with the formula in Eq.~\eqref{eq:proba_td}.

\subsection{Choosing the search space}

In practical scenarios, there is a limit on the level of squeezing that can be achieved, as squeezing determines the amount of energy or photon number in the GBS by Eq.~\eqref{eq:mean_photon}. It is therefore appropriate to restrict the parameter spaces to account for the limitation in squeezing. The restricted search spaces representing the quantum states achievable with squeezing parameters not exceeding $\bar{r}$ are
\begin{gather}
    \label{eq:restricted_parameters_wigner}
    \mathcal{P}_{C,\bar{r}} = [0,\bar{r}]^\ell \times U_\ell(\mathbb{C}) \quad \text{and}\\
    \label{eq:restricted_parameters_bargmann}
    \mathcal{P}_{B,\bar{r}} = \qty{\bm{A} = \bm{A}^{T} \in \mathcal{M}_\ell(\mathbb{C}) ~|~\norm*{\bm{A}} \leq \tanh(\bar{r}) } \, ,
\end{gather}
for respectively the Wigner parametrization and the Bargmann parametrization, where $\bar{r}$ has to be greater than zero.

In Appendix~\ref{sec:expressivity}, we show that restricting the parameter space imposes restrictions on the expressivity of the model in the case where $\bar{r} \leq 0.8814$ and there is a unique minimizer to $H$. Specifically, we show that the maximum attainable fidelity decreases exponentially with the number of 1's in the problem's solution. We believe that the assumption $\bar{r} \leq 0.8814$ is only an artifact of the proof, numerical investigations suggesting that this exponential decay of the expressivity actually holds for any $\bar{r}$. On the other hand, increasing the squeezing upper bound can lead to better performance. Similarly, when prior information suggests that the optimal solution contains more 1's than 0's, it would be advantageous to reverse the encoding of the problem: $0 \longleftrightarrow 1$.

In our simulation, we opt for $\bar{r}=1$, which corresponds to an average number of photons per mode of around $1.38$. This corresponds to about 8.7 dB squeezing, which can be reasonably achieved in experiments~\cite{deng_gaussian_2023}.

\subsection{Methods}
\label{sec:methods}

In the experiments, we employed the Bargmann and Wigner parametrizations, as detailed in Section~\ref{sec:state_preparation}. For both cases, we configured the algorithm using for CVaR parameters $\alpha \in \{0.01, 0.1, 0.25, 1\}$.

Reducing the number of trainable parameters can improve algorithm performance and mitigate problems such as local minima trapping. A pure $\ell$-mode Gaussian state without displacement is determined by $\ell(\ell+1)$ parameters, since $U_\ell(\mathbb{C})$ is a Lie group with real dimension $\ell^2$. However, in the GBS protocol, the output distribution does not depend on both the global phase and the local phases before measurement, reducing the maximum number of trainable parameters to $\ell^2-1$. Through numerical experiments, we found that a linear scaling of the number of trainable parameters relative to the number of binary variables $\ell$ usually yields better results.

In the Bargmann parametrization and for the QUBO case, we train the real parts of the $\ell$ diagonal elements and $2\ell$ off-diagonal elements of the symmetric matrix $\mathbf{A}_{\boldsymbol{\theta}}$. The off-diagonal elements are selected as the $2\ell$ smallest elements of the quadratic matrix $J$ in the QUBO problem. This leads to a total of $3\ell$ trainable parameters. The parameters are initialized as a random perturbation of $J$. In the case of 3-SAT, we trained all $\ell(\ell+1)$ parameters, as we found this led to sufficiently good results. There, the initialization of each parameter is uniformly randomized between 0 and 1.

In the Wigner parametrization for both problem types, we restrict the number of trainable Mach-Zehnder interferometer (MZI) layers in the circuit to two, as illustrated in Fig.~\ref{fig:gbs}. Since the global phase can not be trained, we can further fix the complex phase shift of one MZI, let say $\phi_1$, to zero. Similarly, we do not adjust the local phases. Consequently, this configuration results in a total of $3(\ell-1)$ trainable parameters, which are initialized uniformly in the intervals $[0,2\pi]$, $[0,\pi]$ and $[0,\bar{r}]$ for the $\theta$'s, the $\phi$'s, and the $r$'s respectively. This approach effectively uses a shallow-depth circuit with local interactions. Even though GBS experiments with such optical circuit are easier to implement they are also known to be classically simulable~\cite{qi_efficient_2022} due to their significantly reduced loss rates.

In the VQE approach ($\alpha=1$), we utilize the ADAM optimizer, executing between 2500 and 4000 steps, with learning rates varying from 0.01 to 1.0. Conversely, for CVaR-VQE, we employ the COBYLA optimizer, setting the maximum number of training steps to $70\ell$.

In this study, we evaluate exactly the cost function $\mathcal{C}_{\alpha}(\bm{\theta})$, taking no account of shot noise. In the case of non-shallow-depth circuit, this necessitates computing all the Fock components of the quantum state at each iteration, which is computationally expensive. This explains why we have limited the maximum number of modes to 14.

The simulation was carried out on an MSI MPG Trident AS 12TG-065AT desktop computer equipped with an Intel\textsuperscript{\textregistered} Core\textsuperscript{\texttrademark} i7-12700F processor and 32 GB of DDR4 RAM running Ubuntu 22.04.4 LTS.

\subsection{Results and discussion}
\label{sec:results_comments}

The results are presented in Fig.~\ref{fig:result_main_wigner} and Fig.~\ref{fig:result_main_bargmann} the Wigner parametrization and the Bargmann parametrization respectively. A first general remark is that the numerical experiments reveal one to two orders of magnitude better performance compared to random guessing across most cases. Furthermore, we note that the performance for certain values of $\alpha$ (in particular for $\alpha=0.01$ in most cases) varies favorably with respect to the problem size, as the probability of success decreases slowly. This constitutes an initial validation of our approach. Another general learning is that the Wigner parametrization yields significantly better results on most problems considered. 

For the 3-SAT problem, both parametrizations demonstrate comparable performance, except for $\alpha = 1$, where the Bargmann parametrization significantly outperformed Wigner. Notably, for $\alpha = 0.01$, both parametrizations give an average solution probability closed to the CVaR parameter $\alpha$, indicating that the optimization was efficiently carried out on average in these cases, since CVaR does not specifically reward states whose fidelity is greater than $\alpha$.

Regarding the Graph Partitioning problem, Wigner parametrization generally gives better results than Bargmann parametrization. Specifically, in random partition graphs, the Wigner parametrization shows better scaling for $\alpha = 0.01$ and $\alpha = 0.1$, which was expected due to model optimization on low-energy models. For the Erd\H{o}s--R\'enyi instances, Wigner parametrization outperforms Bargmann at fixed $\alpha$ values, except for $\alpha = 1$ in the non-dense case. The Wigner parametrization shows no significant performance deviation with varying $p$ and $\alpha = 0.01$, $0.1$, or $0.25$, though $\alpha = 0.01$ scaled better, warranting further large-scale testing. We also notice that the performance drops significantly for $\alpha=1$. This discussion indicates that the use of a non-trivial CVaR in conjunction with the Wigner parametrization is well suited to this encoding of the Graph Partitioning problem.

The results are presented in Fig.~\ref{fig:result_main_wigner} and Fig.~\ref{fig:result_main_bargmann} the Wigner parametrization and the Bargmann parametrization respectively. A first general remark is that the numerical experiments reveal one to two orders of magnitude better performance compared to random guessing across most cases. Furthermore, we note that the performance for certain values of $\alpha$ (in particular for $\alpha=0.01$ in most cases) varies favorably with respect to the problem size, as the probability of success decreases slowly. This constitutes an initial validation of our approach. Another general learning is that the Wigner parametrization yields significantly better results on most problems considered.

For the 3-SAT problem, the Bargmann representation performs noticeably better when standard VQE cost function (case $\alpha=1$) is used instead of the CVaR cost function (case $\alpha <1$), while the Wigner representation shows comparable results, except for $\alpha=0.01$. At first glance, this outcome may seem counter-intuitive, as CVaR is generally expected to offer better performance. However, with $\alpha = 1$, the optimization of the ansatz's parameters is carried out analytically using gradients, which is not the case when the CVaR cost function is employed. Therefore, our numerical results indicate that, within our approach, this encoding of the 3-SAT problem is more efficiently addressed using standard VQE, with the use of CVaR unable to counterbalance the disadvantages of gradient-free optimization. However, although our results suggest the Bargmann representation outperforms the Wigner representation for this problem, it is premature to claim its superiority, as we have trained more parameters in the former, $O(\ell^2)$, compared to later, $O(\ell)$. Furthermore, we notice that, for $\alpha = 0.01$, both parametrizations show saturation in the CVaR, indicating that the optimization was efficiently carried out in these cases.

Contrasting with the previous problem, the use of CVaR proves to be crucial for performance when addressing the Graph Partitioning problem. Indeed, when $\alpha = 1$, the average performance barely exceeds random guess, particularly for Erd\H{o}s--R\'enyi graphs. For this encoding of the Graph Partitioning problem, the trade-off between gradient-based descent and CVaR cost function clearly favors the latter.

For the Bargmann representation, poor scaling and low performance are observed for $\alpha = 0.25$ and $0.1$, especially with Erd\H{o}s--R\'enyi graphs. Moreover, except in the case of Random Partition Graph, $\alpha = 0.01$ exhibits no saturation in the CVaR, though it achieves better scaling. This result could be surprising, as one would expect this parametrization to better respect the graph structure of the problem. However, this observation can be explained by the encoding of unbalanced partition penalty in Graph Partitioning problem on a fully connected graph (see the $H_2$ term in Example~\ref{ex:graph_partitioning}), whereas the Bargmann ansatz is trained on a low density graph, with $O(\ell)$ parameters. One way out would be to modify the ratio $c_1/c_2$ defining the balance between the penalty $H_1$ and objective function $H_2$ (see Example~\ref{ex:graph_partitioning}) or to implement the penalty linearly, using for instance slack variables.

In contrast, the Wigner representation performs significantly better, except when CVaR is omitted. For Random Partition Graphs, the results scale well, irrespective of the value of $\alpha <1$. For Erd\H{o}s--R\'enyi graphs, there is no significant performance deviation with varying $p$ and $\alpha = 0.01$, $0.1$, or $0.25$, though $\alpha = 0.01$ seems to scale better for larger $\ell$. The case $\alpha = 0.01$ shows in addition saturation in the CVaR, indicating that the optimization likely converged to a suitable local minimum.

\section{Conclusion and outlook}
\label{sec:conclusions}

In this paper, we introduced an approach using a parametrized Gaussian Boson Sampler (GBS) with threshold detectors combined with the Conditional Value-at-Risk Variational Quantum Eigensolver (CVaR-VQE) cost function to tackle binary optimization problems, specifically PUBO and QUBO. Our method integrates a GBS ansatz with the CVaR-VQE cost function, which is analytically computed in the VQE case, extending the existing literature. We detailed the framework of this approach and proposed two distinct parametrizations for the ansatz. The primary objective was to demonstrate proof of principle for this variational quantum algorithm. We achieved this by conducting numerous numerical experiments on 3-SAT and Graph Partitioning instances, including several random graph generators. Our results indicate that the algorithm significantly outperforms a naive random guess, providing initial validation of its utility.

Looking ahead, we envisage several avenues for further research and improvement. Incorporating shot noise and hardware imperfections such as photon loss into the simulations would provide a more realistic assessment of the algorithm's performance. Exploring specific ansatz, such as the Non-Local Hypercube Structure proposed by Oh \emph{et al}.~\cite{go_exploring_2023}, could be beneficial. This structure aims to produce the maximum entangled state with the fewest parameters, potentially enhancing the efficiency of our approach. Moreover, for problems characterized by small densities, implementing a warm-start strategy could improve convergence and overall performance. Finally, it would be insightful to run the experiments on real quantum hardware of sufficient size, when it becomes accessible.

\acknowledgments
This work is funded by the German Federal Ministry of Education and Research (BMBF) under the project ``PhoQuant”. This work is supported by the European Union’s Horizon Europe Framework Programme (HORIZON) under the ERA Chair scheme with grant agreement no. 101087126. This work is supported with funds from the Ministry of Science, Research and Culture of the State of Brandenburg within the Center for Quantum Technologies and Applications (CQTA).
\begin{figure}[h!]
    \centering
    \includegraphics[width=0.1\textwidth]{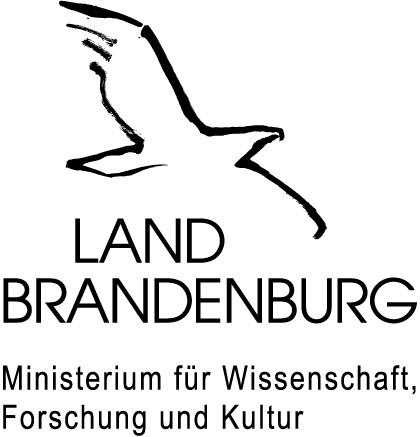}
\end{figure}

J.C.\ and T.S.\ would like to thank the whole Software team at Q.ANT GmbH for their valuable comments, and Niklas Hoppe, Alejandro Lorenzo Ruiz and Andrea Zanzi of the PIC Design and Characterization team at Q.ANT GmbH for their helpful insights on nanophotonics based on the thin-film lithium-niobate.

\appendix

\section{Making the training more efficient}
\label{app:more_efficient}

In this appendix, we detail the algorithm for computing VQE-PUBO cost
function, that is when $\alpha=1$.

We consider a PUBO Hamiltonian $H$ as in Eq.~\eqref{eq:pubo_hamiltonian}, and we want to calculate $\mathcal{C}_1(\bm{\theta}) = \expval*{\hat{H}}_{\bm{\theta}}$. It is important to speed up this calculation as it has to be repeated several times during the training.

We use the same notation as in Section~\ref{sec:vqe_case}. Our algorithm is based on the following observation, see~\cite[Appendix F]{bulmer_threshold_2022}:
\begin{align}
    \notag
    \expval*{\hat{x}_J}_{\bm{\theta}}
     & \coloneqq \mathrm{tr}\qty[\hat{\rho}_{\bm{\theta}} \bigotimes_{j\in J} \hat{\Pi}^{(1)}_j] =  \mathrm{tr}\qty[\hat{\rho}_{\bm{\theta}} \bigotimes_{j\in J} \qty(\hat{\mathbb{I}} - \ket{0_j}\bra{0_{j}})] \\
    \notag
     & = \sum\limits_{J'\subset J} (-1)^{|J'|} \mathrm{tr}\qty[\hat{\rho}^{J'}_{\bm{\theta}} \bigotimes_{j\in J'} \ket{0_j}\bra{0_{j}}]                                                                         \\
    \notag
     & = \sum\limits_{J'\subset J} (-1)^{|J'|} \mathbb{P}\qty(0,\dots,0; \hat{\rho}^{J'}_{\bm{\theta}})                                                                                                         \\
    \label{eq:more_efficient_0}
     & = \sum\limits_{J'\subset J} (-1)^{|J'|} \det(\bm{\Sigma}^{J'}_{\bm{\theta}})^{-1/2} \, .
\end{align}
We insert this in the expression
\begin{align}
    \label{eq:more_efficient_1}
    \expval*{\hat{H}}_{\bm{\theta}} = \sum_{J\in \mathcal{J}_\ell} h_J \expval*{\hat{x}_J}_{\bm{\theta}} \, ,
\end{align}
regroup the determinants together, and we end up with the following formula
\begin{multline*}
    \expval*{\hat{H}}_{\bm{\theta}} = \sum\limits_{J \in \mathcal{J}_\ell} c_J \det(\bm{\Sigma}^{J}_{\bm{\theta}})^{-1/2}  \\\quad \text{where} \quad c_J=(-1)^{|J|} \sum\limits_{J'\supset J} h_{J'} \, .
\end{multline*}
Compared with the naive method where all the terms $\expval*{\hat{x}_J}_{\bm{\theta}}$ are computed separately then added together according to Eq.~\eqref{eq:more_efficient_1}, we have reduced the number of determinants to compute at the expense of the one time computation of the $c_J$'s. Also, at each training steps, the computation of the determinants can be done batch-wise using TensorFlow, which is highly efficient. This method roughly halves the cost of calculating QUBO Hamiltonians with respect to the naive method, and this speed-up increases with the degree of $H$. As our aim was to demonstrate a proof of concept, this calculation was implemented on CPUs, although using GPUs would not require any considerable effort.

\section{Expressivity for finite squeezing}
\label{sec:expressivity}

In this appendix, we consider a QUBO/PUBO energy $H$ with a unique minimizing binary variable assignment $\mathbf{x}\in\{0,1\}^\ell$. We will show that $p_{\bm{\theta}_*}$ as defined in Eq.~\eqref{eq:proba_success} decreases exponentially fast with the number of $1$'s in $\mathbf{x}$ when the search space is restricted to $\mathcal{P}_{C,\bar{r}} \simeq \mathcal{P}_{B,\bar{r}}$ with $\bar{r}<\tanh^{-1}\qty(\tfrac{1}{\sqrt{2}}) \leq 0.8814$.

More precisely, we denote
\begin{align*}
    k \coloneqq \sum_{j=1}^{\ell} x_j \, ,
\end{align*}
and we slightly modify the notation for the search space to make appear the dependence in the number of modes:
\begin{align*}
    \mathcal{P}_{C,\bar{r}}(\ell) = [0,\bar{r}]^\ell \times U_\ell(\mathbb{C})   \, .
\end{align*}
We aim to bound the following quantity
\begin{align}
    \label{eq:app_to_bound}
    \max_{\bm{\theta}\in \mathcal{P}_{C,\bar{r}}(\ell)} \mathbb{P}(\mathbf{x};\hat{\rho}_{\bm{\theta}}) \, ,
\end{align}
where the probability $\mathbb{P}(\mathbf{x};\hat{\rho}_{\bm{\theta}})$ to detect the click pattern $\mathbf{x}$ after preparing $\hat{\rho}_{\bm{\theta}}$ is defined in Eq.~\eqref{eq:proba_td}.

The first step of the proof consists of reducing the search space the set of Gaussian states living only on the modes where $x_j =1$. Indeed, we will show the equality
\begin{align}
    \label{eq:reducing_search_space}
    \max_{\bm{\theta}\in \mathcal{P}_{C,\bar{r}}(\ell)} \mathbb{P}(\mathbf{x};\hat{\rho}_{\bm{\theta}}) = \max_{\bm{\theta}\in \mathcal{P}_{C,\bar{r}}(k)} \mathbb{P}(\overset{k\text{ times}}{\overbrace{1,\dots,1}};\hat{\rho}_{\bm{\theta}}) \, .
\end{align}
To see why this is true, we consider $\bm{\theta} \in  \mathcal{P}_{C,\bar{r}}(\ell)$  and denote $\hat{\rho}^{J_{\mathbf{x}}}_{\bm{\theta}}$ the quantum state $\hat{\rho}_{\bm{\theta}}$ after tracing out the modes in $J_{\mathbf{x}}$. Then, using Eq.~\eqref{eq:proba_td}, we have the following upper bound
\begin{multline*}
    \mathbb{P}(\mathbf{x};\hat{\rho}_{\bm{\theta}})
    = \tr\qty[\qty(\bigotimes_{j\in J_{\mathbf{x}}} \hat{\Pi}_j^{(1)} \bigotimes_{j\notin J_{\mathbf{x}}} \dyad{0_{j}}) \hat{\rho}_{\bm{\theta}}] \\
    \leq \tr\qty(\bigotimes_{j\in J_{\mathbf{x}}} \hat{\Pi}_j^{(1)} \hat{\rho}_{\bm{\theta}}) =  \mathbb{P}(\underset{k\text{ times}}{\underbrace{1,\dots,1}}; \hat{\rho}^{J_{\mathbf{x}}}_{\bm{\theta}}) \, ,
\end{multline*}
which is sufficient to show Eq.~\eqref{eq:reducing_search_space}.

Now, for $\bm{\theta}\in \mathcal{P}_{C,\bar{r}}(k)$, the probability $\mathbb{P}(1,\dots,1;\hat{\rho}_{\bm{\theta}})$ is bounded by the probability of the event ``at least $k$ photons are generated", which only depends on the input squeezing parameters and is maximized when all the squeezing parameters are equal to $\bar{r}$. We denote by $\bm{\theta}^{\star} \in \mathcal{P}_{C,\bar{r}}(k) $ one of such configurations. We also denote by
\begin{align*}
    \hat{N} = \sum_{j=1}^{k} \hat{N}_j = \sum_{j=1}^{k} \hat{a}_j^{\dagger} \hat{a}_j
\end{align*}
the total photon number operator and $Y(\hat{N},\bm{\theta}^\star)$ the random variable associated with $\hat{N}$ in the quantum state $\hat{\rho}_{\bm{\theta}^\star}$. Therefore, in this discussion, we have argued that
\begin{align}
    \label{eq:app_first_bound}
    \max_{\bm{\theta}\in \mathcal{P}_{C,\bar{r}}(k)} \mathbb{P}(1,\dots,1;\hat{\rho}_{\bm{\theta}}) \leq \mathbb{P}\qty(Y(\hat{N},\bm{\theta}^{\star}) \geq k) \, ,
\end{align}
and it remains to bound the right side of this inequality.

The distribution of $Y(\hat{N}/2,\bm{\theta}^{\star})$ follows a negative binomial distribution~\cite[Eq. (32)]{kruse_detailed_2019}
\begin{align*}
    Y(\hat{N}/2,\bm{\theta}^{\star}) \sim \mathrm{NB}(k/2, p) \, ,
\end{align*}
with $p= \sech^{2}(\bar{r}) = 1 - \tanh^{2}(\bar{r})$. More precisely, we have
\begin{align*}
    \mathbb{P}(Y(\hat{N},\bm{\theta}^{\star}) = 2n) = \frac{\Gamma(n+k/2)}{\Gamma(k/2)n!} p^{k/2} (1 - p)^{n} \, ,
\end{align*}
where $\Gamma$ is the Euler Gamma function. Notice that only pairs of photons are generated, which is due to the properties of pure squeezed light~\cite{weedbrook_gaussian_2012, lvovsky_squeezed_2015}. The cumulative distribution function has an explicit form in terms of special functions:
\begin{align}
    \label{eq:tail_Y}
    \mathbb{P}(Y(\hat{N},\bm{\theta}^{\star}) \geq k ) = I_{1-p}\qty(k/2+1,k/2) \, ,
\end{align}
where $I_x(a,b)$ denotes the regularized incomplete beta function with
\begin{align*}
    a = k/2 +1\, , \quad b = k/2 \qq{and} x = 1-p = \tanh^2(\bar{r}) \, .
\end{align*}
Let $m = \tfrac{a}{a+b} = \tfrac{1}{2}\tfrac{k+2}{k+1}$ be the mean of a random variable $X$ following the distribution $\mathrm{Beta}(k/2+1,k/2)$. The beta distribution being sub-Gaussian, we have, for $\epsilon \geq 0$~\cite[Theorem 1]{skorski2023Bernsteintype}:
\begin{align*}
    \mathbb{P}(X \leq m - \epsilon) \leq \exp(-\frac{\epsilon^2}{2(v + c\epsilon/3)}) \, ,
\end{align*}
where
\begin{align*}
    &v = \frac{ab}{(a+b)^2 (a+b+1)} = \frac{1}{4} \frac{k}{(k+1)^2}\, ,\\
    &c = \frac{2(b-a)}{(a+b)(a+b+2)} = - \frac{2}{(k+1)(k+3)} \, .
\end{align*}
Consequently, for any $\bar{r}$ such that 
\begin{align*}
    \epsilon = m + p - 1 = \frac{1}{2} \frac{k+2}{k+1} - \tanh^2(\bar{r}) \geq \epsilon_0 >0 \, ,
\end{align*}
we have, for $k$ large enough:
\begin{multline}
    \label{eq:sub_gaussianity_beta}
    I_{1-p}\qty(k/2+1,k/2) = \mathbb{P}(X - m \leq -\epsilon) \\
     \leq \exp(-\frac{\epsilon^2}{2} \frac{k+1}{(\tfrac{k}{(k+1)^2} - \tfrac{2\epsilon/3}{k+3})}) 
    \leq \exp(-\epsilon_0^2 k) \, .
\end{multline}
Putting together Eq.~\eqref{eq:reducing_search_space}, Eq.~\eqref{eq:app_first_bound}, Eq.~\eqref{eq:tail_Y} and Eq.~\eqref{eq:sub_gaussianity_beta}, we deduce that~\eqref{eq:app_to_bound} decays exponentially fast with $k$ whenever the condition
\begin{align*}
    \tanh^2(\bar{r}) < \frac{1}{2} \frac{k+2}{k+1} \, ,
\end{align*}
is satisfied. In the limit of large $k$, it corresponds to
\begin{align*}
    \bar{r} < \tanh^{-1}\qty(\tfrac{1}{\sqrt{2}}) \simeq 0.8814 \, .
\end{align*}
Notice that in the limit case $\bar{r} = \tanh^{-1}\qty(\tfrac{1}{\sqrt{2}})$, we have $\expval*{\hat{N}_j}_{\bm{\theta}^{\star}} = 1$ for all $j$.

This argument is not sharp and the threshold squeezing value is only an artifact of the proof. Numerical studies suggest that, whatever the maximum squeezing available, the probability $p_{\bm{\theta}_*}$ decreases exponentially fast with $k$ and is saturated by a product of GHZ-like states. More precisely, we denote by $\hat{\rho}_{\mathrm{GHZ},k}$ the $k$-mode GHZ-like state with input squeezing $r_1=r_2=\bar{r}$, see \cite[Section III.B.4]{braunstein_quantum_2005}. Then, we conjecture that
\begin{align*}
    \max_{\bm{\theta}\in \mathcal{P}_{C,\bar{r}}(k)} \mathbb{P}(\overset{k\text{ times}}{\overbrace{1,\dots,1}};\hat{\rho}_{\bm{\theta}})
\end{align*}
is optimized for
\begin{multline*}
    \begin{cases}
        \hat{\rho}_{\mathrm{GHZ},1}                                                       & \text{if} \quad k=1 \, ,              \\
        (\hat{\rho}_{\mathrm{GHZ},2})^{\otimes q}                                         & \text{if} \quad k=2q \, ,             \\
        \hat{\rho}_{\mathrm{GHZ},3} \otimes (\hat{\rho}_{\mathrm{GHZ},2})^{\otimes q - 1} & \text{if} \quad k=2q+1,~ q\geq 1 \, .
    \end{cases}
\end{multline*}
The optimal success-shot probability can be computed exactly, which gives:
\begin{multline*}
    \max_{\bm{\theta}\in \mathcal{P}_{C,\bar{r}}(k)} \mathbb{P}(\overset{k\text{ times}}{\overbrace{1,\dots,1}};\hat{\rho}_{\bm{\theta}}) \\
    = \begin{cases}
        \sech(\bar{r})                 & \text{if} \quad k=1 \, ,              \\
        \tanh(\bar{r})^{k}             & \text{if} \quad k=2q \, ,             \\
        g(\bar{r})\tanh(\bar{r})^{k-3} & \text{if} \quad k=2q+1,~ q\geq 1 \, ,
    \end{cases}
\end{multline*}
with
\begin{align*}
    g(\bar{r}) = 1 - \frac{1}{\cosh(\bar{r})^3} + \frac{9 (\sech(\bar{r})-1) \sech(\bar{r})}{\sqrt{8 \cosh (\bar{r})^2+1}} \, .
\end{align*}
The search for a proof of this conjecture will be the subject of future work.

\bibliography{references}

\begin{thebibliography}{82}%
\makeatletter
\providecommand \@ifxundefined [1]{%
 \@ifx{#1\undefined}
}%
\providecommand \@ifnum [1]{%
 \ifnum #1\expandafter \@firstoftwo
 \else \expandafter \@secondoftwo
 \fi
}%
\providecommand \@ifx [1]{%
 \ifx #1\expandafter \@firstoftwo
 \else \expandafter \@secondoftwo
 \fi
}%
\providecommand \natexlab [1]{#1}%
\providecommand \enquote  [1]{``#1''}%
\providecommand \bibnamefont  [1]{#1}%
\providecommand \bibfnamefont [1]{#1}%
\providecommand \citenamefont [1]{#1}%
\providecommand \href@noop [0]{\@secondoftwo}%
\providecommand \href [0]{\begingroup \@sanitize@url \@href}%
\providecommand \@href[1]{\@@startlink{#1}\@@href}%
\providecommand \@@href[1]{\endgroup#1\@@endlink}%
\providecommand \@sanitize@url [0]{\catcode `\\12\catcode `\$12\catcode `\&12\catcode `\#12\catcode `\^12\catcode `\_12\catcode `\%12\relax}%
\providecommand \@@startlink[1]{}%
\providecommand \@@endlink[0]{}%
\providecommand \url  [0]{\begingroup\@sanitize@url \@url }%
\providecommand \@url [1]{\endgroup\@href {#1}{\urlprefix }}%
\providecommand \urlprefix  [0]{URL }%
\providecommand \Eprint [0]{\href }%
\providecommand \doibase [0]{https://doi.org/}%
\providecommand \selectlanguage [0]{\@gobble}%
\providecommand \bibinfo  [0]{\@secondoftwo}%
\providecommand \bibfield  [0]{\@secondoftwo}%
\providecommand \translation [1]{[#1]}%
\providecommand \BibitemOpen [0]{}%
\providecommand \bibitemStop [0]{}%
\providecommand \bibitemNoStop [0]{.\EOS\space}%
\providecommand \EOS [0]{\spacefactor3000\relax}%
\providecommand \BibitemShut  [1]{\csname bibitem#1\endcsname}%
\let\auto@bib@innerbib\@empty
\bibitem [{\citenamefont {Kim}\ \emph {et~al.}(2023)\citenamefont {Kim}, \citenamefont {Eddins}, \citenamefont {Anand}, \citenamefont {Wei}, \citenamefont {van~den Berg}, \citenamefont {Rosenblatt}, \citenamefont {Nayfeh}, \citenamefont {Wu}, \citenamefont {Zaletel}, \citenamefont {Temme},\ and\ \citenamefont {Kandala}}]{kim_evidence_2023}%
  \BibitemOpen
  \bibfield  {author} {\bibinfo {author} {\bibfnamefont {Y.}~\bibnamefont {Kim}}, \bibinfo {author} {\bibfnamefont {A.}~\bibnamefont {Eddins}}, \bibinfo {author} {\bibfnamefont {S.}~\bibnamefont {Anand}}, \bibinfo {author} {\bibfnamefont {K.~X.}\ \bibnamefont {Wei}}, \bibinfo {author} {\bibfnamefont {E.}~\bibnamefont {van~den Berg}}, \bibinfo {author} {\bibfnamefont {S.}~\bibnamefont {Rosenblatt}}, \bibinfo {author} {\bibfnamefont {H.}~\bibnamefont {Nayfeh}}, \bibinfo {author} {\bibfnamefont {Y.}~\bibnamefont {Wu}}, \bibinfo {author} {\bibfnamefont {M.}~\bibnamefont {Zaletel}}, \bibinfo {author} {\bibfnamefont {K.}~\bibnamefont {Temme}},\ and\ \bibinfo {author} {\bibfnamefont {A.}~\bibnamefont {Kandala}},\ }\href {https://doi.org/10.1038/s41586-023-06096-3} {\bibfield  {journal} {\bibinfo  {journal} {Nature}\ }\textbf {\bibinfo {volume} {618}},\ \bibinfo {pages} {500} (\bibinfo {year} {2023})},\ \bibinfo {note} {number: 7965 Publisher: Nature Publishing Group}\BibitemShut {NoStop}%
\bibitem [{\citenamefont {Dalzell}\ \emph {et~al.}(2023)\citenamefont {Dalzell}, \citenamefont {{McArdle}}, \citenamefont {Berta}, \citenamefont {Bienias}, \citenamefont {Chen}, \citenamefont {Gilyén}, \citenamefont {Hann}, \citenamefont {Kastoryano}, \citenamefont {Khabiboulline}, \citenamefont {Kubica}, \citenamefont {Salton}, \citenamefont {Wang},\ and\ \citenamefont {Brandão}}]{dalzell_quantum_2023}%
  \BibitemOpen
  \bibfield  {author} {\bibinfo {author} {\bibfnamefont {A.~M.}\ \bibnamefont {Dalzell}}, \bibinfo {author} {\bibfnamefont {S.}~\bibnamefont {{McArdle}}}, \bibinfo {author} {\bibfnamefont {M.}~\bibnamefont {Berta}}, \bibinfo {author} {\bibfnamefont {P.}~\bibnamefont {Bienias}}, \bibinfo {author} {\bibfnamefont {C.-F.}\ \bibnamefont {Chen}}, \bibinfo {author} {\bibfnamefont {A.}~\bibnamefont {Gilyén}}, \bibinfo {author} {\bibfnamefont {C.~T.}\ \bibnamefont {Hann}}, \bibinfo {author} {\bibfnamefont {M.~J.}\ \bibnamefont {Kastoryano}}, \bibinfo {author} {\bibfnamefont {E.~T.}\ \bibnamefont {Khabiboulline}}, \bibinfo {author} {\bibfnamefont {A.}~\bibnamefont {Kubica}}, \bibinfo {author} {\bibfnamefont {G.}~\bibnamefont {Salton}}, \bibinfo {author} {\bibfnamefont {S.}~\bibnamefont {Wang}},\ and\ \bibinfo {author} {\bibfnamefont {F.~G. S.~L.}\ \bibnamefont {Brandão}},\ }\href {https://doi.org/10.48550/arXiv.2310.03011} {\bibinfo {title} {Quantum algorithms: A survey of applications and end-to-end complexities}}
  (\bibinfo {year} {2023}),\ \Eprint {https://arxiv.org/abs/2310.03011 [quant-ph]} {2310.03011 [quant-ph]} \BibitemShut {NoStop}%
\bibitem [{\citenamefont {Abbas}\ \emph {et~al.}(2023)\citenamefont {Abbas}, \citenamefont {Ambainis}, \citenamefont {Augustino}, \citenamefont {Bärtschi}, \citenamefont {Buhrman}, \citenamefont {Coffrin}, \citenamefont {Cortiana}, \citenamefont {Dunjko}, \citenamefont {Egger}, \citenamefont {Elmegreen}, \citenamefont {Franco}, \citenamefont {Fratini}, \citenamefont {Fuller}, \citenamefont {Gacon}, \citenamefont {Gonciulea}, \citenamefont {Gribling}, \citenamefont {Gupta}, \citenamefont {Hadfield}, \citenamefont {Heese}, \citenamefont {Kircher}, \citenamefont {Kleinert}, \citenamefont {Koch}, \citenamefont {Korpas}, \citenamefont {Lenk}, \citenamefont {Marecek}, \citenamefont {Markov}, \citenamefont {Mazzola}, \citenamefont {Mensa}, \citenamefont {Mohseni}, \citenamefont {Nannicini}, \citenamefont {O'Meara}, \citenamefont {Tapia}, \citenamefont {Pokutta}, \citenamefont {Proissl}, \citenamefont {Rebentrost}, \citenamefont {Sahin}, \citenamefont {Symons}, \citenamefont {Tornow}, \citenamefont {Valls},
  \citenamefont {Woerner}, \citenamefont {Wolf-Bauwens}, \citenamefont {Yard}, \citenamefont {Yarkoni}, \citenamefont {Zechiel}, \citenamefont {Zhuk},\ and\ \citenamefont {Zoufal}}]{abbas_quantum_2023}%
  \BibitemOpen
  \bibfield  {author} {\bibinfo {author} {\bibfnamefont {A.}~\bibnamefont {Abbas}}, \bibinfo {author} {\bibfnamefont {A.}~\bibnamefont {Ambainis}}, \bibinfo {author} {\bibfnamefont {B.}~\bibnamefont {Augustino}}, \bibinfo {author} {\bibfnamefont {A.}~\bibnamefont {Bärtschi}}, \bibinfo {author} {\bibfnamefont {H.}~\bibnamefont {Buhrman}}, \bibinfo {author} {\bibfnamefont {C.}~\bibnamefont {Coffrin}}, \bibinfo {author} {\bibfnamefont {G.}~\bibnamefont {Cortiana}}, \bibinfo {author} {\bibfnamefont {V.}~\bibnamefont {Dunjko}}, \bibinfo {author} {\bibfnamefont {D.~J.}\ \bibnamefont {Egger}}, \bibinfo {author} {\bibfnamefont {B.~G.}\ \bibnamefont {Elmegreen}}, \bibinfo {author} {\bibfnamefont {N.}~\bibnamefont {Franco}}, \bibinfo {author} {\bibfnamefont {F.}~\bibnamefont {Fratini}}, \bibinfo {author} {\bibfnamefont {B.}~\bibnamefont {Fuller}}, \bibinfo {author} {\bibfnamefont {J.}~\bibnamefont {Gacon}}, \bibinfo {author} {\bibfnamefont {C.}~\bibnamefont {Gonciulea}}, \bibinfo {author} {\bibfnamefont
  {S.}~\bibnamefont {Gribling}}, \bibinfo {author} {\bibfnamefont {S.}~\bibnamefont {Gupta}}, \bibinfo {author} {\bibfnamefont {S.}~\bibnamefont {Hadfield}}, \bibinfo {author} {\bibfnamefont {R.}~\bibnamefont {Heese}}, \bibinfo {author} {\bibfnamefont {G.}~\bibnamefont {Kircher}}, \bibinfo {author} {\bibfnamefont {T.}~\bibnamefont {Kleinert}}, \bibinfo {author} {\bibfnamefont {T.}~\bibnamefont {Koch}}, \bibinfo {author} {\bibfnamefont {G.}~\bibnamefont {Korpas}}, \bibinfo {author} {\bibfnamefont {S.}~\bibnamefont {Lenk}}, \bibinfo {author} {\bibfnamefont {J.}~\bibnamefont {Marecek}}, \bibinfo {author} {\bibfnamefont {V.}~\bibnamefont {Markov}}, \bibinfo {author} {\bibfnamefont {G.}~\bibnamefont {Mazzola}}, \bibinfo {author} {\bibfnamefont {S.}~\bibnamefont {Mensa}}, \bibinfo {author} {\bibfnamefont {N.}~\bibnamefont {Mohseni}}, \bibinfo {author} {\bibfnamefont {G.}~\bibnamefont {Nannicini}}, \bibinfo {author} {\bibfnamefont {C.}~\bibnamefont {O'Meara}}, \bibinfo {author} {\bibfnamefont {E.~P.}\ \bibnamefont
  {Tapia}}, \bibinfo {author} {\bibfnamefont {S.}~\bibnamefont {Pokutta}}, \bibinfo {author} {\bibfnamefont {M.}~\bibnamefont {Proissl}}, \bibinfo {author} {\bibfnamefont {P.}~\bibnamefont {Rebentrost}}, \bibinfo {author} {\bibfnamefont {E.}~\bibnamefont {Sahin}}, \bibinfo {author} {\bibfnamefont {B.~C.~B.}\ \bibnamefont {Symons}}, \bibinfo {author} {\bibfnamefont {S.}~\bibnamefont {Tornow}}, \bibinfo {author} {\bibfnamefont {V.}~\bibnamefont {Valls}}, \bibinfo {author} {\bibfnamefont {S.}~\bibnamefont {Woerner}}, \bibinfo {author} {\bibfnamefont {M.~L.}\ \bibnamefont {Wolf-Bauwens}}, \bibinfo {author} {\bibfnamefont {J.}~\bibnamefont {Yard}}, \bibinfo {author} {\bibfnamefont {S.}~\bibnamefont {Yarkoni}}, \bibinfo {author} {\bibfnamefont {D.}~\bibnamefont {Zechiel}}, \bibinfo {author} {\bibfnamefont {S.}~\bibnamefont {Zhuk}},\ and\ \bibinfo {author} {\bibfnamefont {C.}~\bibnamefont {Zoufal}},\ }\href {https://doi.org/10.48550/arXiv.2312.02279} {\bibinfo {title} {Quantum optimization: Potential, challenges,
  and the path forward}} (\bibinfo {year} {2023}),\ \Eprint {https://arxiv.org/abs/arXiv:2312.02279 [quant-ph]} {arXiv:2312.02279 [quant-ph]} \BibitemShut {NoStop}%
\bibitem [{\citenamefont {Bauer}\ \emph {et~al.}(2020)\citenamefont {Bauer}, \citenamefont {Bravyi}, \citenamefont {Motta},\ and\ \citenamefont {Chan}}]{bauer_quantum_2020}%
  \BibitemOpen
  \bibfield  {author} {\bibinfo {author} {\bibfnamefont {B.}~\bibnamefont {Bauer}}, \bibinfo {author} {\bibfnamefont {S.}~\bibnamefont {Bravyi}}, \bibinfo {author} {\bibfnamefont {M.}~\bibnamefont {Motta}},\ and\ \bibinfo {author} {\bibfnamefont {G.~K.-L.}\ \bibnamefont {Chan}},\ }\href {https://doi.org/10.1021/acs.chemrev.9b00829} {\bibfield  {journal} {\bibinfo  {journal} {Chemical Reviews}\ }\textbf {\bibinfo {volume} {120}},\ \bibinfo {pages} {12685} (\bibinfo {year} {2020})},\ \bibinfo {note} {pMID: 33090772}\BibitemShut {NoStop}%
\bibitem [{\citenamefont {Schuld}\ and\ \citenamefont {Petruccione}(2021)}]{schuld_machine_2021}%
  \BibitemOpen
  \bibfield  {author} {\bibinfo {author} {\bibfnamefont {M.}~\bibnamefont {Schuld}}\ and\ \bibinfo {author} {\bibfnamefont {F.}~\bibnamefont {Petruccione}},\ }\href {https://doi.org/10.1007/978-3-030-83098-4} {\emph {\bibinfo {title} {Machine {{Learning}} with {{Quantum Computers}}}}},\ Quantum {{Science}} and {{Technology}}\ (\bibinfo  {publisher} {Springer International Publishing},\ \bibinfo {address} {Cham},\ \bibinfo {year} {2021})\BibitemShut {NoStop}%
\bibitem [{\citenamefont {Gujju}\ \emph {et~al.}(2024)\citenamefont {Gujju}, \citenamefont {Matsuo},\ and\ \citenamefont {Raymond}}]{gujju_quantum_2024}%
  \BibitemOpen
  \bibfield  {author} {\bibinfo {author} {\bibfnamefont {Y.}~\bibnamefont {Gujju}}, \bibinfo {author} {\bibfnamefont {A.}~\bibnamefont {Matsuo}},\ and\ \bibinfo {author} {\bibfnamefont {R.}~\bibnamefont {Raymond}},\ }\href {https://doi.org/10.1103/PhysRevApplied.21.067001} {\bibfield  {journal} {\bibinfo  {journal} {Phys. Rev. Appl.}\ }\textbf {\bibinfo {volume} {21}},\ \bibinfo {pages} {067001} (\bibinfo {year} {2024})}\BibitemShut {NoStop}%
\bibitem [{\citenamefont {Pirandola}\ \emph {et~al.}(2020)\citenamefont {Pirandola}, \citenamefont {Andersen}, \citenamefont {Banchi}, \citenamefont {Berta}, \citenamefont {Bunandar}, \citenamefont {Colbeck}, \citenamefont {Englund}, \citenamefont {Gehring}, \citenamefont {Lupo}, \citenamefont {Ottaviani}, \citenamefont {Pereira}, \citenamefont {Razavi}, \citenamefont {Shaari}, \citenamefont {Tomamichel}, \citenamefont {Usenko}, \citenamefont {Vallone}, \citenamefont {Villoresi},\ and\ \citenamefont {Wallden}}]{pirandola_advances_2020}%
  \BibitemOpen
  \bibfield  {author} {\bibinfo {author} {\bibfnamefont {S.}~\bibnamefont {Pirandola}}, \bibinfo {author} {\bibfnamefont {U.~L.}\ \bibnamefont {Andersen}}, \bibinfo {author} {\bibfnamefont {L.}~\bibnamefont {Banchi}}, \bibinfo {author} {\bibfnamefont {M.}~\bibnamefont {Berta}}, \bibinfo {author} {\bibfnamefont {D.}~\bibnamefont {Bunandar}}, \bibinfo {author} {\bibfnamefont {R.}~\bibnamefont {Colbeck}}, \bibinfo {author} {\bibfnamefont {D.}~\bibnamefont {Englund}}, \bibinfo {author} {\bibfnamefont {T.}~\bibnamefont {Gehring}}, \bibinfo {author} {\bibfnamefont {C.}~\bibnamefont {Lupo}}, \bibinfo {author} {\bibfnamefont {C.}~\bibnamefont {Ottaviani}}, \bibinfo {author} {\bibfnamefont {J.~L.}\ \bibnamefont {Pereira}}, \bibinfo {author} {\bibfnamefont {M.}~\bibnamefont {Razavi}}, \bibinfo {author} {\bibfnamefont {J.~S.}\ \bibnamefont {Shaari}}, \bibinfo {author} {\bibfnamefont {M.}~\bibnamefont {Tomamichel}}, \bibinfo {author} {\bibfnamefont {V.~C.}\ \bibnamefont {Usenko}}, \bibinfo {author} {\bibfnamefont
  {G.}~\bibnamefont {Vallone}}, \bibinfo {author} {\bibfnamefont {P.}~\bibnamefont {Villoresi}},\ and\ \bibinfo {author} {\bibfnamefont {P.}~\bibnamefont {Wallden}},\ }\href {https://doi.org/10.1364/AOP.361502} {\bibfield  {journal} {\bibinfo  {journal} {Adv. Opt. Photon.}\ }\textbf {\bibinfo {volume} {12}},\ \bibinfo {pages} {1012} (\bibinfo {year} {2020})}\BibitemShut {NoStop}%
\bibitem [{\citenamefont {Cerezo}\ \emph {et~al.}(2021{\natexlab{a}})\citenamefont {Cerezo}, \citenamefont {Arrasmith}, \citenamefont {Babbush}, \citenamefont {Benjamin}, \citenamefont {Endo}, \citenamefont {Fujii}, \citenamefont {McClean}, \citenamefont {Mitarai}, \citenamefont {Yuan}, \citenamefont {Cincio},\ and\ \citenamefont {Coles}}]{cerezo_variational_2021}%
  \BibitemOpen
  \bibfield  {author} {\bibinfo {author} {\bibfnamefont {M.}~\bibnamefont {Cerezo}}, \bibinfo {author} {\bibfnamefont {A.}~\bibnamefont {Arrasmith}}, \bibinfo {author} {\bibfnamefont {R.}~\bibnamefont {Babbush}}, \bibinfo {author} {\bibfnamefont {S.~C.}\ \bibnamefont {Benjamin}}, \bibinfo {author} {\bibfnamefont {S.}~\bibnamefont {Endo}}, \bibinfo {author} {\bibfnamefont {K.}~\bibnamefont {Fujii}}, \bibinfo {author} {\bibfnamefont {J.~R.}\ \bibnamefont {McClean}}, \bibinfo {author} {\bibfnamefont {K.}~\bibnamefont {Mitarai}}, \bibinfo {author} {\bibfnamefont {X.}~\bibnamefont {Yuan}}, \bibinfo {author} {\bibfnamefont {L.}~\bibnamefont {Cincio}},\ and\ \bibinfo {author} {\bibfnamefont {P.~J.}\ \bibnamefont {Coles}},\ }\href {https://doi.org/10.1038/s42254-021-00348-9} {\bibfield  {journal} {\bibinfo  {journal} {Nature Reviews Physics}\ }\textbf {\bibinfo {volume} {3}},\ \bibinfo {pages} {625} (\bibinfo {year} {2021}{\natexlab{a}})},\ \bibinfo {note} {number: 9 Publisher: Nature Publishing Group}\BibitemShut
  {NoStop}%
\bibitem [{\citenamefont {Bharti}\ \emph {et~al.}(2022)\citenamefont {Bharti}, \citenamefont {Cervera-Lierta}, \citenamefont {Kyaw}, \citenamefont {Haug}, \citenamefont {Alperin-Lea}, \citenamefont {Anand}, \citenamefont {Degroote}, \citenamefont {Heimonen}, \citenamefont {Kottmann}, \citenamefont {Menke}, \citenamefont {Mok}, \citenamefont {Sim}, \citenamefont {Kwek},\ and\ \citenamefont {Aspuru-Guzik}}]{bharti_noisy_2022}%
  \BibitemOpen
  \bibfield  {author} {\bibinfo {author} {\bibfnamefont {K.}~\bibnamefont {Bharti}}, \bibinfo {author} {\bibfnamefont {A.}~\bibnamefont {Cervera-Lierta}}, \bibinfo {author} {\bibfnamefont {T.~H.}\ \bibnamefont {Kyaw}}, \bibinfo {author} {\bibfnamefont {T.}~\bibnamefont {Haug}}, \bibinfo {author} {\bibfnamefont {S.}~\bibnamefont {Alperin-Lea}}, \bibinfo {author} {\bibfnamefont {A.}~\bibnamefont {Anand}}, \bibinfo {author} {\bibfnamefont {M.}~\bibnamefont {Degroote}}, \bibinfo {author} {\bibfnamefont {H.}~\bibnamefont {Heimonen}}, \bibinfo {author} {\bibfnamefont {J.~S.}\ \bibnamefont {Kottmann}}, \bibinfo {author} {\bibfnamefont {T.}~\bibnamefont {Menke}}, \bibinfo {author} {\bibfnamefont {W.-K.}\ \bibnamefont {Mok}}, \bibinfo {author} {\bibfnamefont {S.}~\bibnamefont {Sim}}, \bibinfo {author} {\bibfnamefont {L.-C.}\ \bibnamefont {Kwek}},\ and\ \bibinfo {author} {\bibfnamefont {A.}~\bibnamefont {Aspuru-Guzik}},\ }\href {https://doi.org/10.1103/RevModPhys.94.015004} {\bibfield  {journal} {\bibinfo  {journal}
  {Rev. Mod. Phys.}\ }\textbf {\bibinfo {volume} {94}},\ \bibinfo {pages} {015004} (\bibinfo {year} {2022})},\ \bibinfo {note} {publisher: American Physical Society}\BibitemShut {NoStop}%
\bibitem [{\citenamefont {{McClean}}\ \emph {et~al.}(2018)\citenamefont {{McClean}}, \citenamefont {Boixo}, \citenamefont {Smelyanskiy}, \citenamefont {Babbush},\ and\ \citenamefont {Neven}}]{mcclean_barren_2018}%
  \BibitemOpen
  \bibfield  {author} {\bibinfo {author} {\bibfnamefont {J.~R.}\ \bibnamefont {{McClean}}}, \bibinfo {author} {\bibfnamefont {S.}~\bibnamefont {Boixo}}, \bibinfo {author} {\bibfnamefont {V.~N.}\ \bibnamefont {Smelyanskiy}}, \bibinfo {author} {\bibfnamefont {R.}~\bibnamefont {Babbush}},\ and\ \bibinfo {author} {\bibfnamefont {H.}~\bibnamefont {Neven}},\ }\href {https://doi.org/10.1038/s41467-018-07090-4} {\bibfield  {journal} {\bibinfo  {journal} {Nature Communications}\ }\textbf {\bibinfo {volume} {9}},\ \bibinfo {pages} {4812} (\bibinfo {year} {2018})},\ \bibinfo {note} {number: 1 Publisher: Nature Publishing Group}\BibitemShut {NoStop}%
\bibitem [{\citenamefont {Arrasmith}\ \emph {et~al.}(2022)\citenamefont {Arrasmith}, \citenamefont {Holmes}, \citenamefont {Cerezo},\ and\ \citenamefont {Coles}}]{arrasmith_equivalence_2022}%
  \BibitemOpen
  \bibfield  {author} {\bibinfo {author} {\bibfnamefont {A.}~\bibnamefont {Arrasmith}}, \bibinfo {author} {\bibfnamefont {Z.}~\bibnamefont {Holmes}}, \bibinfo {author} {\bibfnamefont {M.}~\bibnamefont {Cerezo}},\ and\ \bibinfo {author} {\bibfnamefont {P.~J.}\ \bibnamefont {Coles}},\ }\href {https://doi.org/10.1088/2058-9565/ac7d06} {\bibfield  {journal} {\bibinfo  {journal} {Quantum Science and Technology}\ }\textbf {\bibinfo {volume} {7}},\ \bibinfo {pages} {045015} (\bibinfo {year} {2022})},\ \bibinfo {note} {publisher: IOP Publishing}\BibitemShut {NoStop}%
\bibitem [{\citenamefont {Larocca}\ \emph {et~al.}(2024)\citenamefont {Larocca}, \citenamefont {Thanasilp}, \citenamefont {Wang}, \citenamefont {Sharma}, \citenamefont {Biamonte}, \citenamefont {Coles}, \citenamefont {Cincio}, \citenamefont {McClean}, \citenamefont {Holmes},\ and\ \citenamefont {Cerezo}}]{Larocca2024}%
  \BibitemOpen
  \bibfield  {author} {\bibinfo {author} {\bibfnamefont {M.}~\bibnamefont {Larocca}}, \bibinfo {author} {\bibfnamefont {S.}~\bibnamefont {Thanasilp}}, \bibinfo {author} {\bibfnamefont {S.}~\bibnamefont {Wang}}, \bibinfo {author} {\bibfnamefont {K.}~\bibnamefont {Sharma}}, \bibinfo {author} {\bibfnamefont {J.}~\bibnamefont {Biamonte}}, \bibinfo {author} {\bibfnamefont {P.~J.}\ \bibnamefont {Coles}}, \bibinfo {author} {\bibfnamefont {L.}~\bibnamefont {Cincio}}, \bibinfo {author} {\bibfnamefont {J.~R.}\ \bibnamefont {McClean}}, \bibinfo {author} {\bibfnamefont {Z.}~\bibnamefont {Holmes}},\ and\ \bibinfo {author} {\bibfnamefont {M.}~\bibnamefont {Cerezo}},\ }\href {https://doi.org/10.48550/ARXIV.2405.00781} {\bibfield  {journal} {\bibinfo  {journal} {arXiv:2405.00781}\ } (\bibinfo {year} {2024})}\BibitemShut {NoStop}%
\bibitem [{\citenamefont {Cerezo}\ \emph {et~al.}(2021{\natexlab{b}})\citenamefont {Cerezo}, \citenamefont {Sone}, \citenamefont {Volkoff}, \citenamefont {Cincio},\ and\ \citenamefont {Coles}}]{Cerezo2021}%
  \BibitemOpen
  \bibfield  {author} {\bibinfo {author} {\bibfnamefont {M.}~\bibnamefont {Cerezo}}, \bibinfo {author} {\bibfnamefont {A.}~\bibnamefont {Sone}}, \bibinfo {author} {\bibfnamefont {T.}~\bibnamefont {Volkoff}}, \bibinfo {author} {\bibfnamefont {L.}~\bibnamefont {Cincio}},\ and\ \bibinfo {author} {\bibfnamefont {P.~J.}\ \bibnamefont {Coles}},\ }\href {https://doi.org/10.1038/s41467-021-21728-w} {\bibfield  {journal} {\bibinfo  {journal} {Nature Communications}\ }\textbf {\bibinfo {volume} {12}},\ \bibinfo {pages} {1791} (\bibinfo {year} {2021}{\natexlab{b}})}\BibitemShut {NoStop}%
\bibitem [{\citenamefont {Wang}\ \emph {et~al.}(2021)\citenamefont {Wang}, \citenamefont {Fontana}, \citenamefont {Cerezo}, \citenamefont {Sharma}, \citenamefont {Sone}, \citenamefont {Cincio},\ and\ \citenamefont {Coles}}]{Wang2021}%
  \BibitemOpen
  \bibfield  {author} {\bibinfo {author} {\bibfnamefont {S.}~\bibnamefont {Wang}}, \bibinfo {author} {\bibfnamefont {E.}~\bibnamefont {Fontana}}, \bibinfo {author} {\bibfnamefont {M.}~\bibnamefont {Cerezo}}, \bibinfo {author} {\bibfnamefont {K.}~\bibnamefont {Sharma}}, \bibinfo {author} {\bibfnamefont {A.}~\bibnamefont {Sone}}, \bibinfo {author} {\bibfnamefont {L.}~\bibnamefont {Cincio}},\ and\ \bibinfo {author} {\bibfnamefont {P.~J.}\ \bibnamefont {Coles}},\ }\href {https://doi.org/10.1038/s41467-021-27045-6} {\bibfield  {journal} {\bibinfo  {journal} {Nature Communications}\ }\textbf {\bibinfo {volume} {12}},\ \bibinfo {pages} {6961} (\bibinfo {year} {2021})}\BibitemShut {NoStop}%
\bibitem [{\citenamefont {Ortiz~Marrero}\ \emph {et~al.}(2021)\citenamefont {Ortiz~Marrero}, \citenamefont {Kieferov\'a},\ and\ \citenamefont {Wiebe}}]{Ortiz2021}%
  \BibitemOpen
  \bibfield  {author} {\bibinfo {author} {\bibfnamefont {C.}~\bibnamefont {Ortiz~Marrero}}, \bibinfo {author} {\bibfnamefont {M.}~\bibnamefont {Kieferov\'a}},\ and\ \bibinfo {author} {\bibfnamefont {N.}~\bibnamefont {Wiebe}},\ }\href {https://doi.org/10.1103/PRXQuantum.2.040316} {\bibfield  {journal} {\bibinfo  {journal} {PRX Quantum}\ }\textbf {\bibinfo {volume} {2}},\ \bibinfo {pages} {040316} (\bibinfo {year} {2021})}\BibitemShut {NoStop}%
\bibitem [{\citenamefont {Barkoutsos}\ \emph {et~al.}(2020)\citenamefont {Barkoutsos}, \citenamefont {Nannicini}, \citenamefont {Robert}, \citenamefont {Tavernelli},\ and\ \citenamefont {Woerner}}]{barkoutsos_improving_2020}%
  \BibitemOpen
  \bibfield  {author} {\bibinfo {author} {\bibfnamefont {P.~K.}\ \bibnamefont {Barkoutsos}}, \bibinfo {author} {\bibfnamefont {G.}~\bibnamefont {Nannicini}}, \bibinfo {author} {\bibfnamefont {A.}~\bibnamefont {Robert}}, \bibinfo {author} {\bibfnamefont {I.}~\bibnamefont {Tavernelli}},\ and\ \bibinfo {author} {\bibfnamefont {S.}~\bibnamefont {Woerner}},\ }\href {https://doi.org/10.22331/q-2020-04-20-256} {\bibfield  {journal} {\bibinfo  {journal} {Quantum}\ }\textbf {\bibinfo {volume} {4}},\ \bibinfo {pages} {256} (\bibinfo {year} {2020})},\ \bibinfo {note} {publisher: Verein zur Förderung des Open Access Publizierens in den Quantenwissenschaften}\BibitemShut {NoStop}%
\bibitem [{\citenamefont {Hamilton}\ \emph {et~al.}(2017)\citenamefont {Hamilton}, \citenamefont {Kruse}, \citenamefont {Sansoni}, \citenamefont {Barkhofen}, \citenamefont {Silberhorn},\ and\ \citenamefont {Jex}}]{hamilton_gaussian_2017}%
  \BibitemOpen
  \bibfield  {author} {\bibinfo {author} {\bibfnamefont {C.~S.}\ \bibnamefont {Hamilton}}, \bibinfo {author} {\bibfnamefont {R.}~\bibnamefont {Kruse}}, \bibinfo {author} {\bibfnamefont {L.}~\bibnamefont {Sansoni}}, \bibinfo {author} {\bibfnamefont {S.}~\bibnamefont {Barkhofen}}, \bibinfo {author} {\bibfnamefont {C.}~\bibnamefont {Silberhorn}},\ and\ \bibinfo {author} {\bibfnamefont {I.}~\bibnamefont {Jex}},\ }\href {https://doi.org/10.1103/PhysRevLett.119.170501} {\bibfield  {journal} {\bibinfo  {journal} {Phys. Rev. Lett.}\ }\textbf {\bibinfo {volume} {119}},\ \bibinfo {pages} {170501} (\bibinfo {year} {2017})},\ \bibinfo {note} {publisher: American Physical Society}\BibitemShut {NoStop}%
\bibitem [{\citenamefont {Kruse}\ \emph {et~al.}(2019)\citenamefont {Kruse}, \citenamefont {Hamilton}, \citenamefont {Sansoni}, \citenamefont {Barkhofen}, \citenamefont {Silberhorn},\ and\ \citenamefont {Jex}}]{kruse_detailed_2019}%
  \BibitemOpen
  \bibfield  {author} {\bibinfo {author} {\bibfnamefont {R.}~\bibnamefont {Kruse}}, \bibinfo {author} {\bibfnamefont {C.~S.}\ \bibnamefont {Hamilton}}, \bibinfo {author} {\bibfnamefont {L.}~\bibnamefont {Sansoni}}, \bibinfo {author} {\bibfnamefont {S.}~\bibnamefont {Barkhofen}}, \bibinfo {author} {\bibfnamefont {C.}~\bibnamefont {Silberhorn}},\ and\ \bibinfo {author} {\bibfnamefont {I.}~\bibnamefont {Jex}},\ }\href {https://doi.org/10.1103/PhysRevA.100.032326} {\bibfield  {journal} {\bibinfo  {journal} {Phys. Rev. A}\ }\textbf {\bibinfo {volume} {100}},\ \bibinfo {pages} {032326} (\bibinfo {year} {2019})},\ \bibinfo {note} {publisher: American Physical Society}\BibitemShut {NoStop}%
\bibitem [{\citenamefont {Hangleiter}\ and\ \citenamefont {Eisert}(2023)}]{hangleiter_computational_2022}%
  \BibitemOpen
  \bibfield  {author} {\bibinfo {author} {\bibfnamefont {D.}~\bibnamefont {Hangleiter}}\ and\ \bibinfo {author} {\bibfnamefont {J.}~\bibnamefont {Eisert}},\ }\href {https://doi.org/10.1103/RevModPhys.95.035001} {\bibfield  {journal} {\bibinfo  {journal} {Rev. Mod. Phys.}\ }\textbf {\bibinfo {volume} {95}},\ \bibinfo {pages} {035001} (\bibinfo {year} {2023})}\BibitemShut {NoStop}%
\bibitem [{\citenamefont {Zhong}\ \emph {et~al.}(2019)\citenamefont {Zhong}, \citenamefont {Peng}, \citenamefont {Li}, \citenamefont {Hu}, \citenamefont {Li}, \citenamefont {Qin}, \citenamefont {Wu}, \citenamefont {Zhang}, \citenamefont {Li}, \citenamefont {Zhang}, \citenamefont {Wang}, \citenamefont {You}, \citenamefont {Jiang}, \citenamefont {Li}, \citenamefont {Liu}, \citenamefont {Dowling}, \citenamefont {Lu},\ and\ \citenamefont {Pan}}]{zhong_experimental_2019}%
  \BibitemOpen
  \bibfield  {author} {\bibinfo {author} {\bibfnamefont {H.-S.}\ \bibnamefont {Zhong}}, \bibinfo {author} {\bibfnamefont {L.-C.}\ \bibnamefont {Peng}}, \bibinfo {author} {\bibfnamefont {Y.}~\bibnamefont {Li}}, \bibinfo {author} {\bibfnamefont {Y.}~\bibnamefont {Hu}}, \bibinfo {author} {\bibfnamefont {W.}~\bibnamefont {Li}}, \bibinfo {author} {\bibfnamefont {J.}~\bibnamefont {Qin}}, \bibinfo {author} {\bibfnamefont {D.}~\bibnamefont {Wu}}, \bibinfo {author} {\bibfnamefont {W.}~\bibnamefont {Zhang}}, \bibinfo {author} {\bibfnamefont {H.}~\bibnamefont {Li}}, \bibinfo {author} {\bibfnamefont {L.}~\bibnamefont {Zhang}}, \bibinfo {author} {\bibfnamefont {Z.}~\bibnamefont {Wang}}, \bibinfo {author} {\bibfnamefont {L.}~\bibnamefont {You}}, \bibinfo {author} {\bibfnamefont {X.}~\bibnamefont {Jiang}}, \bibinfo {author} {\bibfnamefont {L.}~\bibnamefont {Li}}, \bibinfo {author} {\bibfnamefont {N.-L.}\ \bibnamefont {Liu}}, \bibinfo {author} {\bibfnamefont {J.~P.}\ \bibnamefont {Dowling}}, \bibinfo {author} {\bibfnamefont
  {C.-Y.}\ \bibnamefont {Lu}},\ and\ \bibinfo {author} {\bibfnamefont {J.-W.}\ \bibnamefont {Pan}},\ }\href {https://doi.org/10.1016/j.scib.2019.04.007} {\bibfield  {journal} {\bibinfo  {journal} {Science Bulletin}\ }\textbf {\bibinfo {volume} {64}},\ \bibinfo {pages} {511} (\bibinfo {year} {2019})}\BibitemShut {NoStop}%
\bibitem [{\citenamefont {Zhong}\ \emph {et~al.}(2020)\citenamefont {Zhong}, \citenamefont {Wang}, \citenamefont {Deng}, \citenamefont {Chen}, \citenamefont {Peng}, \citenamefont {Luo}, \citenamefont {Qin}, \citenamefont {Wu}, \citenamefont {Ding}, \citenamefont {Hu}, \citenamefont {Hu}, \citenamefont {Yang}, \citenamefont {Zhang}, \citenamefont {Li}, \citenamefont {Li}, \citenamefont {Jiang}, \citenamefont {Gan}, \citenamefont {Yang}, \citenamefont {You}, \citenamefont {Wang}, \citenamefont {Li}, \citenamefont {Liu}, \citenamefont {Lu},\ and\ \citenamefont {Pan}}]{zhong_quantum_2020}%
  \BibitemOpen
  \bibfield  {author} {\bibinfo {author} {\bibfnamefont {H.-S.}\ \bibnamefont {Zhong}}, \bibinfo {author} {\bibfnamefont {H.}~\bibnamefont {Wang}}, \bibinfo {author} {\bibfnamefont {Y.-H.}\ \bibnamefont {Deng}}, \bibinfo {author} {\bibfnamefont {M.-C.}\ \bibnamefont {Chen}}, \bibinfo {author} {\bibfnamefont {L.-C.}\ \bibnamefont {Peng}}, \bibinfo {author} {\bibfnamefont {Y.-H.}\ \bibnamefont {Luo}}, \bibinfo {author} {\bibfnamefont {J.}~\bibnamefont {Qin}}, \bibinfo {author} {\bibfnamefont {D.}~\bibnamefont {Wu}}, \bibinfo {author} {\bibfnamefont {X.}~\bibnamefont {Ding}}, \bibinfo {author} {\bibfnamefont {Y.}~\bibnamefont {Hu}}, \bibinfo {author} {\bibfnamefont {P.}~\bibnamefont {Hu}}, \bibinfo {author} {\bibfnamefont {X.-Y.}\ \bibnamefont {Yang}}, \bibinfo {author} {\bibfnamefont {W.-J.}\ \bibnamefont {Zhang}}, \bibinfo {author} {\bibfnamefont {H.}~\bibnamefont {Li}}, \bibinfo {author} {\bibfnamefont {Y.}~\bibnamefont {Li}}, \bibinfo {author} {\bibfnamefont {X.}~\bibnamefont {Jiang}}, \bibinfo {author}
  {\bibfnamefont {L.}~\bibnamefont {Gan}}, \bibinfo {author} {\bibfnamefont {G.}~\bibnamefont {Yang}}, \bibinfo {author} {\bibfnamefont {L.}~\bibnamefont {You}}, \bibinfo {author} {\bibfnamefont {Z.}~\bibnamefont {Wang}}, \bibinfo {author} {\bibfnamefont {L.}~\bibnamefont {Li}}, \bibinfo {author} {\bibfnamefont {N.-L.}\ \bibnamefont {Liu}}, \bibinfo {author} {\bibfnamefont {C.-Y.}\ \bibnamefont {Lu}},\ and\ \bibinfo {author} {\bibfnamefont {J.-W.}\ \bibnamefont {Pan}},\ }\href {https://doi.org/10.1126/science.abe8770} {\bibfield  {journal} {\bibinfo  {journal} {Science}\ }\textbf {\bibinfo {volume} {370}},\ \bibinfo {pages} {1460} (\bibinfo {year} {2020})},\ \bibinfo {note} {publisher: American Association for the Advancement of Science}\BibitemShut {NoStop}%
\bibitem [{\citenamefont {Zhong}\ \emph {et~al.}(2021)\citenamefont {Zhong}, \citenamefont {Deng}, \citenamefont {Qin}, \citenamefont {Wang}, \citenamefont {Chen}, \citenamefont {Peng}, \citenamefont {Luo}, \citenamefont {Wu}, \citenamefont {Gong}, \citenamefont {Su}, \citenamefont {Hu}, \citenamefont {Hu}, \citenamefont {Yang}, \citenamefont {Zhang}, \citenamefont {Li}, \citenamefont {Li}, \citenamefont {Jiang}, \citenamefont {Gan}, \citenamefont {Yang}, \citenamefont {You}, \citenamefont {Wang}, \citenamefont {Li}, \citenamefont {Liu}, \citenamefont {Renema}, \citenamefont {Lu},\ and\ \citenamefont {Pan}}]{zhong_phase-programmable_2021}%
  \BibitemOpen
  \bibfield  {author} {\bibinfo {author} {\bibfnamefont {H.-S.}\ \bibnamefont {Zhong}}, \bibinfo {author} {\bibfnamefont {Y.-H.}\ \bibnamefont {Deng}}, \bibinfo {author} {\bibfnamefont {J.}~\bibnamefont {Qin}}, \bibinfo {author} {\bibfnamefont {H.}~\bibnamefont {Wang}}, \bibinfo {author} {\bibfnamefont {M.-C.}\ \bibnamefont {Chen}}, \bibinfo {author} {\bibfnamefont {L.-C.}\ \bibnamefont {Peng}}, \bibinfo {author} {\bibfnamefont {Y.-H.}\ \bibnamefont {Luo}}, \bibinfo {author} {\bibfnamefont {D.}~\bibnamefont {Wu}}, \bibinfo {author} {\bibfnamefont {S.-Q.}\ \bibnamefont {Gong}}, \bibinfo {author} {\bibfnamefont {H.}~\bibnamefont {Su}}, \bibinfo {author} {\bibfnamefont {Y.}~\bibnamefont {Hu}}, \bibinfo {author} {\bibfnamefont {P.}~\bibnamefont {Hu}}, \bibinfo {author} {\bibfnamefont {X.-Y.}\ \bibnamefont {Yang}}, \bibinfo {author} {\bibfnamefont {W.-J.}\ \bibnamefont {Zhang}}, \bibinfo {author} {\bibfnamefont {H.}~\bibnamefont {Li}}, \bibinfo {author} {\bibfnamefont {Y.}~\bibnamefont {Li}}, \bibinfo {author}
  {\bibfnamefont {X.}~\bibnamefont {Jiang}}, \bibinfo {author} {\bibfnamefont {L.}~\bibnamefont {Gan}}, \bibinfo {author} {\bibfnamefont {G.}~\bibnamefont {Yang}}, \bibinfo {author} {\bibfnamefont {L.}~\bibnamefont {You}}, \bibinfo {author} {\bibfnamefont {Z.}~\bibnamefont {Wang}}, \bibinfo {author} {\bibfnamefont {L.}~\bibnamefont {Li}}, \bibinfo {author} {\bibfnamefont {N.-L.}\ \bibnamefont {Liu}}, \bibinfo {author} {\bibfnamefont {J.~J.}\ \bibnamefont {Renema}}, \bibinfo {author} {\bibfnamefont {C.-Y.}\ \bibnamefont {Lu}},\ and\ \bibinfo {author} {\bibfnamefont {J.-W.}\ \bibnamefont {Pan}},\ }\href {https://doi.org/10.1103/PhysRevLett.127.180502} {\bibfield  {journal} {\bibinfo  {journal} {Phys. Rev. Lett.}\ }\textbf {\bibinfo {volume} {127}},\ \bibinfo {pages} {180502} (\bibinfo {year} {2021})},\ \bibinfo {note} {publisher: American Physical Society}\BibitemShut {NoStop}%
\bibitem [{\citenamefont {Madsen}\ \emph {et~al.}(2022)\citenamefont {Madsen}, \citenamefont {Laudenbach}, \citenamefont {Askarani}, \citenamefont {Rortais}, \citenamefont {Vincent}, \citenamefont {Bulmer}, \citenamefont {Miatto}, \citenamefont {Neuhaus}, \citenamefont {Helt}, \citenamefont {Collins}, \citenamefont {Lita}, \citenamefont {Gerrits}, \citenamefont {Nam}, \citenamefont {Vaidya}, \citenamefont {Menotti}, \citenamefont {Dhand}, \citenamefont {Vernon}, \citenamefont {Quesada},\ and\ \citenamefont {Lavoie}}]{madsen_quantum_2022}%
  \BibitemOpen
  \bibfield  {author} {\bibinfo {author} {\bibfnamefont {L.~S.}\ \bibnamefont {Madsen}}, \bibinfo {author} {\bibfnamefont {F.}~\bibnamefont {Laudenbach}}, \bibinfo {author} {\bibfnamefont {M.~F.}\ \bibnamefont {Askarani}}, \bibinfo {author} {\bibfnamefont {F.}~\bibnamefont {Rortais}}, \bibinfo {author} {\bibfnamefont {T.}~\bibnamefont {Vincent}}, \bibinfo {author} {\bibfnamefont {J.~F.~F.}\ \bibnamefont {Bulmer}}, \bibinfo {author} {\bibfnamefont {F.~M.}\ \bibnamefont {Miatto}}, \bibinfo {author} {\bibfnamefont {L.}~\bibnamefont {Neuhaus}}, \bibinfo {author} {\bibfnamefont {L.~G.}\ \bibnamefont {Helt}}, \bibinfo {author} {\bibfnamefont {M.~J.}\ \bibnamefont {Collins}}, \bibinfo {author} {\bibfnamefont {A.~E.}\ \bibnamefont {Lita}}, \bibinfo {author} {\bibfnamefont {T.}~\bibnamefont {Gerrits}}, \bibinfo {author} {\bibfnamefont {S.~W.}\ \bibnamefont {Nam}}, \bibinfo {author} {\bibfnamefont {V.~D.}\ \bibnamefont {Vaidya}}, \bibinfo {author} {\bibfnamefont {M.}~\bibnamefont {Menotti}}, \bibinfo {author}
  {\bibfnamefont {I.}~\bibnamefont {Dhand}}, \bibinfo {author} {\bibfnamefont {Z.}~\bibnamefont {Vernon}}, \bibinfo {author} {\bibfnamefont {N.}~\bibnamefont {Quesada}},\ and\ \bibinfo {author} {\bibfnamefont {J.}~\bibnamefont {Lavoie}},\ }\href {https://doi.org/10.1038/s41586-022-04725-x} {\bibfield  {journal} {\bibinfo  {journal} {Nature}\ }\textbf {\bibinfo {volume} {606}},\ \bibinfo {pages} {75} (\bibinfo {year} {2022})},\ \bibinfo {note} {number: 7912 Publisher: Nature Publishing Group}\BibitemShut {NoStop}%
\bibitem [{\citenamefont {Deng}\ \emph {et~al.}(2023)\citenamefont {Deng}, \citenamefont {Gu}, \citenamefont {Liu}, \citenamefont {Gong}, \citenamefont {Su}, \citenamefont {Zhang}, \citenamefont {Tang}, \citenamefont {Jia}, \citenamefont {Xu}, \citenamefont {Chen}, \citenamefont {Qin}, \citenamefont {Peng}, \citenamefont {Yan}, \citenamefont {Hu}, \citenamefont {Huang}, \citenamefont {Li}, \citenamefont {Li}, \citenamefont {Chen}, \citenamefont {Jiang}, \citenamefont {Gan}, \citenamefont {Yang}, \citenamefont {You}, \citenamefont {Li}, \citenamefont {Zhong}, \citenamefont {Wang}, \citenamefont {Liu}, \citenamefont {Renema}, \citenamefont {Lu},\ and\ \citenamefont {Pan}}]{deng_gaussian_2023}%
  \BibitemOpen
  \bibfield  {author} {\bibinfo {author} {\bibfnamefont {Y.-H.}\ \bibnamefont {Deng}}, \bibinfo {author} {\bibfnamefont {Y.-C.}\ \bibnamefont {Gu}}, \bibinfo {author} {\bibfnamefont {H.-L.}\ \bibnamefont {Liu}}, \bibinfo {author} {\bibfnamefont {S.-Q.}\ \bibnamefont {Gong}}, \bibinfo {author} {\bibfnamefont {H.}~\bibnamefont {Su}}, \bibinfo {author} {\bibfnamefont {Z.-J.}\ \bibnamefont {Zhang}}, \bibinfo {author} {\bibfnamefont {H.-Y.}\ \bibnamefont {Tang}}, \bibinfo {author} {\bibfnamefont {M.-H.}\ \bibnamefont {Jia}}, \bibinfo {author} {\bibfnamefont {J.-M.}\ \bibnamefont {Xu}}, \bibinfo {author} {\bibfnamefont {M.-C.}\ \bibnamefont {Chen}}, \bibinfo {author} {\bibfnamefont {J.}~\bibnamefont {Qin}}, \bibinfo {author} {\bibfnamefont {L.-C.}\ \bibnamefont {Peng}}, \bibinfo {author} {\bibfnamefont {J.}~\bibnamefont {Yan}}, \bibinfo {author} {\bibfnamefont {Y.}~\bibnamefont {Hu}}, \bibinfo {author} {\bibfnamefont {J.}~\bibnamefont {Huang}}, \bibinfo {author} {\bibfnamefont {H.}~\bibnamefont {Li}}, \bibinfo
  {author} {\bibfnamefont {Y.}~\bibnamefont {Li}}, \bibinfo {author} {\bibfnamefont {Y.}~\bibnamefont {Chen}}, \bibinfo {author} {\bibfnamefont {X.}~\bibnamefont {Jiang}}, \bibinfo {author} {\bibfnamefont {L.}~\bibnamefont {Gan}}, \bibinfo {author} {\bibfnamefont {G.}~\bibnamefont {Yang}}, \bibinfo {author} {\bibfnamefont {L.}~\bibnamefont {You}}, \bibinfo {author} {\bibfnamefont {L.}~\bibnamefont {Li}}, \bibinfo {author} {\bibfnamefont {H.-S.}\ \bibnamefont {Zhong}}, \bibinfo {author} {\bibfnamefont {H.}~\bibnamefont {Wang}}, \bibinfo {author} {\bibfnamefont {N.-L.}\ \bibnamefont {Liu}}, \bibinfo {author} {\bibfnamefont {J.~J.}\ \bibnamefont {Renema}}, \bibinfo {author} {\bibfnamefont {C.-Y.}\ \bibnamefont {Lu}},\ and\ \bibinfo {author} {\bibfnamefont {J.-W.}\ \bibnamefont {Pan}},\ }\href {https://doi.org/10.1103/PhysRevLett.131.150601} {\bibfield  {journal} {\bibinfo  {journal} {Phys. Rev. Lett.}\ }\textbf {\bibinfo {volume} {131}},\ \bibinfo {pages} {150601} (\bibinfo {year} {2023})}\BibitemShut {NoStop}%
\bibitem [{\citenamefont {Oh}\ \emph {et~al.}(2024)\citenamefont {Oh}, \citenamefont {Liu}, \citenamefont {Alexeev}, \citenamefont {Fefferman},\ and\ \citenamefont {Jiang}}]{oh_classical_2023}%
  \BibitemOpen
  \bibfield  {author} {\bibinfo {author} {\bibfnamefont {C.}~\bibnamefont {Oh}}, \bibinfo {author} {\bibfnamefont {M.}~\bibnamefont {Liu}}, \bibinfo {author} {\bibfnamefont {Y.}~\bibnamefont {Alexeev}}, \bibinfo {author} {\bibfnamefont {B.}~\bibnamefont {Fefferman}},\ and\ \bibinfo {author} {\bibfnamefont {L.}~\bibnamefont {Jiang}},\ }\href {https://doi.org/10.1038/s41567-024-02535-8} {\bibfield  {journal} {\bibinfo  {journal} {Nat. Phys.}\ }\textbf {\bibinfo {volume} {20}},\ \bibinfo {pages} {1461} (\bibinfo {year} {2024})}\BibitemShut {NoStop}%
\bibitem [{\citenamefont {Chai}\ \emph {et~al.}(2023{\natexlab{a}})\citenamefont {Chai}, \citenamefont {Funcke}, \citenamefont {Hartung}, \citenamefont {Jansen}, \citenamefont {K\"uhn}, \citenamefont {Stornati},\ and\ \citenamefont {Stollenwerk}}]{chai_towards_2023}%
  \BibitemOpen
  \bibfield  {author} {\bibinfo {author} {\bibfnamefont {Y.}~\bibnamefont {Chai}}, \bibinfo {author} {\bibfnamefont {L.}~\bibnamefont {Funcke}}, \bibinfo {author} {\bibfnamefont {T.}~\bibnamefont {Hartung}}, \bibinfo {author} {\bibfnamefont {K.}~\bibnamefont {Jansen}}, \bibinfo {author} {\bibfnamefont {S.}~\bibnamefont {K\"uhn}}, \bibinfo {author} {\bibfnamefont {P.}~\bibnamefont {Stornati}},\ and\ \bibinfo {author} {\bibfnamefont {T.}~\bibnamefont {Stollenwerk}},\ }\href {https://doi.org/10.1103/PhysRevApplied.20.064025} {\bibfield  {journal} {\bibinfo  {journal} {Phys. Rev. Appl.}\ }\textbf {\bibinfo {volume} {20}},\ \bibinfo {pages} {064025} (\bibinfo {year} {2023}{\natexlab{a}})}\BibitemShut {NoStop}%
\bibitem [{\citenamefont {Cazalis}\ \emph {et~al.}(2023)\citenamefont {Cazalis}, \citenamefont {Chai}, \citenamefont {Jansen}, \citenamefont {Kühn},\ and\ \citenamefont {Shah}}]{cazalis_gaussian_2023}%
  \BibitemOpen
  \bibfield  {author} {\bibinfo {author} {\bibfnamefont {J.}~\bibnamefont {Cazalis}}, \bibinfo {author} {\bibfnamefont {Y.}~\bibnamefont {Chai}}, \bibinfo {author} {\bibfnamefont {K.}~\bibnamefont {Jansen}}, \bibinfo {author} {\bibfnamefont {S.}~\bibnamefont {Kühn}},\ and\ \bibinfo {author} {\bibfnamefont {T.}~\bibnamefont {Shah}},\ }in\ \href {https://doi.org/10.1109/QCE57702.2023.10268} {\emph {\bibinfo {booktitle} {2023 {IEEE} International Conference on Quantum Computing and Engineering ({QCE})}}},\ Vol.~\bibinfo {volume} {02}\ (\bibinfo {year} {2023})\ pp.\ \bibinfo {pages} {332--333}\BibitemShut {NoStop}%
\bibitem [{\citenamefont {Banchi}\ \emph {et~al.}(2020)\citenamefont {Banchi}, \citenamefont {Quesada},\ and\ \citenamefont {Arrazola}}]{banchi_training_2020}%
  \BibitemOpen
  \bibfield  {author} {\bibinfo {author} {\bibfnamefont {L.}~\bibnamefont {Banchi}}, \bibinfo {author} {\bibfnamefont {N.}~\bibnamefont {Quesada}},\ and\ \bibinfo {author} {\bibfnamefont {J.~M.}\ \bibnamefont {Arrazola}},\ }\href {https://doi.org/10.1103/PhysRevA.102.012417} {\bibfield  {journal} {\bibinfo  {journal} {Phys. Rev. A}\ }\textbf {\bibinfo {volume} {102}},\ \bibinfo {pages} {012417} (\bibinfo {year} {2020})},\ \bibinfo {note} {publisher: American Physical Society}\BibitemShut {NoStop}%
\bibitem [{\citenamefont {Quesada}\ \emph {et~al.}(2018)\citenamefont {Quesada}, \citenamefont {Arrazola},\ and\ \citenamefont {Killoran}}]{quesada_gaussian_2018}%
  \BibitemOpen
  \bibfield  {author} {\bibinfo {author} {\bibfnamefont {N.}~\bibnamefont {Quesada}}, \bibinfo {author} {\bibfnamefont {J.~M.}\ \bibnamefont {Arrazola}},\ and\ \bibinfo {author} {\bibfnamefont {N.}~\bibnamefont {Killoran}},\ }\href {https://doi.org/10.1103/PhysRevA.98.062322} {\bibfield  {journal} {\bibinfo  {journal} {Phys. Rev. A}\ }\textbf {\bibinfo {volume} {98}},\ \bibinfo {pages} {062322} (\bibinfo {year} {2018})},\ \bibinfo {note} {publisher: American Physical Society}\BibitemShut {NoStop}%
\bibitem [{\citenamefont {Bulmer}\ \emph {et~al.}(2022{\natexlab{a}})\citenamefont {Bulmer}, \citenamefont {Paesani}, \citenamefont {Chadwick},\ and\ \citenamefont {Quesada}}]{bulmer_threshold_2022}%
  \BibitemOpen
  \bibfield  {author} {\bibinfo {author} {\bibfnamefont {J.~F.~F.}\ \bibnamefont {Bulmer}}, \bibinfo {author} {\bibfnamefont {S.}~\bibnamefont {Paesani}}, \bibinfo {author} {\bibfnamefont {R.~S.}\ \bibnamefont {Chadwick}},\ and\ \bibinfo {author} {\bibfnamefont {N.}~\bibnamefont {Quesada}},\ }\href {https://doi.org/10.1103/PhysRevA.106.043712} {\bibfield  {journal} {\bibinfo  {journal} {Phys. Rev. A}\ }\textbf {\bibinfo {volume} {106}},\ \bibinfo {pages} {043712} (\bibinfo {year} {2022}{\natexlab{a}})},\ \bibinfo {note} {publisher: American Physical Society}\BibitemShut {NoStop}%
\bibitem [{\citenamefont {Clements}\ \emph {et~al.}(2016)\citenamefont {Clements}, \citenamefont {Humphreys}, \citenamefont {Metcalf}, \citenamefont {Kolthammer},\ and\ \citenamefont {Walmsley}}]{clements_optimal_2016}%
  \BibitemOpen
  \bibfield  {author} {\bibinfo {author} {\bibfnamefont {W.~R.}\ \bibnamefont {Clements}}, \bibinfo {author} {\bibfnamefont {P.~C.}\ \bibnamefont {Humphreys}}, \bibinfo {author} {\bibfnamefont {B.~J.}\ \bibnamefont {Metcalf}}, \bibinfo {author} {\bibfnamefont {W.~S.}\ \bibnamefont {Kolthammer}},\ and\ \bibinfo {author} {\bibfnamefont {I.~A.}\ \bibnamefont {Walmsley}},\ }\href {https://doi.org/10.1364/OPTICA.3.001460} {\bibfield  {journal} {\bibinfo  {journal} {Optica}\ }\textbf {\bibinfo {volume} {3}},\ \bibinfo {pages} {1460} (\bibinfo {year} {2016})}\BibitemShut {NoStop}%
\bibitem [{\citenamefont {Thekkadath}\ \emph {et~al.}(2022)\citenamefont {Thekkadath}, \citenamefont {Sempere-Llagostera}, \citenamefont {Bell}, \citenamefont {Patel}, \citenamefont {Kim},\ and\ \citenamefont {Walmsley}}]{thekkadath_experimental_2022}%
  \BibitemOpen
  \bibfield  {author} {\bibinfo {author} {\bibfnamefont {G.}~\bibnamefont {Thekkadath}}, \bibinfo {author} {\bibfnamefont {S.}~\bibnamefont {Sempere-Llagostera}}, \bibinfo {author} {\bibfnamefont {B.}~\bibnamefont {Bell}}, \bibinfo {author} {\bibfnamefont {R.}~\bibnamefont {Patel}}, \bibinfo {author} {\bibfnamefont {M.}~\bibnamefont {Kim}},\ and\ \bibinfo {author} {\bibfnamefont {I.}~\bibnamefont {Walmsley}},\ }\href {https://doi.org/10.1103/PRXQuantum.3.020336} {\bibfield  {journal} {\bibinfo  {journal} {PRX Quantum}\ }\textbf {\bibinfo {volume} {3}},\ \bibinfo {pages} {020336} (\bibinfo {year} {2022})},\ \bibinfo {note} {publisher: American Physical Society}\BibitemShut {NoStop}%
\bibitem [{Note1()}]{Note1}%
  \BibitemOpen
  \bibinfo {note} {Google Sycamore's repetition rate can reach a maximum of 15 kHz, which provides an order of magnitude for the upper limit of sampling rates for non-photonic quantum computers, since superconducting qubits have the shortest gate time in this category.}\BibitemShut {Stop}%
\bibitem [{\citenamefont {Bell}\ \emph {et~al.}(2019)\citenamefont {Bell}, \citenamefont {Thekkadath}, \citenamefont {Ge}, \citenamefont {Cai},\ and\ \citenamefont {Walmsley}}]{bell_testing_2019}%
  \BibitemOpen
  \bibfield  {author} {\bibinfo {author} {\bibfnamefont {B.~A.}\ \bibnamefont {Bell}}, \bibinfo {author} {\bibfnamefont {G.~S.}\ \bibnamefont {Thekkadath}}, \bibinfo {author} {\bibfnamefont {R.}~\bibnamefont {Ge}}, \bibinfo {author} {\bibfnamefont {X.}~\bibnamefont {Cai}},\ and\ \bibinfo {author} {\bibfnamefont {I.~A.}\ \bibnamefont {Walmsley}},\ }\href {https://doi.org/10.1364/OE.27.035646} {\bibfield  {journal} {\bibinfo  {journal} {Optics Express}\ }\textbf {\bibinfo {volume} {27}},\ \bibinfo {pages} {35646} (\bibinfo {year} {2019})},\ \bibinfo {note} {publisher: Optica Publishing Group}\BibitemShut {NoStop}%
\bibitem [{\citenamefont {Paesani}\ \emph {et~al.}(2019)\citenamefont {Paesani}, \citenamefont {Ding}, \citenamefont {Santagati}, \citenamefont {Chakhmakhchyan}, \citenamefont {Vigliar}, \citenamefont {Rottwitt}, \citenamefont {Oxenløwe}, \citenamefont {Wang}, \citenamefont {Thompson},\ and\ \citenamefont {Laing}}]{paesani_generation_2019}%
  \BibitemOpen
  \bibfield  {author} {\bibinfo {author} {\bibfnamefont {S.}~\bibnamefont {Paesani}}, \bibinfo {author} {\bibfnamefont {Y.}~\bibnamefont {Ding}}, \bibinfo {author} {\bibfnamefont {R.}~\bibnamefont {Santagati}}, \bibinfo {author} {\bibfnamefont {L.}~\bibnamefont {Chakhmakhchyan}}, \bibinfo {author} {\bibfnamefont {C.}~\bibnamefont {Vigliar}}, \bibinfo {author} {\bibfnamefont {K.}~\bibnamefont {Rottwitt}}, \bibinfo {author} {\bibfnamefont {L.~K.}\ \bibnamefont {Oxenløwe}}, \bibinfo {author} {\bibfnamefont {J.}~\bibnamefont {Wang}}, \bibinfo {author} {\bibfnamefont {M.~G.}\ \bibnamefont {Thompson}},\ and\ \bibinfo {author} {\bibfnamefont {A.}~\bibnamefont {Laing}},\ }\href {https://doi.org/10.1038/s41567-019-0567-8} {\bibfield  {journal} {\bibinfo  {journal} {Nat. Phys.}\ }\textbf {\bibinfo {volume} {15}},\ \bibinfo {pages} {925} (\bibinfo {year} {2019})},\ \bibinfo {note} {number: 9 Publisher: Nature Publishing Group}\BibitemShut {NoStop}%
\bibitem [{\citenamefont {Arrazola}\ \emph {et~al.}(2021)\citenamefont {Arrazola}, \citenamefont {Bergholm}, \citenamefont {Brádler}, \citenamefont {Bromley}, \citenamefont {Collins}, \citenamefont {Dhand}, \citenamefont {Fumagalli}, \citenamefont {Gerrits}, \citenamefont {Goussev}, \citenamefont {Helt}, \citenamefont {Hundal}, \citenamefont {Isacsson}, \citenamefont {Israel}, \citenamefont {Izaac}, \citenamefont {Jahangiri}, \citenamefont {Janik}, \citenamefont {Killoran}, \citenamefont {Kumar}, \citenamefont {Lavoie}, \citenamefont {Lita}, \citenamefont {Mahler}, \citenamefont {Menotti}, \citenamefont {Morrison}, \citenamefont {Nam}, \citenamefont {Neuhaus}, \citenamefont {Qi}, \citenamefont {Quesada}, \citenamefont {Repingon}, \citenamefont {Sabapathy}, \citenamefont {Schuld}, \citenamefont {Su}, \citenamefont {Swinarton}, \citenamefont {Száva}, \citenamefont {Tan}, \citenamefont {Tan}, \citenamefont {Vaidya}, \citenamefont {Vernon}, \citenamefont {Zabaneh},\ and\ \citenamefont
  {Zhang}}]{arrazola_quantum_2021}%
  \BibitemOpen
  \bibfield  {author} {\bibinfo {author} {\bibfnamefont {J.~M.}\ \bibnamefont {Arrazola}}, \bibinfo {author} {\bibfnamefont {V.}~\bibnamefont {Bergholm}}, \bibinfo {author} {\bibfnamefont {K.}~\bibnamefont {Brádler}}, \bibinfo {author} {\bibfnamefont {T.~R.}\ \bibnamefont {Bromley}}, \bibinfo {author} {\bibfnamefont {M.~J.}\ \bibnamefont {Collins}}, \bibinfo {author} {\bibfnamefont {I.}~\bibnamefont {Dhand}}, \bibinfo {author} {\bibfnamefont {A.}~\bibnamefont {Fumagalli}}, \bibinfo {author} {\bibfnamefont {T.}~\bibnamefont {Gerrits}}, \bibinfo {author} {\bibfnamefont {A.}~\bibnamefont {Goussev}}, \bibinfo {author} {\bibfnamefont {L.~G.}\ \bibnamefont {Helt}}, \bibinfo {author} {\bibfnamefont {J.}~\bibnamefont {Hundal}}, \bibinfo {author} {\bibfnamefont {T.}~\bibnamefont {Isacsson}}, \bibinfo {author} {\bibfnamefont {R.~B.}\ \bibnamefont {Israel}}, \bibinfo {author} {\bibfnamefont {J.}~\bibnamefont {Izaac}}, \bibinfo {author} {\bibfnamefont {S.}~\bibnamefont {Jahangiri}}, \bibinfo {author} {\bibfnamefont
  {R.}~\bibnamefont {Janik}}, \bibinfo {author} {\bibfnamefont {N.}~\bibnamefont {Killoran}}, \bibinfo {author} {\bibfnamefont {S.~P.}\ \bibnamefont {Kumar}}, \bibinfo {author} {\bibfnamefont {J.}~\bibnamefont {Lavoie}}, \bibinfo {author} {\bibfnamefont {A.~E.}\ \bibnamefont {Lita}}, \bibinfo {author} {\bibfnamefont {D.~H.}\ \bibnamefont {Mahler}}, \bibinfo {author} {\bibfnamefont {M.}~\bibnamefont {Menotti}}, \bibinfo {author} {\bibfnamefont {B.}~\bibnamefont {Morrison}}, \bibinfo {author} {\bibfnamefont {S.~W.}\ \bibnamefont {Nam}}, \bibinfo {author} {\bibfnamefont {L.}~\bibnamefont {Neuhaus}}, \bibinfo {author} {\bibfnamefont {H.~Y.}\ \bibnamefont {Qi}}, \bibinfo {author} {\bibfnamefont {N.}~\bibnamefont {Quesada}}, \bibinfo {author} {\bibfnamefont {A.}~\bibnamefont {Repingon}}, \bibinfo {author} {\bibfnamefont {K.~K.}\ \bibnamefont {Sabapathy}}, \bibinfo {author} {\bibfnamefont {M.}~\bibnamefont {Schuld}}, \bibinfo {author} {\bibfnamefont {D.}~\bibnamefont {Su}}, \bibinfo {author} {\bibfnamefont
  {J.}~\bibnamefont {Swinarton}}, \bibinfo {author} {\bibfnamefont {A.}~\bibnamefont {Száva}}, \bibinfo {author} {\bibfnamefont {K.}~\bibnamefont {Tan}}, \bibinfo {author} {\bibfnamefont {P.}~\bibnamefont {Tan}}, \bibinfo {author} {\bibfnamefont {V.~D.}\ \bibnamefont {Vaidya}}, \bibinfo {author} {\bibfnamefont {Z.}~\bibnamefont {Vernon}}, \bibinfo {author} {\bibfnamefont {Z.}~\bibnamefont {Zabaneh}},\ and\ \bibinfo {author} {\bibfnamefont {Y.}~\bibnamefont {Zhang}},\ }\href {https://doi.org/10.1038/s41586-021-03202-1} {\bibfield  {journal} {\bibinfo  {journal} {Nature}\ }\textbf {\bibinfo {volume} {591}},\ \bibinfo {pages} {54} (\bibinfo {year} {2021})},\ \bibinfo {note} {number: 7848 Publisher: Nature Publishing Group}\BibitemShut {NoStop}%
\bibitem [{\citenamefont {Weedbrook}\ \emph {et~al.}(2012)\citenamefont {Weedbrook}, \citenamefont {Pirandola}, \citenamefont {García-Patrón}, \citenamefont {Cerf}, \citenamefont {Ralph}, \citenamefont {Shapiro},\ and\ \citenamefont {Lloyd}}]{weedbrook_gaussian_2012}%
  \BibitemOpen
  \bibfield  {author} {\bibinfo {author} {\bibfnamefont {C.}~\bibnamefont {Weedbrook}}, \bibinfo {author} {\bibfnamefont {S.}~\bibnamefont {Pirandola}}, \bibinfo {author} {\bibfnamefont {R.}~\bibnamefont {García-Patrón}}, \bibinfo {author} {\bibfnamefont {N.~J.}\ \bibnamefont {Cerf}}, \bibinfo {author} {\bibfnamefont {T.~C.}\ \bibnamefont {Ralph}}, \bibinfo {author} {\bibfnamefont {J.~H.}\ \bibnamefont {Shapiro}},\ and\ \bibinfo {author} {\bibfnamefont {S.}~\bibnamefont {Lloyd}},\ }\href {https://doi.org/10.1103/RevModPhys.84.621} {\bibfield  {journal} {\bibinfo  {journal} {Rev. Mod. Phys.}\ }\textbf {\bibinfo {volume} {84}},\ \bibinfo {pages} {621} (\bibinfo {year} {2012})},\ \bibinfo {note} {publisher: American Physical Society}\BibitemShut {NoStop}%
\bibitem [{\citenamefont {Serafini}(2023)}]{serafini_quantum_2023}%
  \BibitemOpen
  \bibfield  {author} {\bibinfo {author} {\bibfnamefont {A.}~\bibnamefont {Serafini}},\ }\href {https://doi.org/10.1201/9781003250975} {\emph {\bibinfo {title} {Quantum {Continuous} {Variables}: {A} {Primer} of {Theoretical} {Methods}}}},\ \bibinfo {edition} {2nd}\ ed.\ (\bibinfo  {publisher} {CRC Press},\ \bibinfo {address} {Boca Raton},\ \bibinfo {year} {2023})\BibitemShut {NoStop}%
\bibitem [{\citenamefont {Yao}\ \emph {et~al.}(2024)\citenamefont {Yao}, \citenamefont {Miatto},\ and\ \citenamefont {Quesada}}]{yao_design_2023}%
  \BibitemOpen
  \bibfield  {author} {\bibinfo {author} {\bibfnamefont {Y.}~\bibnamefont {Yao}}, \bibinfo {author} {\bibfnamefont {F.}~\bibnamefont {Miatto}},\ and\ \bibinfo {author} {\bibfnamefont {N.}~\bibnamefont {Quesada}},\ }\href {https://doi.org/10.21468/SciPostPhys.17.3.082} {\bibfield  {journal} {\bibinfo  {journal} {SciPost Phys.}\ }\textbf {\bibinfo {volume} {17}},\ \bibinfo {pages} {082} (\bibinfo {year} {2024})}\BibitemShut {NoStop}%
\bibitem [{\citenamefont {Gagatsos}\ and\ \citenamefont {Guha}(2019)}]{gagatsos_efficient_2019}%
  \BibitemOpen
  \bibfield  {author} {\bibinfo {author} {\bibfnamefont {C.~N.}\ \bibnamefont {Gagatsos}}\ and\ \bibinfo {author} {\bibfnamefont {S.}~\bibnamefont {Guha}},\ }\href {https://doi.org/10.1103/PhysRevA.99.053816} {\bibfield  {journal} {\bibinfo  {journal} {Phys. Rev. A}\ }\textbf {\bibinfo {volume} {99}},\ \bibinfo {pages} {053816} (\bibinfo {year} {2019})},\ \bibinfo {note} {publisher: American Physical Society}\BibitemShut {NoStop}%
\bibitem [{\citenamefont {Bromley}\ \emph {et~al.}(2020)\citenamefont {Bromley}, \citenamefont {Arrazola}, \citenamefont {Jahangiri}, \citenamefont {Izaac}, \citenamefont {Quesada}, \citenamefont {Gran}, \citenamefont {Schuld}, \citenamefont {Swinarton}, \citenamefont {Zabaneh},\ and\ \citenamefont {Killoran}}]{bromley_applications_2020}%
  \BibitemOpen
  \bibfield  {author} {\bibinfo {author} {\bibfnamefont {T.~R.}\ \bibnamefont {Bromley}}, \bibinfo {author} {\bibfnamefont {J.~M.}\ \bibnamefont {Arrazola}}, \bibinfo {author} {\bibfnamefont {S.}~\bibnamefont {Jahangiri}}, \bibinfo {author} {\bibfnamefont {J.}~\bibnamefont {Izaac}}, \bibinfo {author} {\bibfnamefont {N.}~\bibnamefont {Quesada}}, \bibinfo {author} {\bibfnamefont {A.~D.}\ \bibnamefont {Gran}}, \bibinfo {author} {\bibfnamefont {M.}~\bibnamefont {Schuld}}, \bibinfo {author} {\bibfnamefont {J.}~\bibnamefont {Swinarton}}, \bibinfo {author} {\bibfnamefont {Z.}~\bibnamefont {Zabaneh}},\ and\ \bibinfo {author} {\bibfnamefont {N.}~\bibnamefont {Killoran}},\ }\href {https://doi.org/10.1088/2058-9565/ab8504} {\bibfield  {journal} {\bibinfo  {journal} {Quantum Science and Technology}\ }\textbf {\bibinfo {volume} {5}},\ \bibinfo {pages} {034010} (\bibinfo {year} {2020})},\ \bibinfo {note} {publisher: IOP Publishing}\BibitemShut {NoStop}%
\bibitem [{\citenamefont {Houde}\ \emph {et~al.}(2024)\citenamefont {Houde}, \citenamefont {McCutcheon},\ and\ \citenamefont {Quesada}}]{houde_matrix_2024}%
  \BibitemOpen
  \bibfield  {author} {\bibinfo {author} {\bibfnamefont {M.}~\bibnamefont {Houde}}, \bibinfo {author} {\bibfnamefont {W.}~\bibnamefont {McCutcheon}},\ and\ \bibinfo {author} {\bibfnamefont {N.}~\bibnamefont {Quesada}},\ }\href {https://doi.org/10.48550/arXiv.2403.04596} {\bibinfo {title} {Matrix decompositions in {{Quantum Optics}}: {{Takagi}}/{{Autonne}}, {{Bloch-Messiah}}/{{Euler}}, {{Iwasawa}}, and {{Williamson}}}} (\bibinfo {year} {2024}),\ \Eprint {https://arxiv.org/abs/2403.04596} {arxiv:2403.04596 [physics, physics:quant-ph]} \BibitemShut {NoStop}%
\bibitem [{\citenamefont {Bulmer}\ \emph {et~al.}(2022{\natexlab{b}})\citenamefont {Bulmer}, \citenamefont {Bell}, \citenamefont {Chadwick}, \citenamefont {Jones}, \citenamefont {Moise}, \citenamefont {Rigazzi}, \citenamefont {Thorbecke}, \citenamefont {Haus}, \citenamefont {Van~Vaerenbergh}, \citenamefont {Patel}, \citenamefont {Walmsley},\ and\ \citenamefont {Laing}}]{bulmer_boundary_2022}%
  \BibitemOpen
  \bibfield  {author} {\bibinfo {author} {\bibfnamefont {J.~F.~F.}\ \bibnamefont {Bulmer}}, \bibinfo {author} {\bibfnamefont {B.~A.}\ \bibnamefont {Bell}}, \bibinfo {author} {\bibfnamefont {R.~S.}\ \bibnamefont {Chadwick}}, \bibinfo {author} {\bibfnamefont {A.~E.}\ \bibnamefont {Jones}}, \bibinfo {author} {\bibfnamefont {D.}~\bibnamefont {Moise}}, \bibinfo {author} {\bibfnamefont {A.}~\bibnamefont {Rigazzi}}, \bibinfo {author} {\bibfnamefont {J.}~\bibnamefont {Thorbecke}}, \bibinfo {author} {\bibfnamefont {U.-U.}\ \bibnamefont {Haus}}, \bibinfo {author} {\bibfnamefont {T.}~\bibnamefont {Van~Vaerenbergh}}, \bibinfo {author} {\bibfnamefont {R.~B.}\ \bibnamefont {Patel}}, \bibinfo {author} {\bibfnamefont {I.~A.}\ \bibnamefont {Walmsley}},\ and\ \bibinfo {author} {\bibfnamefont {A.}~\bibnamefont {Laing}},\ }\href {https://doi.org/10.1126/sciadv.abl9236} {\bibfield  {journal} {\bibinfo  {journal} {Science Advances}\ }\textbf {\bibinfo {volume} {8}},\ \bibinfo {pages} {eabl9236} (\bibinfo {year}
  {2022}{\natexlab{b}})},\ \bibinfo {note} {publisher: American Association for the Advancement of Science}\BibitemShut {NoStop}%
\bibitem [{\citenamefont {Kaposi}\ \emph {et~al.}(2022)\citenamefont {Kaposi}, \citenamefont {Kolarovszki}, \citenamefont {Kozsik}, \citenamefont {Zimbor\'as},\ and\ \citenamefont {Rakyta}}]{kaposi_polynomial_2022}%
  \BibitemOpen
  \bibfield  {author} {\bibinfo {author} {\bibfnamefont {A.}~\bibnamefont {Kaposi}}, \bibinfo {author} {\bibfnamefont {Z.}~\bibnamefont {Kolarovszki}}, \bibinfo {author} {\bibfnamefont {T.}~\bibnamefont {Kozsik}}, \bibinfo {author} {\bibfnamefont {Z.}~\bibnamefont {Zimbor\'as}},\ and\ \bibinfo {author} {\bibfnamefont {P.}~\bibnamefont {Rakyta}},\ }\href {https://doi.org/10.48550/arXiv.2109.04528} {\bibinfo {title} {Polynomial speedup in {Torontonian} calculation by a scalable recursive algorithm}} (\bibinfo {year} {2022})\BibitemShut {NoStop}%
\bibitem [{\citenamefont {Deshpande}\ \emph {et~al.}(2022)\citenamefont {Deshpande}, \citenamefont {Mehta}, \citenamefont {Vincent}, \citenamefont {Quesada}, \citenamefont {Hinsche}, \citenamefont {Ioannou}, \citenamefont {Madsen}, \citenamefont {Lavoie}, \citenamefont {Qi}, \citenamefont {Eisert}, \citenamefont {Hangleiter}, \citenamefont {Fefferman},\ and\ \citenamefont {Dhand}}]{deshpande_quantum_2022}%
  \BibitemOpen
  \bibfield  {author} {\bibinfo {author} {\bibfnamefont {A.}~\bibnamefont {Deshpande}}, \bibinfo {author} {\bibfnamefont {A.}~\bibnamefont {Mehta}}, \bibinfo {author} {\bibfnamefont {T.}~\bibnamefont {Vincent}}, \bibinfo {author} {\bibfnamefont {N.}~\bibnamefont {Quesada}}, \bibinfo {author} {\bibfnamefont {M.}~\bibnamefont {Hinsche}}, \bibinfo {author} {\bibfnamefont {M.}~\bibnamefont {Ioannou}}, \bibinfo {author} {\bibfnamefont {L.}~\bibnamefont {Madsen}}, \bibinfo {author} {\bibfnamefont {J.}~\bibnamefont {Lavoie}}, \bibinfo {author} {\bibfnamefont {H.}~\bibnamefont {Qi}}, \bibinfo {author} {\bibfnamefont {J.}~\bibnamefont {Eisert}}, \bibinfo {author} {\bibfnamefont {D.}~\bibnamefont {Hangleiter}}, \bibinfo {author} {\bibfnamefont {B.}~\bibnamefont {Fefferman}},\ and\ \bibinfo {author} {\bibfnamefont {I.}~\bibnamefont {Dhand}},\ }\href {https://doi.org/10.1126/sciadv.abi7894} {\bibfield  {journal} {\bibinfo  {journal} {Science Advances}\ }\textbf {\bibinfo {volume} {8}},\ \bibinfo {pages} {eabi7894}
  (\bibinfo {year} {2022})},\ \bibinfo {note} {publisher: American Association for the Advancement of Science}\BibitemShut {NoStop}%
\bibitem [{\citenamefont {Papadimitriou}\ and\ \citenamefont {Steiglitz}(1982)}]{papadimitriou_combinatorial_1982}%
  \BibitemOpen
  \bibfield  {author} {\bibinfo {author} {\bibfnamefont {C.~H.}\ \bibnamefont {Papadimitriou}}\ and\ \bibinfo {author} {\bibfnamefont {K.}~\bibnamefont {Steiglitz}},\ }\href@noop {} {\emph {\bibinfo {title} {Combinatorial optimization: algorithms and complexity}}}\ (\bibinfo  {publisher} {Prentice Hall},\ \bibinfo {address} {Englewood Cliffs, N.J},\ \bibinfo {year} {1982})\BibitemShut {NoStop}%
\bibitem [{\citenamefont {Nemhauser}\ and\ \citenamefont {Wolsey}(1988)}]{nemhauser_integer_1988}%
  \BibitemOpen
  \bibfield  {author} {\bibinfo {author} {\bibfnamefont {G.~L.}\ \bibnamefont {Nemhauser}}\ and\ \bibinfo {author} {\bibfnamefont {L.~A.}\ \bibnamefont {Wolsey}},\ }\href@noop {} {\emph {\bibinfo {title} {Integer and combinatorial optimization}}},\ Wiley-{Interscience} series in discrete mathematics and optimization\ (\bibinfo  {publisher} {Wiley},\ \bibinfo {address} {New York},\ \bibinfo {year} {1988})\BibitemShut {NoStop}%
\bibitem [{\citenamefont {Korte}\ and\ \citenamefont {Vygen}(2018)}]{korte_combinatorial_2018}%
  \BibitemOpen
  \bibfield  {author} {\bibinfo {author} {\bibfnamefont {B.}~\bibnamefont {Korte}}\ and\ \bibinfo {author} {\bibfnamefont {J.}~\bibnamefont {Vygen}},\ }\href {https://doi.org/10.1007/978-3-662-56039-6} {\emph {\bibinfo {title} {Combinatorial {Optimization}: {Theory} and {Algorithms}}}},\ \bibinfo {series} {Algorithms and {Combinatorics}}, Vol.~\bibinfo {volume} {21}\ (\bibinfo  {publisher} {Springer},\ \bibinfo {address} {Berlin, Heidelberg},\ \bibinfo {year} {2018})\BibitemShut {NoStop}%
\bibitem [{\citenamefont {Kochenberger}\ \emph {et~al.}(2014)\citenamefont {Kochenberger}, \citenamefont {Hao}, \citenamefont {Glover}, \citenamefont {Lewis}, \citenamefont {Lü}, \citenamefont {Wang},\ and\ \citenamefont {Wang}}]{kochenberger_unconstrained_2014}%
  \BibitemOpen
  \bibfield  {author} {\bibinfo {author} {\bibfnamefont {G.}~\bibnamefont {Kochenberger}}, \bibinfo {author} {\bibfnamefont {J.-K.}\ \bibnamefont {Hao}}, \bibinfo {author} {\bibfnamefont {F.}~\bibnamefont {Glover}}, \bibinfo {author} {\bibfnamefont {M.}~\bibnamefont {Lewis}}, \bibinfo {author} {\bibfnamefont {Z.}~\bibnamefont {Lü}}, \bibinfo {author} {\bibfnamefont {H.}~\bibnamefont {Wang}},\ and\ \bibinfo {author} {\bibfnamefont {Y.}~\bibnamefont {Wang}},\ }\href {https://doi.org/10.1007/s10878-014-9734-0} {\bibfield  {journal} {\bibinfo  {journal} {Journal of Combinatorial Optimization}\ }\textbf {\bibinfo {volume} {28}},\ \bibinfo {pages} {58} (\bibinfo {year} {2014})}\BibitemShut {NoStop}%
\bibitem [{\citenamefont {Glover}\ \emph {et~al.}(2022)\citenamefont {Glover}, \citenamefont {Kochenberger}, \citenamefont {Hennig},\ and\ \citenamefont {Du}}]{glover_quantum_2022}%
  \BibitemOpen
  \bibfield  {author} {\bibinfo {author} {\bibfnamefont {F.}~\bibnamefont {Glover}}, \bibinfo {author} {\bibfnamefont {G.}~\bibnamefont {Kochenberger}}, \bibinfo {author} {\bibfnamefont {R.}~\bibnamefont {Hennig}},\ and\ \bibinfo {author} {\bibfnamefont {Y.}~\bibnamefont {Du}},\ }\href {https://doi.org/10.1007/s10479-022-04634-2} {\bibfield  {journal} {\bibinfo  {journal} {Annals of Operations Research}\ }\textbf {\bibinfo {volume} {314}},\ \bibinfo {pages} {141} (\bibinfo {year} {2022})}\BibitemShut {NoStop}%
\bibitem [{\citenamefont {Lucas}(2014)}]{lucas_ising_2014}%
  \BibitemOpen
  \bibfield  {author} {\bibinfo {author} {\bibfnamefont {A.}~\bibnamefont {Lucas}},\ }\href {https://www.frontiersin.org/articles/10.3389/fphy.2014.00005} {\bibfield  {journal} {\bibinfo  {journal} {Frontiers in Physics}\ }\textbf {\bibinfo {volume} {2}} (\bibinfo {year} {2014})}\BibitemShut {NoStop}%
\bibitem [{\citenamefont {Babbush}\ \emph {et~al.}(2013)\citenamefont {Babbush}, \citenamefont {O'Gorman},\ and\ \citenamefont {Aspuru-Guzik}}]{babbush_resource_2013}%
  \BibitemOpen
  \bibfield  {author} {\bibinfo {author} {\bibfnamefont {R.}~\bibnamefont {Babbush}}, \bibinfo {author} {\bibfnamefont {B.}~\bibnamefont {O'Gorman}},\ and\ \bibinfo {author} {\bibfnamefont {A.}~\bibnamefont {Aspuru-Guzik}},\ }\href {https://doi.org/10.1002/andp.201300120} {\bibfield  {journal} {\bibinfo  {journal} {Annalen der Physik}\ }\textbf {\bibinfo {volume} {525}},\ \bibinfo {pages} {877} (\bibinfo {year} {2013})}\BibitemShut {NoStop}%
\bibitem [{\citenamefont {Stein}\ \emph {et~al.}(2023)\citenamefont {Stein}, \citenamefont {Chamanian}, \citenamefont {Zorn}, \citenamefont {Nüßlein}, \citenamefont {Zielinski}, \citenamefont {Kölle},\ and\ \citenamefont {Linnhoff-Popien}}]{stein_evidence_2023}%
  \BibitemOpen
  \bibfield  {author} {\bibinfo {author} {\bibfnamefont {J.}~\bibnamefont {Stein}}, \bibinfo {author} {\bibfnamefont {F.}~\bibnamefont {Chamanian}}, \bibinfo {author} {\bibfnamefont {M.}~\bibnamefont {Zorn}}, \bibinfo {author} {\bibfnamefont {J.}~\bibnamefont {Nüßlein}}, \bibinfo {author} {\bibfnamefont {S.}~\bibnamefont {Zielinski}}, \bibinfo {author} {\bibfnamefont {M.}~\bibnamefont {Kölle}},\ and\ \bibinfo {author} {\bibfnamefont {C.}~\bibnamefont {Linnhoff-Popien}},\ }in\ \href {https://doi.org/10.1145/3583133.3596358} {\emph {\bibinfo {booktitle} {Proceedings of the {Companion} {Conference} on {Genetic} and {Evolutionary} {Computation}}}}\ (\bibinfo {year} {2023})\ pp.\ \bibinfo {pages} {2254--2262},\ \bibinfo {note} {arXiv:2305.03390 [quant-ph]}\BibitemShut {NoStop}%
\bibitem [{\citenamefont {Salehi}\ \emph {et~al.}(2022)\citenamefont {Salehi}, \citenamefont {Glos},\ and\ \citenamefont {Miszczak}}]{salehi_unconstrained_2022}%
  \BibitemOpen
  \bibfield  {author} {\bibinfo {author} {\bibfnamefont {O.}~\bibnamefont {Salehi}}, \bibinfo {author} {\bibfnamefont {A.}~\bibnamefont {Glos}},\ and\ \bibinfo {author} {\bibfnamefont {J.~A.}\ \bibnamefont {Miszczak}},\ }\href {https://doi.org/10.1007/s11128-021-03405-5} {\bibfield  {journal} {\bibinfo  {journal} {Quantum Information Processing}\ }\textbf {\bibinfo {volume} {21}},\ \bibinfo {pages} {67} (\bibinfo {year} {2022})}\BibitemShut {NoStop}%
\bibitem [{\citenamefont {Tabi}\ \emph {et~al.}(2020)\citenamefont {Tabi}, \citenamefont {El-Safty}, \citenamefont {Kallus}, \citenamefont {Hága}, \citenamefont {Kozsik}, \citenamefont {Glos},\ and\ \citenamefont {Zimborás}}]{tabi_quantum_2020}%
  \BibitemOpen
  \bibfield  {author} {\bibinfo {author} {\bibfnamefont {Z.}~\bibnamefont {Tabi}}, \bibinfo {author} {\bibfnamefont {K.~H.}\ \bibnamefont {El-Safty}}, \bibinfo {author} {\bibfnamefont {Z.}~\bibnamefont {Kallus}}, \bibinfo {author} {\bibfnamefont {P.}~\bibnamefont {Hága}}, \bibinfo {author} {\bibfnamefont {T.}~\bibnamefont {Kozsik}}, \bibinfo {author} {\bibfnamefont {A.}~\bibnamefont {Glos}},\ and\ \bibinfo {author} {\bibfnamefont {Z.}~\bibnamefont {Zimborás}},\ }in\ \href {https://doi.org/10.1109/QCE49297.2020.00018} {\emph {\bibinfo {booktitle} {2020 {IEEE} {International} {Conference} on {Quantum} {Computing} and {Engineering} ({QCE})}}}\ (\bibinfo {year} {2020})\ pp.\ \bibinfo {pages} {56--62}\BibitemShut {NoStop}%
\bibitem [{\citenamefont {Chai}\ \emph {et~al.}(2023{\natexlab{b}})\citenamefont {Chai}, \citenamefont {Epifanovsky}, \citenamefont {Jansen}, \citenamefont {Kaushik},\ and\ \citenamefont {Kühn}}]{chai_simulating_2023}%
  \BibitemOpen
  \bibfield  {author} {\bibinfo {author} {\bibfnamefont {Y.}~\bibnamefont {Chai}}, \bibinfo {author} {\bibfnamefont {E.}~\bibnamefont {Epifanovsky}}, \bibinfo {author} {\bibfnamefont {K.}~\bibnamefont {Jansen}}, \bibinfo {author} {\bibfnamefont {A.}~\bibnamefont {Kaushik}},\ and\ \bibinfo {author} {\bibfnamefont {S.}~\bibnamefont {Kühn}},\ }\href {https://doi.org/10.48550/arXiv.2309.09686} {\bibinfo {title} {Simulating the flight gate assignment problem on a trapped ion quantum computer}} (\bibinfo {year} {2023}{\natexlab{b}}),\ \bibinfo {note} {arXiv:2309.09686 [quant-ph]}\BibitemShut {NoStop}%
\bibitem [{\citenamefont {Buluç}\ \emph {et~al.}(2016)\citenamefont {Buluç}, \citenamefont {Meyerhenke}, \citenamefont {Safro}, \citenamefont {Sanders},\ and\ \citenamefont {Schulz}}]{buluc_recent_2016}%
  \BibitemOpen
  \bibfield  {author} {\bibinfo {author} {\bibfnamefont {A.}~\bibnamefont {Buluç}}, \bibinfo {author} {\bibfnamefont {H.}~\bibnamefont {Meyerhenke}}, \bibinfo {author} {\bibfnamefont {I.}~\bibnamefont {Safro}}, \bibinfo {author} {\bibfnamefont {P.}~\bibnamefont {Sanders}},\ and\ \bibinfo {author} {\bibfnamefont {C.}~\bibnamefont {Schulz}},\ }in\ \href {https://doi.org/10.1007/978-3-319-49487-6_4} {\emph {\bibinfo {booktitle} {Algorithm {Engineering}: {Selected} {Results} and {Surveys}}}},\ \bibinfo {series and number} {Lecture {Notes} in {Computer} {Science}},\ \bibinfo {editor} {edited by\ \bibinfo {editor} {\bibfnamefont {L.}~\bibnamefont {Kliemann}}\ and\ \bibinfo {editor} {\bibfnamefont {P.}~\bibnamefont {Sanders}}}\ (\bibinfo  {publisher} {Springer International Publishing},\ \bibinfo {address} {Cham},\ \bibinfo {year} {2016})\ pp.\ \bibinfo {pages} {117--158}\BibitemShut {NoStop}%
\bibitem [{\citenamefont {Howson}(1997)}]{howson_logic_1997}%
  \BibitemOpen
  \bibfield  {author} {\bibinfo {author} {\bibfnamefont {C.}~\bibnamefont {Howson}},\ }\href {https://doi.org/10.4324/9780203976739} {\emph {\bibinfo {title} {Logic with {Trees}: {An} {Introduction} to {Symbolic} {Logic}}}}\ (\bibinfo  {publisher} {Routledge},\ \bibinfo {address} {London},\ \bibinfo {year} {1997})\BibitemShut {NoStop}%
\bibitem [{\citenamefont {Karp}(2010)}]{karp_reducibility_2010}%
  \BibitemOpen
  \bibfield  {author} {\bibinfo {author} {\bibfnamefont {R.~M.}\ \bibnamefont {Karp}},\ }in\ \href {https://doi.org/10.1007/978-3-540-68279-0_8} {\emph {\bibinfo {booktitle} {50 {Years} of {Integer} {Programming} 1958-2008: {From} the {Early} {Years} to the {State}-of-the-{Art}}}},\ \bibinfo {editor} {edited by\ \bibinfo {editor} {\bibfnamefont {M.}~\bibnamefont {Jünger}}, \bibinfo {editor} {\bibfnamefont {T.~M.}\ \bibnamefont {Liebling}}, \bibinfo {editor} {\bibfnamefont {D.}~\bibnamefont {Naddef}}, \bibinfo {editor} {\bibfnamefont {G.~L.}\ \bibnamefont {Nemhauser}}, \bibinfo {editor} {\bibfnamefont {W.~R.}\ \bibnamefont {Pulleyblank}}, \bibinfo {editor} {\bibfnamefont {G.}~\bibnamefont {Reinelt}}, \bibinfo {editor} {\bibfnamefont {G.}~\bibnamefont {Rinaldi}},\ and\ \bibinfo {editor} {\bibfnamefont {L.~A.}\ \bibnamefont {Wolsey}}}\ (\bibinfo  {publisher} {Springer},\ \bibinfo {address} {Berlin, Heidelberg},\ \bibinfo {year} {2010})\ pp.\ \bibinfo {pages} {219--241}\BibitemShut {NoStop}%
\bibitem [{\citenamefont {Tilly}\ \emph {et~al.}(2022)\citenamefont {Tilly}, \citenamefont {Chen}, \citenamefont {Cao}, \citenamefont {Picozzi}, \citenamefont {Setia}, \citenamefont {Li}, \citenamefont {Grant}, \citenamefont {Wossnig}, \citenamefont {Rungger}, \citenamefont {Booth},\ and\ \citenamefont {Tennyson}}]{tilly_variational_2022}%
  \BibitemOpen
  \bibfield  {author} {\bibinfo {author} {\bibfnamefont {J.}~\bibnamefont {Tilly}}, \bibinfo {author} {\bibfnamefont {H.}~\bibnamefont {Chen}}, \bibinfo {author} {\bibfnamefont {S.}~\bibnamefont {Cao}}, \bibinfo {author} {\bibfnamefont {D.}~\bibnamefont {Picozzi}}, \bibinfo {author} {\bibfnamefont {K.}~\bibnamefont {Setia}}, \bibinfo {author} {\bibfnamefont {Y.}~\bibnamefont {Li}}, \bibinfo {author} {\bibfnamefont {E.}~\bibnamefont {Grant}}, \bibinfo {author} {\bibfnamefont {L.}~\bibnamefont {Wossnig}}, \bibinfo {author} {\bibfnamefont {I.}~\bibnamefont {Rungger}}, \bibinfo {author} {\bibfnamefont {G.~H.}\ \bibnamefont {Booth}},\ and\ \bibinfo {author} {\bibfnamefont {J.}~\bibnamefont {Tennyson}},\ }\href {https://doi.org/10.1016/j.physrep.2022.08.003} {\bibfield  {journal} {\bibinfo  {journal} {Physics Reports}\ }\bibinfo {series} {The {Variational} {Quantum} {Eigensolver}: a review of methods and best practices},\ \textbf {\bibinfo {volume} {986}},\ \bibinfo {pages} {1} (\bibinfo {year} {2022})}\BibitemShut
  {NoStop}%
\bibitem [{\citenamefont {Mitarai}\ \emph {et~al.}(2018{\natexlab{a}})\citenamefont {Mitarai}, \citenamefont {Negoro}, \citenamefont {Kitagawa},\ and\ \citenamefont {Fujii}}]{mitarai_quantum_2018}%
  \BibitemOpen
  \bibfield  {author} {\bibinfo {author} {\bibfnamefont {K.}~\bibnamefont {Mitarai}}, \bibinfo {author} {\bibfnamefont {M.}~\bibnamefont {Negoro}}, \bibinfo {author} {\bibfnamefont {M.}~\bibnamefont {Kitagawa}},\ and\ \bibinfo {author} {\bibfnamefont {K.}~\bibnamefont {Fujii}},\ }\href {https://doi.org/10.1103/PhysRevA.98.032309} {\bibfield  {journal} {\bibinfo  {journal} {Phys. Rev. A}\ }\textbf {\bibinfo {volume} {98}},\ \bibinfo {pages} {032309} (\bibinfo {year} {2018}{\natexlab{a}})},\ \bibinfo {note} {publisher: American Physical Society}\BibitemShut {NoStop}%
\bibitem [{\citenamefont {Schuld}\ \emph {et~al.}(2019{\natexlab{a}})\citenamefont {Schuld}, \citenamefont {Bergholm}, \citenamefont {Gogolin}, \citenamefont {Izaac},\ and\ \citenamefont {Killoran}}]{schuld_evaluating_2019}%
  \BibitemOpen
  \bibfield  {author} {\bibinfo {author} {\bibfnamefont {M.}~\bibnamefont {Schuld}}, \bibinfo {author} {\bibfnamefont {V.}~\bibnamefont {Bergholm}}, \bibinfo {author} {\bibfnamefont {C.}~\bibnamefont {Gogolin}}, \bibinfo {author} {\bibfnamefont {J.}~\bibnamefont {Izaac}},\ and\ \bibinfo {author} {\bibfnamefont {N.}~\bibnamefont {Killoran}},\ }\href {https://doi.org/10.1103/PhysRevA.99.032331} {\bibfield  {journal} {\bibinfo  {journal} {Phys. Rev. A}\ }\textbf {\bibinfo {volume} {99}},\ \bibinfo {pages} {032331} (\bibinfo {year} {2019}{\natexlab{a}})},\ \bibinfo {note} {publisher: American Physical Society}\BibitemShut {NoStop}%
\bibitem [{\citenamefont {Cao}\ \emph {et~al.}(2019)\citenamefont {Cao}, \citenamefont {Romero}, \citenamefont {Olson}, \citenamefont {Degroote}, \citenamefont {Johnson}, \citenamefont {Kieferová}, \citenamefont {Kivlichan}, \citenamefont {Menke}, \citenamefont {Peropadre}, \citenamefont {Sawaya}, \citenamefont {Sim}, \citenamefont {Veis},\ and\ \citenamefont {Aspuru-Guzik}}]{cao_quantum_2019}%
  \BibitemOpen
  \bibfield  {author} {\bibinfo {author} {\bibfnamefont {Y.}~\bibnamefont {Cao}}, \bibinfo {author} {\bibfnamefont {J.}~\bibnamefont {Romero}}, \bibinfo {author} {\bibfnamefont {J.~P.}\ \bibnamefont {Olson}}, \bibinfo {author} {\bibfnamefont {M.}~\bibnamefont {Degroote}}, \bibinfo {author} {\bibfnamefont {P.~D.}\ \bibnamefont {Johnson}}, \bibinfo {author} {\bibfnamefont {M.}~\bibnamefont {Kieferová}}, \bibinfo {author} {\bibfnamefont {I.~D.}\ \bibnamefont {Kivlichan}}, \bibinfo {author} {\bibfnamefont {T.}~\bibnamefont {Menke}}, \bibinfo {author} {\bibfnamefont {B.}~\bibnamefont {Peropadre}}, \bibinfo {author} {\bibfnamefont {N.~P.~D.}\ \bibnamefont {Sawaya}}, \bibinfo {author} {\bibfnamefont {S.}~\bibnamefont {Sim}}, \bibinfo {author} {\bibfnamefont {L.}~\bibnamefont {Veis}},\ and\ \bibinfo {author} {\bibfnamefont {A.}~\bibnamefont {Aspuru-Guzik}},\ }\href {https://doi.org/10.1021/acs.chemrev.8b00803} {\bibfield  {journal} {\bibinfo  {journal} {Chemical Reviews}\ }\textbf {\bibinfo {volume} {119}},\
  \bibinfo {pages} {10856} (\bibinfo {year} {2019})}\BibitemShut {NoStop}%
\bibitem [{\citenamefont {Nannicini}(2019)}]{nannicini_performance_2019}%
  \BibitemOpen
  \bibfield  {author} {\bibinfo {author} {\bibfnamefont {G.}~\bibnamefont {Nannicini}},\ }\href {https://doi.org/10.1103/PhysRevE.99.013304} {\bibfield  {journal} {\bibinfo  {journal} {Phys. Rev. E}\ }\textbf {\bibinfo {volume} {99}},\ \bibinfo {pages} {013304} (\bibinfo {year} {2019})},\ \bibinfo {note} {publisher: American Physical Society}\BibitemShut {NoStop}%
\bibitem [{\citenamefont {Díez-Valle}\ \emph {et~al.}(2021)\citenamefont {Díez-Valle}, \citenamefont {Porras},\ and\ \citenamefont {García-Ripoll}}]{diez-valle_quantum_2021}%
  \BibitemOpen
  \bibfield  {author} {\bibinfo {author} {\bibfnamefont {P.}~\bibnamefont {Díez-Valle}}, \bibinfo {author} {\bibfnamefont {D.}~\bibnamefont {Porras}},\ and\ \bibinfo {author} {\bibfnamefont {J.~J.}\ \bibnamefont {García-Ripoll}},\ }\href {https://doi.org/10.1103/PhysRevA.104.062426} {\bibfield  {journal} {\bibinfo  {journal} {Phys. Rev. A}\ }\textbf {\bibinfo {volume} {104}},\ \bibinfo {pages} {062426} (\bibinfo {year} {2021})},\ \bibinfo {note} {publisher: American Physical Society}\BibitemShut {NoStop}%
\bibitem [{\citenamefont {Conti}(2021)}]{conti_training_2021}%
  \BibitemOpen
  \bibfield  {author} {\bibinfo {author} {\bibfnamefont {C.}~\bibnamefont {Conti}},\ }\href {https://doi.org/10.1007/s42484-021-00052-y} {\bibfield  {journal} {\bibinfo  {journal} {Quantum Machine Intelligence}\ }\textbf {\bibinfo {volume} {3}},\ \bibinfo {pages} {26} (\bibinfo {year} {2021})}\BibitemShut {NoStop}%
\bibitem [{\citenamefont {Conti}(2022)}]{conti_variational_2022}%
  \BibitemOpen
  \bibfield  {author} {\bibinfo {author} {\bibfnamefont {C.}~\bibnamefont {Conti}},\ }\href {https://doi.org/10.1103/PhysRevA.106.013518} {\bibfield  {journal} {\bibinfo  {journal} {Phys. Rev. A}\ }\textbf {\bibinfo {volume} {106}},\ \bibinfo {pages} {013518} (\bibinfo {year} {2022})},\ \bibinfo {note} {publisher: American Physical Society}\BibitemShut {NoStop}%
\bibitem [{\citenamefont {Abadi}\ \emph {et~al.}(2015)\citenamefont {Abadi}, \citenamefont {Agarwal}, \citenamefont {Barham}, \citenamefont {Brevdo}, \citenamefont {Chen}, \citenamefont {Citro}, \citenamefont {Corrado}, \citenamefont {Davis}, \citenamefont {Dean}, \citenamefont {Devin}, \citenamefont {Ghemawat}, \citenamefont {Goodfellow}, \citenamefont {Harp}, \citenamefont {Irving}, \citenamefont {Isard}, \citenamefont {Jia}, \citenamefont {Jozefowicz}, \citenamefont {Kaiser}, \citenamefont {Kudlur}, \citenamefont {Levenberg}, \citenamefont {Man\'{e}}, \citenamefont {Monga}, \citenamefont {Moore}, \citenamefont {Murray}, \citenamefont {Olah}, \citenamefont {Schuster}, \citenamefont {Shlens}, \citenamefont {Steiner}, \citenamefont {Sutskever}, \citenamefont {Talwar}, \citenamefont {Tucker}, \citenamefont {Vanhoucke}, \citenamefont {Vasudevan}, \citenamefont {Vi\'{e}gas}, \citenamefont {Vinyals}, \citenamefont {Warden}, \citenamefont {Wattenberg}, \citenamefont {Wicke}, \citenamefont {Yu},\ and\ \citenamefont
  {Zheng}}]{tensorflow2015-whitepaper}%
  \BibitemOpen
  \bibfield  {author} {\bibinfo {author} {\bibfnamefont {M.}~\bibnamefont {Abadi}}, \bibinfo {author} {\bibfnamefont {A.}~\bibnamefont {Agarwal}}, \bibinfo {author} {\bibfnamefont {P.}~\bibnamefont {Barham}}, \bibinfo {author} {\bibfnamefont {E.}~\bibnamefont {Brevdo}}, \bibinfo {author} {\bibfnamefont {Z.}~\bibnamefont {Chen}}, \bibinfo {author} {\bibfnamefont {C.}~\bibnamefont {Citro}}, \bibinfo {author} {\bibfnamefont {G.~S.}\ \bibnamefont {Corrado}}, \bibinfo {author} {\bibfnamefont {A.}~\bibnamefont {Davis}}, \bibinfo {author} {\bibfnamefont {J.}~\bibnamefont {Dean}}, \bibinfo {author} {\bibfnamefont {M.}~\bibnamefont {Devin}}, \bibinfo {author} {\bibfnamefont {S.}~\bibnamefont {Ghemawat}}, \bibinfo {author} {\bibfnamefont {I.}~\bibnamefont {Goodfellow}}, \bibinfo {author} {\bibfnamefont {A.}~\bibnamefont {Harp}}, \bibinfo {author} {\bibfnamefont {G.}~\bibnamefont {Irving}}, \bibinfo {author} {\bibfnamefont {M.}~\bibnamefont {Isard}}, \bibinfo {author} {\bibfnamefont {Y.}~\bibnamefont {Jia}}, \bibinfo
  {author} {\bibfnamefont {R.}~\bibnamefont {Jozefowicz}}, \bibinfo {author} {\bibfnamefont {L.}~\bibnamefont {Kaiser}}, \bibinfo {author} {\bibfnamefont {M.}~\bibnamefont {Kudlur}}, \bibinfo {author} {\bibfnamefont {J.}~\bibnamefont {Levenberg}}, \bibinfo {author} {\bibfnamefont {D.}~\bibnamefont {Man\'{e}}}, \bibinfo {author} {\bibfnamefont {R.}~\bibnamefont {Monga}}, \bibinfo {author} {\bibfnamefont {S.}~\bibnamefont {Moore}}, \bibinfo {author} {\bibfnamefont {D.}~\bibnamefont {Murray}}, \bibinfo {author} {\bibfnamefont {C.}~\bibnamefont {Olah}}, \bibinfo {author} {\bibfnamefont {M.}~\bibnamefont {Schuster}}, \bibinfo {author} {\bibfnamefont {J.}~\bibnamefont {Shlens}}, \bibinfo {author} {\bibfnamefont {B.}~\bibnamefont {Steiner}}, \bibinfo {author} {\bibfnamefont {I.}~\bibnamefont {Sutskever}}, \bibinfo {author} {\bibfnamefont {K.}~\bibnamefont {Talwar}}, \bibinfo {author} {\bibfnamefont {P.}~\bibnamefont {Tucker}}, \bibinfo {author} {\bibfnamefont {V.}~\bibnamefont {Vanhoucke}}, \bibinfo {author}
  {\bibfnamefont {V.}~\bibnamefont {Vasudevan}}, \bibinfo {author} {\bibfnamefont {F.}~\bibnamefont {Vi\'{e}gas}}, \bibinfo {author} {\bibfnamefont {O.}~\bibnamefont {Vinyals}}, \bibinfo {author} {\bibfnamefont {P.}~\bibnamefont {Warden}}, \bibinfo {author} {\bibfnamefont {M.}~\bibnamefont {Wattenberg}}, \bibinfo {author} {\bibfnamefont {M.}~\bibnamefont {Wicke}}, \bibinfo {author} {\bibfnamefont {Y.}~\bibnamefont {Yu}},\ and\ \bibinfo {author} {\bibfnamefont {X.}~\bibnamefont {Zheng}},\ }\href {https://www.tensorflow.org/} {\bibinfo {title} {{TensorFlow}: Large-scale machine learning on heterogeneous systems}} (\bibinfo {year} {2015}),\ \bibinfo {note} {software available from tensorflow.org}\BibitemShut {NoStop}%
\bibitem [{\citenamefont {Kingma}\ and\ \citenamefont {Ba}(2017)}]{kingma_adam_2017}%
  \BibitemOpen
  \bibfield  {author} {\bibinfo {author} {\bibfnamefont {D.~P.}\ \bibnamefont {Kingma}}\ and\ \bibinfo {author} {\bibfnamefont {J.}~\bibnamefont {Ba}},\ }\href {https://doi.org/10.48550/arXiv.1412.6980} {\bibinfo {title} {Adam: {A} {Method} for {Stochastic} {Optimization}}} (\bibinfo {year} {2017}),\ \bibinfo {note} {arXiv:1412.6980 [cs]}\BibitemShut {NoStop}%
\bibitem [{\citenamefont {Bradler}\ and\ \citenamefont {Wallner}(2021)}]{bradler_certain_2021}%
  \BibitemOpen
  \bibfield  {author} {\bibinfo {author} {\bibfnamefont {K.}~\bibnamefont {Bradler}}\ and\ \bibinfo {author} {\bibfnamefont {H.}~\bibnamefont {Wallner}},\ }\href {https://doi.org/10.48550/arXiv.2112.09766} {\bibinfo {title} {Certain properties and applications of shallow bosonic circuits}} (\bibinfo {year} {2021}),\ \bibinfo {note} {arXiv:2112.09766 [math-ph, physics:physics, physics:quant-ph]}\BibitemShut {NoStop}%
\bibitem [{\citenamefont {Brod}\ \emph {et~al.}(2019)\citenamefont {Brod}, \citenamefont {Galvão}, \citenamefont {Crespi}, \citenamefont {Osellame}, \citenamefont {Spagnolo},\ and\ \citenamefont {Sciarrino}}]{brod_photonic_2019}%
  \BibitemOpen
  \bibfield  {author} {\bibinfo {author} {\bibfnamefont {D.~J.}\ \bibnamefont {Brod}}, \bibinfo {author} {\bibfnamefont {E.~F.}\ \bibnamefont {Galvão}}, \bibinfo {author} {\bibfnamefont {A.}~\bibnamefont {Crespi}}, \bibinfo {author} {\bibfnamefont {R.}~\bibnamefont {Osellame}}, \bibinfo {author} {\bibfnamefont {N.}~\bibnamefont {Spagnolo}},\ and\ \bibinfo {author} {\bibfnamefont {F.}~\bibnamefont {Sciarrino}},\ }\href {https://doi.org/10.1117/1.AP.1.3.034001} {\bibfield  {journal} {\bibinfo  {journal} {Advanced Photonics}\ }\textbf {\bibinfo {volume} {1}},\ \bibinfo {pages} {034001} (\bibinfo {year} {2019})},\ \bibinfo {note} {publisher: SPIE}\BibitemShut {NoStop}%
\bibitem [{\citenamefont {Mari}\ \emph {et~al.}(2021)\citenamefont {Mari}, \citenamefont {Bromley},\ and\ \citenamefont {Killoran}}]{mari_estimating_2021}%
  \BibitemOpen
  \bibfield  {author} {\bibinfo {author} {\bibfnamefont {A.}~\bibnamefont {Mari}}, \bibinfo {author} {\bibfnamefont {T.~R.}\ \bibnamefont {Bromley}},\ and\ \bibinfo {author} {\bibfnamefont {N.}~\bibnamefont {Killoran}},\ }\href {https://doi.org/10.1103/PhysRevA.103.012405} {\bibfield  {journal} {\bibinfo  {journal} {Phys. Rev. A}\ }\textbf {\bibinfo {volume} {103}},\ \bibinfo {pages} {012405} (\bibinfo {year} {2021})},\ \bibinfo {note} {publisher: American Physical Society}\BibitemShut {NoStop}%
\bibitem [{\citenamefont {Mitarai}\ \emph {et~al.}(2018{\natexlab{b}})\citenamefont {Mitarai}, \citenamefont {Negoro}, \citenamefont {Kitagawa},\ and\ \citenamefont {Fujii}}]{mitarai2018}%
  \BibitemOpen
  \bibfield  {author} {\bibinfo {author} {\bibfnamefont {K.}~\bibnamefont {Mitarai}}, \bibinfo {author} {\bibfnamefont {M.}~\bibnamefont {Negoro}}, \bibinfo {author} {\bibfnamefont {M.}~\bibnamefont {Kitagawa}},\ and\ \bibinfo {author} {\bibfnamefont {K.}~\bibnamefont {Fujii}},\ }\href {https://doi.org/10.1103/PhysRevA.98.032309} {\bibfield  {journal} {\bibinfo  {journal} {Phys. Rev. A}\ }\textbf {\bibinfo {volume} {98}},\ \bibinfo {pages} {032309} (\bibinfo {year} {2018}{\natexlab{b}})}\BibitemShut {NoStop}%
\bibitem [{\citenamefont {Schuld}\ \emph {et~al.}(2019{\natexlab{b}})\citenamefont {Schuld}, \citenamefont {Bergholm}, \citenamefont {Gogolin}, \citenamefont {Izaac},\ and\ \citenamefont {Killoran}}]{schuld2019evaluating}%
  \BibitemOpen
  \bibfield  {author} {\bibinfo {author} {\bibfnamefont {M.}~\bibnamefont {Schuld}}, \bibinfo {author} {\bibfnamefont {V.}~\bibnamefont {Bergholm}}, \bibinfo {author} {\bibfnamefont {C.}~\bibnamefont {Gogolin}}, \bibinfo {author} {\bibfnamefont {J.}~\bibnamefont {Izaac}},\ and\ \bibinfo {author} {\bibfnamefont {N.}~\bibnamefont {Killoran}},\ }\href {https://doi.org/10.1103/PhysRevA.99.032331} {\bibfield  {journal} {\bibinfo  {journal} {Phys. Rev. A}\ }\textbf {\bibinfo {volume} {99}},\ \bibinfo {pages} {032331} (\bibinfo {year} {2019}{\natexlab{b}})}\BibitemShut {NoStop}%
\bibitem [{\citenamefont {Powell}(1994)}]{powell_direct_1994}%
  \BibitemOpen
  \bibfield  {author} {\bibinfo {author} {\bibfnamefont {M.~J.~D.}\ \bibnamefont {Powell}},\ }in\ \href {https://doi.org/10.1007/978-94-015-8330-5_4} {\emph {\bibinfo {booktitle} {Advances in {Optimization} and {Numerical} {Analysis}}}},\ \bibinfo {series and number} {Mathematics and {Its} {Applications}},\ \bibinfo {editor} {edited by\ \bibinfo {editor} {\bibfnamefont {S.}~\bibnamefont {Gomez}}\ and\ \bibinfo {editor} {\bibfnamefont {J.-P.}\ \bibnamefont {Hennart}}}\ (\bibinfo  {publisher} {Springer Netherlands},\ \bibinfo {address} {Dordrecht},\ \bibinfo {year} {1994})\ pp.\ \bibinfo {pages} {51--67}\BibitemShut {NoStop}%
\bibitem [{\citenamefont {Selman}\ \emph {et~al.}(1996)\citenamefont {Selman}, \citenamefont {Mitchell},\ and\ \citenamefont {Levesque}}]{selman_generating_1996}%
  \BibitemOpen
  \bibfield  {author} {\bibinfo {author} {\bibfnamefont {B.}~\bibnamefont {Selman}}, \bibinfo {author} {\bibfnamefont {D.~G.}\ \bibnamefont {Mitchell}},\ and\ \bibinfo {author} {\bibfnamefont {H.~J.}\ \bibnamefont {Levesque}},\ }\href {https://doi.org/10.1016/0004-3702(95)00045-3} {\bibfield  {journal} {\bibinfo  {journal} {Artificial Intelligence}\ }\bibinfo {series} {Frontiers in {Problem} {Solving}: {Phase} {Transitions} and {Complexity}},\ \textbf {\bibinfo {volume} {81}},\ \bibinfo {pages} {17} (\bibinfo {year} {1996})}\BibitemShut {NoStop}%
\bibitem [{\citenamefont {Hagberg}\ \emph {et~al.}(2008)\citenamefont {Hagberg}, \citenamefont {Swart},\ and\ \citenamefont {Schult}}]{Hagberg2008}%
  \BibitemOpen
  \bibfield  {author} {\bibinfo {author} {\bibfnamefont {A.}~\bibnamefont {Hagberg}}, \bibinfo {author} {\bibfnamefont {P.~J.}\ \bibnamefont {Swart}},\ and\ \bibinfo {author} {\bibfnamefont {D.~A.}\ \bibnamefont {Schult}},\ }\href@noop {} {\emph {\bibinfo {title} {Exploring Network Structure, Dynamics, and Function Using {{NetworkX}}}}},\ \bibinfo {type} {Tech. Rep.}\ \bibinfo {number} {LA-UR-08-05495; LA-UR-08-5495}\ (\bibinfo  {institution} {Los Alamos National Laboratory (LANL), Los Alamos, NM (United States)},\ \bibinfo {year} {2008})\BibitemShut {NoStop}%
\bibitem [{\citenamefont {Qi}\ \emph {et~al.}(2022)\citenamefont {Qi}, \citenamefont {Cifuentes}, \citenamefont {Br\'adler}, \citenamefont {Israel}, \citenamefont {Kalajdzievski},\ and\ \citenamefont {Quesada}}]{qi_efficient_2022}%
  \BibitemOpen
  \bibfield  {author} {\bibinfo {author} {\bibfnamefont {H.}~\bibnamefont {Qi}}, \bibinfo {author} {\bibfnamefont {D.}~\bibnamefont {Cifuentes}}, \bibinfo {author} {\bibfnamefont {K.}~\bibnamefont {Br\'adler}}, \bibinfo {author} {\bibfnamefont {R.}~\bibnamefont {Israel}}, \bibinfo {author} {\bibfnamefont {T.}~\bibnamefont {Kalajdzievski}},\ and\ \bibinfo {author} {\bibfnamefont {N.}~\bibnamefont {Quesada}},\ }\href {https://doi.org/10.1103/PhysRevA.105.052412} {\bibfield  {journal} {\bibinfo  {journal} {Phys. Rev. A}\ }\textbf {\bibinfo {volume} {105}},\ \bibinfo {pages} {052412} (\bibinfo {year} {2022})}\BibitemShut {NoStop}%
\bibitem [{\citenamefont {Go}\ \emph {et~al.}(2024)\citenamefont {Go}, \citenamefont {Oh}, \citenamefont {Jiang},\ and\ \citenamefont {Jeong}}]{go_exploring_2023}%
  \BibitemOpen
  \bibfield  {author} {\bibinfo {author} {\bibfnamefont {B.}~\bibnamefont {Go}}, \bibinfo {author} {\bibfnamefont {C.}~\bibnamefont {Oh}}, \bibinfo {author} {\bibfnamefont {L.}~\bibnamefont {Jiang}},\ and\ \bibinfo {author} {\bibfnamefont {H.}~\bibnamefont {Jeong}},\ }\href {https://doi.org/10.1103/PhysRevA.109.052613} {\bibfield  {journal} {\bibinfo  {journal} {Phys. Rev. A}\ }\textbf {\bibinfo {volume} {109}},\ \bibinfo {pages} {052613} (\bibinfo {year} {2024})}\BibitemShut {NoStop}%
\bibitem [{\citenamefont {Lvovsky}(2015)}]{lvovsky_squeezed_2015}%
  \BibitemOpen
  \bibfield  {author} {\bibinfo {author} {\bibfnamefont {A.~I.}\ \bibnamefont {Lvovsky}},\ }in\ \href {https://doi.org/10.1002/9781119009719.ch5} {\emph {\bibinfo {booktitle} {Photonics}}}\ (\bibinfo  {publisher} {John Wiley \& Sons, Ltd},\ \bibinfo {year} {2015})\ pp.\ \bibinfo {pages} {121--163}\BibitemShut {NoStop}%
\bibitem [{\citenamefont {Skorski}(2023)}]{skorski2023Bernsteintype}%
  \BibitemOpen
  \bibfield  {author} {\bibinfo {author} {\bibfnamefont {M.}~\bibnamefont {Skorski}},\ }\href {https://doi.org/10.15559/23-VMSTA223} {\bibfield  {journal} {\bibinfo  {journal} {Modern Stochastics: Theory and Applications}\ }\textbf {\bibinfo {volume} {10}},\ \bibinfo {pages} {211} (\bibinfo {year} {2023})}\BibitemShut {NoStop}%
\bibitem [{\citenamefont {Braunstein}\ and\ \citenamefont {van Loock}(2005)}]{braunstein_quantum_2005}%
  \BibitemOpen
  \bibfield  {author} {\bibinfo {author} {\bibfnamefont {S.~L.}\ \bibnamefont {Braunstein}}\ and\ \bibinfo {author} {\bibfnamefont {P.}~\bibnamefont {van Loock}},\ }\href {https://doi.org/10.1103/RevModPhys.77.513} {\bibfield  {journal} {\bibinfo  {journal} {Rev. Mod. Phys.}\ }\textbf {\bibinfo {volume} {77}},\ \bibinfo {pages} {513} (\bibinfo {year} {2005})}\BibitemShut {NoStop}%
\end{thebibliography}%
\end{document}